\documentclass[12pt]{iopart}
\usepackage{amsmath,amsthm,amsfonts,amssymb}
\usepackage{bm,epsfig,revsymb,url}
\usepackage{colonequals}
\usepackage{cite}
\usepackage[colorlinks=true,dvipdfm]{hyperref}

\begin{document}

\title[Control of quantum phenomena: Past, present, and
  future]{Control of quantum phenomena: Past, present, and future}

\author{Constantin~Brif, Raj~Chakrabarti\footnote{Present address:
    School of Chemical Engineering, Purdue University, Forney Hall of
    Chemical Engineering, 480 Stadium Mall Drive, West Lafayette,
    Indiana 47907} and Herschel~Rabitz}

\address{Department of Chemistry, Princeton University, Princeton,
  New Jersey 08544}

\eads{\mailto{cbrif@princeton.edu}, \mailto{rchakra@purdue.edu} and
  \mailto{hrabitz@princeton.edu}}

\date{\today}

\begin{abstract}
Quantum control is concerned with active manipulation of physical and
chemical processes on the atomic and molecular scale.
This work presents a perspective of progress in the field of control
over quantum phenomena, tracing the evolution of theoretical concepts
and experimental methods from early developments to the most recent
advances. Among numerous theoretical insights and technological
improvements that produced the present state-of-the-art in quantum
control, there have been several breakthroughs of foremost
importance. On the technology side, the current experimental successes
would be impossible without the development of intense femtosecond
laser sources and pulse shapers. On the theory side, the two most
critical insights were (1) realizing that ultrafast atomic and
molecular dynamics can be controlled via manipulation of quantum
interferences and (2) understanding that optimally shaped ultrafast
laser pulses are the most effective means for producing the desired
quantum interference patterns in the controlled system. Finally, these
theoretical and experimental advances were brought together by the
crucial concept of adaptive feedback control, which is a laboratory
procedure employing measurement-driven, closed-loop optimization to
identify the best shapes of femtosecond laser control pulses for
steering quantum dynamics towards the desired objective.  Optimization
in adaptive feedback control experiments is guided by a learning
algorithm, with stochastic methods proving to be especially effective.
Adaptive feedback control of quantum phenomena has found numerous
applications in many areas of the physical and chemical sciences, and
this paper reviews the extensive experiments. Other subjects discussed
include quantum optimal control theory, quantum control landscapes,
the role of theoretical control designs in experimental realizations,
and real-time quantum feedback control. The paper concludes with a
prospective of open research directions that are likely to attract
significant attention in the future.
\end{abstract}


\submitto{\NJP}

\maketitle


\section{Introduction}
\label{sec:Intro}

For many decades, physicists and chemists have employed various
spectroscopic methods to carefully observe quantum systems on the
atomic and molecular scale. The fascinating feature of quantum control
is the ability to not just observe but actively manipulate the course
of physical and chemical processes, thereby providing hitherto
unattainable means to explore quantum dynamics. This remarkable
capability along with a multitude of possible practical applications
have attracted enormous attention to the field of control over quantum
phenomena. This area of research has experienced extensive development
during the last two decades and continues to grow rapidly. A notable
feature of this development is the fruitful interplay between
theoretical and experimental advances. 

Various theoretical and experimental aspects of quantum control have
been reviewed in a number of articles and books
\cite{BrumerShapiro1992, WarrenRabitzDahleh1993, KohlerKrause1995,
  Kawashima1995, Weiner1995, Manz1996, GordonRice1997,
  ShapiroBrumer1997, GaspardBurghardt1997, Bergmann1998, RiceZhao2000,
  Weiner2000, Rabitz2000, RabitzZhu2000, Brixner2001, Vitanov2001,
  BrownRabitz2002, LevisRabitz2002, WeinachtBucksbaum2002,
  Bandrauk2002, BrumerShapiro2003, BrifRabitz2003, Rabitz2003,
  ShapiroBrumer2003, Brixner2003, BrixnerGerber2003, Goswami2003,
  WalmsleyRabitz2003, BrixnerGerber2004, MarcosLozovoy2004,
  BrixnerPfeifer2005, Carley2005, Wollenhaupt2005, Wohlleben2005,
  PfeiferSpielmann2006, DAlessandro2007, Nuernberger2007,
  WerschnikGross2007, ChakrabartiRabitz2007review, BalintKurti2008,
  Winterfeldt2008, Ohmori2009, Rego2009, Silberberg2009,
  KrauszIvanov2009}. This paper starts with a short review of
historical developments as a basis for evaluating the current status
of the field and forecasting future directions of research. We try to
identify important trends, follow their evolution from the past
through the present, and cautiously project them into the future. This
paper is not intended to be a complete review of quantum control, but
rather a perspective and prospective on the field. 

In section~\ref{sec:Early}, we discuss the historical evolution of
relevant key ideas from the first attempts to use monochromatic laser
fields for selective excitation of molecular bonds, through the
inception of the crucial concept of control via manipulation of
quantum interferences, and to the emergence of advanced contemporary
methods that employ specially tailored ultrafast laser pulses to
control quantum dynamics of a wide variety of physical and chemical
systems in a precise and effective manner. After this historical
summary, we review in more detail the recent progress in the field,
focusing on significant theoretical concepts, experimental methods,
and practical advances that have shaped the development of quantum
control during the last decade. Section~\ref{sec:OCT} is devoted to
quantum optimal control theory (QOCT), which is currently the leading
theoretical approach for identifying the structure of controls (e.g.,
the shape of laser pulses) that enable attaining the quantum dynamical
objective in the best possible way. We present the formalism of QOCT
(i.e., the types of objective functionals used in various problems and
methods employed to search for optimal controls), consider the issues
of controllability and existence of optimal control solutions, survey
applications, and discuss the advantages and limitations of this
approach. In section~\ref{sec:Landscapes}, we review the theory of
quantum control landscapes, which provides a basis to analyze the
complexity of finding optimal solutions. Topics discussed in that
section include the landscape topology (i.e., the characterization of
critical points), optimality conditions for control solutions, Pareto
optimality for multi-objective control, homotopy trajectory control
methods, and the practical implications of control landscape analysis.
The important theoretical advances in the field of quantum control
have laid the foundation for the fascinating discoveries occurring in
laboratories where closed-loop optimizations guided by learning
algorithms alter quantum dynamics of real physical and chemical
systems in dramatic and often unexpected way.  Section~\ref{sec:AFC},
which constitutes a very significant portion of this paper, is devoted
to laboratory implementations of adaptive feedback control (AFC) of
quantum phenomena. We review numerous AFC experiments that have been
performed during the last decade in areas ranging from photochemistry
to quantum information sciences. These experimental studies (most of
which employ shaped femtosecond laser pulses) clearly demonstrate the
capability of AFC to manipulate dynamics of a broad variety of quantum
systems and explore the underlying physical mechanisms. The role of
theoretical control designs in experimental realizations is discussed
in section~\ref{sec:Designs}. In particular, we emphasize the
importance of theoretical studies for the feasibility analysis of
quantum control experiments.
Section~\ref{sec:RTFC} presents concepts and potential applications of
real-time feedback control (RTFC). Both measurement-based and coherent
types of RTFC are described, along with current technological
obstacles limiting more extensive use of these approaches in the
laboratory. Future directions of quantum control are considered in
section~\ref{sec:Future}, including important unsolved problems and
some emerging new trends and applications. Finally, concluding remarks
are given in section~\ref{sec:Concl}.


\section{Early developments of quantum control}
\label{sec:Early}

The historical origins of quantum control lie in early attempts to use
lasers for manipulation of chemical reactions, in particular,
selective breaking of bonds in molecules. Lasers, with their tight
frequency control and high intensity, were considered ideal for the
role of molecular-scale `scissors' to precisely cut an identified
bond, without damage to others. In the 1960s, when the remarkable
characteristics of lasers were initially realized, it was thought that
transforming this dream into reality would be relatively simple. These
hopes were based on intuitive, appealing logic. The procedure involved
tuning the monochromatic laser radiation to the characteristic
frequency of a particular chemical bond in a molecule. It was
suggested that the energy of the laser would naturally be absorbed in
a selective way, causing excitation and, ultimately, breakage of the
targeted bond. Numerous attempts were made in the 1970s to implement
this idea \cite{Letokhov1977, BloemYabl1978, Zewail1980}.  However, it
was soon realized that intramolecular vibrational redistribution of
the deposited energy rapidly dissipates the initial local excitation
and thus generally prevents selective bond breaking
\cite{BloemZewail1984, ElsaesserKaiser1991, Zewail1996}. This process
effectively increases the rovibrational temperature in the molecule in
the same manner as incoherent heating does, often resulting in
breakage of the weakest bond(s), which is usually not the target of
interest.

\subsection{Control via two-pathway quantum interference}
\label{sec:Early-TPI}

Several important steps towards modern quantum control were made in
the late 1980s. Brumer and Shapiro \cite{BrumerShapiro1986a,
  BrumerShapiro1986b, Shapiro1988, BrumerShapiro1989} identified the
role of quantum interference in optical control of molecular
systems. They proposed to use two monochromatic laser beams with
commensurate frequencies and tunable intensities and phases for
creating quantum interference between two reaction pathways. The
theoretical analysis showed that by tuning the phase difference
between the two laser fields it would be possible to control branching
ratios of molecular reactions \cite{ChanBrumerShapiro1991,
  ChenBrumerShapiro1993, Lee1998}. The method of two-pathway quantum
interference can be also used for controlling population transfer
between bound states \cite{ChenYinElliott1990, ChenElliott1990} (in
this case, the number of photons absorbed along two pathways often
must be either all even or all odd to ensure that the wave functions
excited by the two lasers have the same parity; most commonly, one-
and three-photon excitations were considered).

The principle of coherent control via two-pathway quantum interference
was demonstrated during the 1990s in a number of experiments,
including control of population transfer in bound-to-bound transitions
in atoms and molecules \cite{ChenYinElliott1990, ChenElliott1990,
  ParkLuGordon1991, LuParkXieGordon1992, XingBersohn1996,
  WangBersohn1996}, control of energy and angular distributions of
photoionized electrons \cite{MullerBucksbaum1990,
  SchumacherBucksbaum1994, YinChenElliottSmith1992,
  YinElliottShehadehGrant1995} and photodissociation products
\cite{Sheehy1995} in bound-to-continuum transitions, control of
cross-sections of photochemical reactions
\cite{KleimanZhuLiGordon1995, KleimanZhuAllenGordon1995,
  ZhuKleimanLiLuTrentelmanGordon1995}, and control of photocurrents in
semiconductors \cite{DupontCorkum1995, Hache1997}. However, practical
applications of this method are limited by a number of factors. In
particular, it is quite difficult in practice to match excitation
rates along the two pathways, either because one of the absorption
cross-sections is very small or because other competing processes
intervene. Another practical limitation, characteristic of experiments
in optically dense media, is undesirable phase and amplitude locking
of the two laser fields \cite{ChenElliott1996}. Due to these factors
and other technical issues (e.g., imperfect focusing and alignment of
the two laser beams), modulation depths achieved in two-pathway
interference experiments were modest: typically, about 25--50\% for
control of population transfer between bound states
\cite{ChenElliott1990, ParkLuGordon1991, LuParkXieGordon1992,
  WangBersohn1996} (the highest reported value was about 75\% in one
experiment \cite{XingBersohn1996}), and about 15--25\% for control of
dissociation and ionization branching ratios in molecules
\cite{KleimanZhuLiGordon1995, KleimanZhuAllenGordon1995}. Two-pathway
interference control is a nascent form of full multi-pathway control
offered by operating with broad-bandwidth optimally shaped pulses.

\subsection{Pump-dump control}
\label{sec:Early-pumpdump}

In the 1980s, Tannor, Kosloff, and Rice \cite{TannorRice1985,
  TannorKosloffRice1986} proposed a method for selectively controlling
intramolecular reactions by using two successive femtosecond laser
pulses with a tunable time delay between them. The first laser pulse
(the ``pump'') generates a vibrational wave packet on an
electronically excited potential-energy surface of the molecule. After
the initial excitation, the wave packet evolves freely until the
second laser pulse (the ``dump'') transfers some of the population
back to the ground potential-energy surface into the desired reaction
channel. Reaction selectivity is achieved by using the time delay
between the two laser pulses to control the location at which the
excited wave packet is dumped to the ground potential-energy surface
\cite{GordonRice1997, RiceZhao2000}. For example, it may be possible
to use this method to move the ground-state wave-function beyond a
barrier obstructing the target reaction channel. In some cases, the
second pulse transfers the population to an electronic state other
than the ground state (e.g., to a higher excited state) in a
pump-repump scheme. 

The feasibility of the pump-dump control method was demonstrated in a
number of experiments \cite{Baumert1991a, Baumert1991b,
  BaumertGerber1994, Potter1992, Herek1994}. The pump-dump scheme can
be also used as a time-resolved spectroscopy technique to explore
transient molecular states and thus obtain new information about the
dynamics of the molecule at various stages of a reaction
\cite{GaiMcDonaldAnfinrud1997, Logunov2001, Ruhman2002, Larsen2004a,
  Larsen2004b, Larsen2005, Vengris2005, Vengris2004}. In pump-dump
control experiments, the system dynamics often can be explained in the
time domain in a simple and intuitive way to provide a satisfactory
qualitative interpretation of the control mechanism. The pump-dump
method gained considerable popularity \cite{GordonRice1997,
  RiceZhao2000, MarcosLozovoy2004} due to its capabilities to control
and investigate molecular dynamics. However, the employment of
transform-limited laser pulses significantly restricts the
effectiveness of this technique as a practical control tool. More
effective control of the wave-packet dynamics and, consequently,
higher reaction selectivity can be achieved by optimally shaping one
or both of the pulses. For example, even a chirp of the pump pulse may
improve the effectiveness of control by producing more localized
wave packets (the use of pulse chirping will be discussed in
section~\ref{sec:Early-chirp} in more detail). Recent experimental
applications of the pump-dump scheme with shaped laser pulses
(optimized using adaptive methods) will be discussed in
section~\ref{sec:AFC}.

\subsection{Control via stimulated Raman adiabatic passage}
\label{sec:Early-STIRAP}

In the late 1980s, Bergmann and collaborators \cite{Gaubatz1988,
  Kuklinski1989, Gaubatz1990, ShoreBergmann1991} demonstrated a very
efficient adiabatic method for population transfer between discrete
quantum states in atoms or molecules. In this approach known as
stimulated Raman adiabatic passage (STIRAP), two time-delayed laser
pulses (typically, of nanosecond duration) are applied to a
three-level $\Lambda$-type configuration to achieve complete
population transfer between the two lower levels via the intermediate
upper level.  Interestingly, the pulse sequence employed in the STIRAP
method is counter-intuitive, i.e., the Stokes laser pulse that couples
the intermediate and final states precedes (but overlaps) the pump
laser pulse that couples the initial and intermediate states. The
laser electric fields should be sufficiently strong to generate many
cycles of Rabi oscillations. The laser-induced coherence between the
quantum states is controlled by tuning the time delay, so that the
transient population in the intermediate state remains almost zero,
thus avoiding losses by radiative decay. Detailed reviews of STIRAP
and related adiabatic passage techniques can be found in
\cite{Bergmann1998, Vitanov2001}. While the efficiency of the STIRAP
method, under appropriate conditions, is very high, its applicability
is restricted to control of population transfer between a few discrete
states as arise in atoms and small (diatomic and triatomic)
molecules. In larger polyatomic molecules, the very high density of
levels 
generally prevents successful adiabatic passage \cite{Bergmann1998,
  Vitanov2001}. 

\subsection{Control via wave-packet interferometry}
\label{sec:Early-WPI}

Another two-pulse approach for control of population transfer between
bound states employs Ramsey interference of optically excited wave
packets \cite{Salour1977, Teets1977}. In this method, referred to as
wave-packet interferometry (WPI) \cite{Ohmori2009}, two time-delayed
laser pulses excite an atomic, molecular, or quantum-dot transition,
resulting in two wave packets on an excited state. Quantum
interference between the two coherent wave packets can be controlled
by tuning the time delay between the laser pulses. For control of
population transfer, constructive or destructive interference between
the excited wave packets gives rise to larger or smaller excited-state
population, respectively. The same control mechanism is also
applicable to other problems such as control of atomic radial
wave-functions and control of molecular alignment. WPI was
demonstrated with Rydberg \cite{Noordam1992, Jones1993, Jones1995a}
and fine-structure \cite{Blanchet1997, Bouchene1998} wave packets in
atoms, vibrational \cite{Scherer1991, Scherer1992, Blanchet1998,
  Doule2001, Ohmori2003} and rotational \cite{Hertz2000} wave packets
in molecules, and exciton fine-structure wave packets in semiconductor
quantum dots \cite{Bonadeo1998, Flissikowski2004} (for a detailed
review of coherent control applications of WPI, see \cite{Ohmori2009};
the use of WPI for molecular state reconstruction is reviewed in
\cite{Cina2008}). Once again, much more effective manipulation of
quantum interferences is possible in this control scheme when shaped
laser pulses are used instead of transform-limited ones (see
section~\ref{sec:AFC} for details).

\subsection{Introduction of QOCT}
\label{sec:Early-QOCT}

Although the control approaches discussed in
sections~\ref{sec:Early-TPI}--\ref{sec:Early-WPI} were initially
perceived as quite different, it is now clear that on a fundamental
level all of them employ the mechanism of quantum interference induced
by control laser fields. A common feature of these methods is that
they generally attempt to manipulate the evolution of quantum systems
by controlling just one parameter: the phase difference between two
laser fields in control via two-pathway quantum interference; the time
delay between two laser pulses in pump-dump control, STIRAP, and
WPI. While single-parameter control may be relatively effective in
some simple systems, more complex systems and applications
require more flexible and capable control resources. The
single-parameter control schemes have been unified and generalized by
the concept of control with specially tailored ultrashort laser
pulses. Rabitz and co-workers \cite{Shi1988, Peirce1988,
  ShiRabitz1989} and others \cite{Kosloff1989, Jakubetz1990} suggested
that it would be possible to steer the quantum evolution to a desired
product channel by specifically designing and tailoring the
time-dependent electric field of the laser pulse to the
characteristics of the system.  Specifically, QOCT may be used to
design laser pulse shapes which are best suited for achieving the
desired goal \cite{Shi1988, Peirce1988, ShiRabitz1989, Kosloff1989,
  Jakubetz1990, ShiRabitz1990a, ShiRabitz1990b, Dahleh1990,
  ShiRabitz1991, Gross1991, Kaluza1994, Sugawara1994}. An optimally
shaped laser pulse typically has a complex form, both temporally and
spectrally. The phases and amplitudes of different frequency
components are optimized to excite an interference pattern amongst
distinct quantum pathways, to best achieve the desired dynamics. The
first optimal fields for quantum control were computed by Shi, Woody,
and Rabitz \cite{Shi1988} who showed that the amplitudes of the
interfering vibrational modes of a laser-driven molecule could add up
constructively in a given bond. We will review QOCT and its
applications in more detail in section~\ref{sec:OCT} (for earlier
reviews of QOCT, see \cite{GordonRice1997, RiceZhao2000,
  RabitzZhu2000, WerschnikGross2007, BalintKurti2008}).

\subsection{Control with linearly chirped pulses}
\label{sec:Early-chirp}

Laser pulse-shaping technology rapidly developed during the early
1990s \cite{Kawashima1995, Weiner1995, Weiner2000}. However, the
capabilities of pulse shaping were not fully exploited in quantum
control until the first experimental demonstrations of adaptive
feedback control (AFC) in 1997--1998 \cite{BardeenYakovlevWilson1997,
  Assion1998}. Initially, ultrashort laser pulses with time-varying
photon frequencies were used to tune just the linear chirp, which
represents an increase or decrease of the instantaneous frequency as a
function of time under the pulse envelope.\footnote{The instantaneous
  frequency $\omega(t)$ of a linearly chirped pulse with a carrier
  frequency $\omega_0$ is given at time $t$ by $\omega(t) = \omega_0 +
  2 b t$, where $b$ is the chirp parameter that can be negative or
  positive.} Linearly chirped femtosecond laser pulses were
successfully applied for control of various atomic and molecular
processes, including control of vibrational wave packets
\cite{BardeenWangShank1995, Kohler1995, BardeenCheWilson1997a,
  BardeenCheWilson1997b, BardeenWangShank1998, Misawa2000,
  Malkmus2005}, control of population transfer between atomic states
\cite{Melinger1992, Broers1992, Balling1994} and between molecular
vibrational levels \cite{KleimanArrivo1998, Witte2003, Witte2004} via
``ladder-climbing'' processes, control of electronic excitations in
molecules \cite{AssionBaumert1996, CerulloBardeen1996,
  BardeenYakovlev1998, Brakenhoff1999, VogtNuernbergerSelle2006},
selective excitation of vibrational modes in coherent anti-Stokes
Raman scattering (CARS) \cite{ChenMaterny2000}, improvement of the
resolution of CARS spectroscopy \cite{Hellerer2004, Knutsen2004}, and
control of photoelectron spectra \cite{WollenhauptPrakelt2006} and
transitions through multiple highly excited states
\cite{KrugBayerWollenhaupt2009} in strong-field ionization of atoms.
In particular, when the emission and absorption bands of a molecule
strongly overlap, pulses with negative and positive chirp excite
vibrational modes predominately in the ground and excited electronic
states, respectively \cite{BardeenWangShank1995, BardeenWangShank1998,
  Misawa2000, Malkmus2005}. Chirped pulses can be also used to control
the localization of vibrational wave packets in diatomic molecules,
with the negative and positive chirp increasing and decreasing the
localization, respectively \cite{Kohler1995, BardeenCheWilson1997a,
  BardeenCheWilson1997b}. Based on this effect, pump pulses with
negative chirp were used to enhance selectivity in pump-dump control
of photodissociation reactions \cite{BardeenCheWilson1997b}. Recently,
the localization effect of negatively chirped pulses was used to
protect vibrational wave packets against rotationally-induced
decoherence \cite{Branderhorst2008}. Due to their effectiveness in
various applications, chirped laser pulses are widely used in quantum
control. However, by the end of the 1990s, many experimenters realized
that more sophisticated pulse shapes, beyond just linear chirp,
provide a much more powerful and flexible tool for control of quantum
phenomena in complex physical and chemical systems. Femtosecond
pulse-shaping technology is utilized to the fullest extent in AFC
experiments where laser pulses are optimally tailored to meet the
needs of complex quantum dynamics objectives \cite{Rabitz2000,
  Brixner2001, LevisRabitz2002, WeinachtBucksbaum2002, BrifRabitz2003,
  Brixner2003, BrixnerGerber2003, Goswami2003, WalmsleyRabitz2003,
  BrixnerGerber2004, BrixnerPfeifer2005, Nuernberger2007,
  Winterfeldt2008}. The enormous growth of this field during the last
decade is reviewed in section~\ref{sec:AFC}.

\subsection{Control via non-resonant dynamic Stark effect}
\label{sec:Early-Stark}

Optimal control of quantum phenomena in atoms and molecules usually
operates at laser intensities sufficient to be in the non-perturbative
regime. Thus, controlled dynamics will naturally utilize the dynamic
Stark shift amongst other available physical processes in order to
reach the target. In a recent quantum control development, Stolow and
co-workers proposed and experimentally demonstrated manipulation of
molecular processes exclusively employing the non-resonant dynamic
Stark effect (NRDSE) \cite{UnderwoodSpannerIvanov2003PRL,
  SussmanUnderwood2006PRA, SussmanIvanovStolow2005PRA,
  SussmanTownsendIvanov2006S}. In this approach, a quantum system is
controlled by an infrared laser pulse in the intermediate
field-strength regime (non-perturbative but non-ionizing). Laser
frequency and intensity are chosen to eliminate the complex competing
processes (e.g., multiphoton resonances and strong-field ionization),
so that only the NRDSE contributes to the control mechanism. By
utilizing Raman coupling, control via NRDSE reversibly modifies the
effective Hamiltonian during system evolution, thus making it possible
to affect the course of intramolecular dynamic processes. For example,
a suitably timed infrared laser pulse can act as a ``photonic
catalyst'' by reversibly modifying potential energy barriers during a
chemical reaction without inducing any real electronic transitions
\cite{SussmanTownsendIvanov2006S}. Control via NRDSE was successfully
applied to create field-free ``switched'' wave packets (which can be
employed, e.g., for molecular axis alignment)
\cite{UnderwoodSpannerIvanov2003PRL, SussmanUnderwood2006PRA} and
modify branching ratios in non-adiabatic molecular photodissociation
\cite{SussmanIvanovStolow2005PRA, SussmanTownsendIvanov2006S}.

\subsection{Control of nuclear spins with radiofrequency fields}
\label{sec:Early-NMR}

One of the earliest examples of coherent control of quantum dynamics
is manipulation of nuclear spin ensembles using radiofrequency (RF)
fields \cite{Abragam1983}. The main application of nuclear magnetic
resonance (NMR) control techniques is high-resolution spectroscopy of
polyatomic molecules (e.g., protein structure determination)
\cite{Ernst1990, Freeman1998, Levitt2008, Slichter2010}. While control
of an isolated spin by a time-dependent magnetic field is a simple
quantum problem, in reality, NMR spectroscopy of molecules containing
tens or even hundreds of nuclei involves many complex issues such as
the effect of interactions between the spins, thermal relaxation,
instrumental noise, and influence of the solvent. Therefore, modern
NMR spectroscopy often employs thousands of precisely sequenced and
phase-modulated pulses. Among important NMR control techniques are
composite pulses, refocusing, and pulse shaping. In particular, the
use of shaped RF pulses in NMR makes it possible to improve the
frequency selectivity, suppress the solvent contribution, simplify
high-resolution spectra, and reduce the size and duration of
experiments \cite{Freeman1998PNMRS}.  In recent years, NMR became an
important testbed for developing control methods for applications in
quantum information sciences \cite{Cory2000FP, Jones2000FP,
  Jones2001PNMRS, VandersypenChuang2005, RyanNegrevergne2008}. In
order to perform fault-tolerant quantum computations, the system
dynamics must be controlled with an unprecedented level of precision,
which requires even more sophisticated designs of control pulses than
in high-resolution spectroscopy. In particular, QOCT was recently
applied to identify optimal sequences of RF pulses for operation of
NMR quantum information processors \cite{RyanNegrevergne2008,
  KhanejaReiss2005}.


\section{Quantum optimal control theory}
\label{sec:OCT}

The most comprehensive means for coherently controlling the evolution
of a quantum system (e.g., a molecule) undergoing a complex dynamical
process is through the coordinated interaction between the system and
the electromagnetic field whose spectral content and temporal profile
may be continuously altered throughout the process. For a specified
control objective, and with restrictions imposed by many possible
constraints, the time-dependent field required to manipulate the
system in a desired way can be designed using QOCT
\cite{RabitzZhu2000, WerschnikGross2007, BalintKurti2008}. This
general formulation encompasses both the weak and strong field limits
and incorporates as special cases the one-parameter methods such as
control via two-pathway quantum interference and pump-dump control.

\subsection{Description of controlled quantum dynamics}
\label{sec:OCT-dynamics}

Optimal control theory has an extensive history in traditional
engineering applications \cite{Bryson1975, Stengel1994}, but control
of quantum phenomena imposes special features. Consider the time
evolution of a controlled quantum system,
\begin{equation}
\label{schrodinger}
\frac{\rmd}{\rmd t}U(t) = - \frac{\rmi}{\hbar} 
[H_0 - \varepsilon(t) \cdot \mu] U(t), \ \ \ U(0) = I .
\end{equation}
Here, $U(t)$ is the unitary evolution operator of the system at time
$t$, $I$ is the identity operator, $H_0$ is the free Hamiltonian,
$\mu$ is the dipole operator, and $\varepsilon(t)$ is the control
function at time $t$. For the control function, we will use the
notation $\varepsilon(\cdot) \in \mathbb{K}$ (where $\mathbb{K}$ is
the space of locally bounded, sufficiently smooth, square integrable
functions of time defined on some interval $[0,T]$, with $T$ fixed).
Equation (\ref{schrodinger}) adequately describes the coherent quantum
dynamics of a molecular system interacting with a laser electric field
in the dipole approximation or a spin system interacting with a
time-dependent magnetic field. We will consider finite-level quantum
systems and denote the dimension of the system's Hilbert space as
$N$.\footnote{If one is not concerned with the physically irrelevant
  global phase of the evolution operator, the control problem can be
  restricted to the Hamiltonian represented by a traceless Hermitian
  matrix, and $U \in \mathrm{SU}(N)$, where $\mathrm{SU}(N)$ is the
  special unitary group.}

For quantum systems undergoing environmentally-induced decoherence,
there are many dynamical models depending on the character of the
system-environment coupling. If the system and environment are
initially uncoupled, the evolution of the system's reduced density
matrix $\rho$ from $t = 0$ to $t$ is described by a completely
positive, trace preserving map: $\rho(t) = \Phi(\rho_0)$ where $\rho_0
= \rho(0)$. This map (which is often called the Kraus map) can be
expressed using the operator-sum representation (OSR)
\cite{Kraus1983}:
\begin{equation}
\label{eq:Kraus}
\rho(t) = \Phi(\rho_0) = \sum_{j=1}^{n} K_j(t) \rho_0 K_j^{\dag}(t) ,
\end{equation}
where $K_j$ are Kraus operators ($N \times N$ complex matrices), which
satisfy the condition $\sum_{j=1}^n K_j^{\dag} K_j = I_N$, and $I_N$
denotes the identity operator on the Hilbert space of dimension
$N$. There exist infinitely many OSRs (with different sets of Kraus
operators) for the same Kraus map. Various types of quantum master
equations can be derived from the Kraus OSR under additional
assumptions \cite{BreuerPetruccione2002, Gardiner2004}. In particular,
for the generic class of Markovian environments, the dynamics of an
open quantum system can be described using the quantum master equation
of the Lindblad type \cite{BreuerPetruccione2002, Gardiner2004,
  Lindblad1976}:
\begin{equation}
\label{eq:master}
\frac{\rmd}{\rmd t}\rho(t) = 
-\frac{\rmi}{\hbar}[H_0 - \varepsilon(t) \cdot \mu, \rho(t)]
+ \sum_{i=1}^{N^2 - 1} \gamma_i \left[ L_i \rho(t) L_i^{\dag}  
- \frac{1}{2} L_i^{\dag} L_i \rho(t) 
- \frac{1}{2} \rho(t) L_i^{\dag} L_i 
\right] ,
\end{equation}
where $\gamma_i$ are non-negative constants and $L_i$ are Lindblad
operators ($N \times N$ complex matrices) that represent the
non-unitary effect of coupling to the environment. For a closed
system, (\ref{eq:master}) reduces to the von Neumann equation,
$\dot{\rho}(t) = - (\rmi / \hbar) [H_0 - \varepsilon(t) \cdot \mu,
  \rho(t)]$. For simplicity, we will formulate QOCT below using the
Schr\"{o}dinger equation (\ref{schrodinger}) for the unitary evolution
operator; analogous formulations using the von Neumann equation or the
Lindblad master equation for the density matrix are available in the
literature \cite{Gross1991, YanGillilan1993, BartanaKosloffTannor1993,
  BartanaKosloffTannor1997, OhtsukiZhuRabitz1999}.

\subsection{Control objective functionals}
\label{sec:OCT-objectives}

The general class of control objective functionals (cost functionals
of the Bolza type) can be written as
\begin{equation}\label{cost}
J[U(\cdot) , \varepsilon(\cdot)] = F(U(T)) 
+ \int_0^T G(U(t), \varepsilon(t)) \rmd t ,
\end{equation}
where $F$ 
is a continuously differentiable function on $\mathrm{U}(N)$, and $G$
is a continuously differentiable function on $\mathrm{U}(N) \times
\mathbb{R}$. The optimal control problem may be stated as the search
for 
\begin{equation}\label{max}
J_{\mathrm{opt}} = \max_{\varepsilon(\cdot)} 
J [U(\cdot),\varepsilon(\cdot)] ,
\end{equation}
subject to the dynamical constraint (\ref{schrodinger}). If only the
term $\int_0^T G(U(t), \varepsilon(t)) \rmd t$ is present, the cost
functional is said to be of the Lagrange type, whereas if only the
term $F(U(T))$ is present, the functional is said to be of the Mayer
type \cite{Stengel1994}. Three classes of problems corresponding to
different choices of $F(U(T))$ have received the most attention in
the quantum control community to date: (i) evolution-operator control,
(ii) state control, and (iii) observable control. 

For evolution-operator control, the goal is to generate $U(T)$ such
that it is as close as possible to the target unitary transformation
$W$. The Mayer-type cost functional in this case can be generally
expressed as
\begin{equation}
\label{gatefn-general}
F_1(U(T)) = 1 - \|  W - U(T) \| ,
\end{equation}
where $\| \cdot \|$ is an appropriate normalized matrix norm, i.e.,
$F_1(U(T))$ is maximized when the distance between $U(T)$ and $W$ is
minimized.  This type of objective is common in quantum computing
applications \cite{NielsenChuang2000}, where $F_1(U(T))$ is the
fidelity of a quantum gate \cite{PalaoKosloff2002,
  PalaoKosloff2003}. One frequently used form of the objective
functional $F_1(U(T))$ is obtained utilizing the squared Hilbert-Schmidt
norm \cite{HornJohnson1990} in (\ref{gatefn-general}) with an
appropriate normalization (i.e., $\| X \| = (2 N)^{-1} \Tr (X^{\dag}
X)$) \cite{RabitzHsiehRosenthal2005PRA, HsiehRabitz2008PRA,
  HoDominyRabitz2009PRA}:
\begin{equation}
\label{gatefn}
F_1(U(T)) = \frac{1}{N} \Re \Tr\left[ W^{\dag} U(T) \right] .
\end{equation}
Other forms of the objective functional, which employ different matrix
norms in (\ref{gatefn-general}), are possible as well
\cite{HoDominyRabitz2009PRA, Gilchrist2005, KosutGrace2006,
  GraceDominy2010NJP}. For example, a modification of (\ref{gatefn}),
$F_1(U(T)) = N^{-1} \left| \Tr\left[ W^{\dag} U(T) \right] \right|$,
which is independent of the global phase of $U(T)$, can be used. Note
that $F_1(U(T))$ is independent of the initial state, as the quantum
gate must produce the same unitary transformation for any input state
of the qubit system \cite{NielsenChuang2000}.

For state control, the goal is to transform the initial state $\rho_0$
into a final state $\rho(T) = U(T) \rho_0 U^{\dag}(T)$ that is as
close as possible to the target state $\rho_{\mathrm{f}}$. The
corresponding Mayer-type cost functional is
\begin{equation}
\label{statefn}
F_2(U(T)) = 1 - \| U(T)\rho_0 U^{\dag}(T) - \rho_{\mathrm{f}} \| ,
\end{equation}
where $\| \cdot \|$ is an appropriate normalized matrix norm (e.g.,
the Hilbert-Schmidt norm can be used) \cite{DemiralpRabitz1993,
  Jozsa1994, FuchsGraaf1999, JirariPotz2005}.

For observable control, the goal is typically to maximize the
expectation value of a target quantum observable $\Theta$ (represented
by a Hermitian operator). The corresponding Mayer-type cost functional
is \cite{ShiRabitz1990a, RabitzHsiehRosenthal2006JCP,
  ShenHsiehRabitz2006JCP, WuRabitzHsieh2008JPA, HsiehWuRabitz2009JCP}
\begin{equation}
\label{obsfn}
F_3(U(T)) = \Tr\left[ U(T) \rho_0 U^{\dag}(T) \Theta \right] .
\end{equation}
An important special case is state-transition control (also known as
population transfer control), for which $\rho_0 = |\psi_{\mathrm{i}}
\rangle \langle \psi_{\mathrm{i}}|$ and $\Theta = |\psi_{\mathrm{f}}
\rangle \langle \psi_{\mathrm{f}}|$, where $|\psi_{\mathrm{i}}\rangle$
and $|\psi_{\mathrm{f}}\rangle$ are eigenstates of the free
Hamiltonian $H_0$; in this case, the objective functional
(\ref{obsfn}) has the form $F_3(U(T)) = P_{\mathrm{i} \rightarrow
  \mathrm{f}} = |\langle \psi_{\mathrm{f}} | U(T) | \psi_{\mathrm{i}}
\rangle |^2$, which is the probability of transition (i.e., the
population transfer yield) between the energy levels of the quantum
system \cite{RabitzHsiehRosenthal2004, RabitzHoHsieh2006PRA}. In many
chemical and physical applications of quantum control, absolute yields
are not known, and therefore maximizing the expectation value of an
observable (e.g., the population transfer yield) is a more appropriate
laboratory goal than minimizing the distance to a target expectation
value.

Also, in quantum control experiments (see section~\ref{sec:AFC}),
measuring the expectation value of an observable is much easier than
estimating the quantum state or evolution operator. Existing methods
of quantum-state and evolution-operator estimation rely on tomographic
techniques \cite{BertrandBertrand1987, VogelRisken1989, Leonhardt1997,
  Buzek1998, BrifMann1999PRA, BrifMann2000JOB245, Rehacek2008,
  DarianoPresti2001} which are extremely expensive in terms of the
number of required measurements (e.g., in quantum computing
applications, standard methods of state and process tomography require
numbers of measurements that scale exponentially in the number of
qubits \cite{NielsenChuang2000, DarianoPresti2001,
  KosutWalmsleyRabitz2004, Branderhorst2008JPB, Mohseni2008PRA,
  YoungWhaley2009}). Therefore, virtually all quantum control
experiments so far have used observable control with objective
functionals of the form (\ref{obsfn}). For example, in an AFC
experiment \cite{Branderhorst2008}, in which the goal was to maximize
the degree of coherence, the expectation value of an observable
representing the degree of quantum state localization was used as a
coherence ``surrogate,'' instead of state purity or von Neumann
entropy which are nonlinear functions of the density matrix and hence
would require state estimation. Nevertheless, future laboratory
applications of quantum control, in particular in the field of quantum
information sciences, will require evolution-operator control and
state control, with the use of objective functionals of the types
(\ref{gatefn-general}) and (\ref{statefn}), respectively, together
with novel state and process estimation methods \cite{Mohseni2008PRA,
  YoungWhaley2009, EmersonSilva2007, Kosut2008L1norm,
  BranderhorstNunn2009NJP, ShabaniKosutRabitz2009,
  BenderskyPastawskiPaz2008, SchmiegelowLarotondaPaz2010,
  CramerPlenio2010, FlammiaGross2010}.

Recently, attention has turned to problems requiring simultaneous
maximization of several control objectives \cite{OhtsukiNakagami2001,
  Shir2007, RajWu2008a, RajWu2008b, BeltraniGhosh2009}. In the
framework of QOCT, these optimization problems are sometimes handled
through the use of a weighted-sum objective functional, such as
\cite{Shir2007, RajWu2008a}
\begin{equation}
\label{multiobsfn}
F_4(U(T)) = \sum_{k=1}^n \alpha_k
\mathrm{Tr}\left[ U(T) \rho_0 U^{\dag}(T) \Theta_k \right] , 
\end{equation}
which extends (\ref{obsfn}) to multiple quantum observables. Also,
general methods of multi-objective optimization
\cite{ChankongHaimes1983, Steuer1986, Miettinen1998} have been
recently applied to various quantum control problems \cite{RajWu2008a,
  RajWu2008b, BeltraniGhosh2009}.

Another common goal in quantum control  is to maximize a Lagrange-type
cost functional subject to a constraint on $U(T)$
\cite{Jurdjevic1997, DAlessandro2001a}. For example, this type of
control problem can be formulated as follows:
\begin{equation}
\label{canonical}
\max_{\varepsilon(\cdot)} \int_0^T G(\varepsilon(t))~\rmd t,
\ \ \ \text{subject~to}~ F(U(T)) = \chi , 
\end{equation}
where $F(U(T))$ is the Mayer-type cost functional for
evolution-operator, state, or observable control (as described above),
and $\chi$ is a constant that corresponds to the target value of
$F$. Often, the goal is to minimize the total field fluence, in which
case $G(\varepsilon(t)) = -\frac{1}{2}\varepsilon^2(t)$ is used.

\subsection{Controllability of quantum systems}
\label{sec:OCT-controllability}

One of the fundamental issues of quantum control is to assess the
system's controllability. A quantum system is controllable in a set of
configurations, $\mathcal{S} = \{ \zeta \}$, if for any pair of
configurations $\zeta_1, \zeta_2 \in \mathcal{S}$ there exists a
time-dependent control $\varepsilon(\cdot)$ that can drive the system
from the initial configuration $\zeta_1$ to the final configuration
$\zeta_2$ in a finite time $T$. Here, the notion of configuration
means either the state of the system $\rho$, the expectation value of
an observable $\Tr(\rho \Theta)$, the evolution operator $U$, or the
Kraus map $\Phi$, depending on the specific control
problem. Controllability of closed quantum systems with unitary
dynamics has been well studied \cite{HuangTarnClark1983,
  Ramakrishna1995, TuriniciRabitz2001, TuriniciRabitz2003,
  Albertini2001, FuSchirmerSolomon2001, SchirmerFuSolomon2001,
  Altafini2002, GirardeauKoch1998, SchirmerLeahy2001,
  SchirmerSolomon2002a, SchirmerSolomon2002b, Albertini2003,
  ShahTannorRice2002PRA, GongRice2004PRA, SchirmerPullen2005,
  TuriniciRabitz2010JPA, ClarkLucarelliTarn2003, WuTarnLi2006,
  VilelaMendesManko2010}. Controllability analysis was also extended
to open quantum systems \cite{LloydViola2001,
  SolomonSchirmer2004eprint, Altafini2003JMP, Altafini2004PRA,
  Romano2005, WuPechenBrif2007, VilelaMendes2009, DirrHelmke2009}. 

Controllability is determined by the equation of motion as well as
properties of the Hamiltonian. For a closed quantum system with
unitary dynamics (\ref{schrodinger}), evolution-operator
controllability implies that for any unitary operator $W$ there exists
a finite time $T$ and a control $\varepsilon(\cdot)$, such that $W =
U(T)$, where $U(T)$ is the solution of~(\ref{schrodinger}). For an
$N$-level closed system, a necessary and sufficient condition for
evolution-operator controllability is that the dynamical Lie group
$\mathfrak{G}$ of the system (i.e., the Lie group generated by the
system's Hamiltonian) be $\mathrm{U}(N)$ (or $\mathrm{SU}(N)$ for a
traceless Hamiltonian) \cite{SchirmerSolomon2002a,
  SchirmerSolomon2002b, Albertini2003}.

Unitary evolution preserves the spectrum of the quantum state (i.e.,
the eigenvalues of the density matrix). All density matrices that have
the same eigenvalues form a set of unitarily equivalent states (e.g.,
the set of all pure states). Therefore, under unitary evolution, a
quantum system can be state controllable only within a set of
unitarily equivalent states \cite{SchirmerSolomon2002a,
  SchirmerSolomon2002b}. Density-matrix controllability means that for
any pair of unitarily equivalent density matrices $\rho_1$ and
$\rho_2$ there exists a control $\varepsilon(\cdot)$ that drives
$\rho_1$ into $\rho_2$ (in a finite time). It has been shown
\cite{SchirmerSolomon2002a, SchirmerSolomon2002b, Albertini2003} that
density-matrix controllability is equivalent to evolution-operator
controllability. For specific classes of density matrices, the
requirements for controllability are weaker \cite{Albertini2001,
  FuSchirmerSolomon2001, SchirmerFuSolomon2001}. For example,
pure-state controllability requires that the system's dynamical Lie
group $\mathfrak{G}$ is transitive on the sphere $\mathbb{S}^{2N-1}$.
For infinite-level quantum systems evolving on non-compact Lie groups,
such as those arising in quantum optics, the conditions for
controllability are more stringent \cite{WuTarnLi2006,
  VilelaMendesManko2010, VilelaMendes2009, WuRaj2008}.

\subsection{Searching for optimal controls}
\label{sec:OCT-formalism}

To solve for optimal controls that maximize an objective functional
(of the types discussed in section~\ref{sec:OCT-objectives}), it is
convenient to define a functional $\tilde{J}$ that explicitly
incorporates the dynamical constraint (\ref{schrodinger}):
\begin{eqnarray}
\label{jbar}
\fl \tilde{J}[U(\cdot) , \phi(\cdot) , \varepsilon(\cdot)] 
& = & F(U(T)) 
+ \lambda \int_0^T  G(U(t) , \varepsilon(t)) \rmd t \nonumber \\ 
& & - 2 \Re \int_0^T \Tr\left\{ \phi^{\dag}(t) 
\left[ \frac{\rmd U(t)}{\rmd t} + \frac{\rmi}{\hbar}
  \left( H_0-\varepsilon(t)\cdot\mu \right) U(t) \right] 
\right\} \rmd t .
\end{eqnarray}
Here, $\lambda$ is a scalar weight and an auxiliary operator $\phi(t)$
is a Lagrange multiplier employed to enforce satisfaction of
Eq.~(\ref{schrodinger}).

Various modifications of the objective functional (\ref{jbar}) are
possible, for example, QOCT can be formulated for open systems with
non-unitary dynamics \cite{Gross1991, YanGillilan1993,
  BartanaKosloffTannor1993, BartanaKosloffTannor1997,
  OhtsukiZhuRabitz1999, OhtsukiNakagami2003, XuYanOhtsuki2004,
  BeyversSaalfrank2008, CuiXiPan2008PRA, MohseniRezakhani2009PRA}.
Modified objective functionals can also comprise additional spectral
and fluence constraints on the control field \cite{WerschnikGross2005,
  LapertTehini2009}, take into account nonlinear interactions with the
control field \cite{LapertTehini2008, OhtsukiNakagami2008}, deal with
time-dependent and time-averaged targets \cite{BeyversSaalfrank2008,
  SerbanWerschnikGross2005, KaiserMay2004, GrigorenkoGarcia2002}, and
include the final time as a free control parameter
\cite{MishimaYamashita2009a, MishimaYamashita2009b}. It is also
possible to formulate QOCT with time minimization as a control goal
(time optimal control) \cite{Khaneja2001, Khaneja2002,
  ReissKhaneja2002, YuanKhaneja2005}. As we mentioned earlier, QOCT
can be also extended to incorporate optimization of multiple
objectives \cite{OhtsukiNakagami2001, Shir2007, RajWu2008a,
  RajWu2008b, BeltraniGhosh2009}.

A necessary condition for a solution of the optimization problem
(\ref{max}) subject to the dynamical constraint (\ref{schrodinger}) is
that the first-order functional derivatives of $\tilde{J}$ with
respect to $U(\cdot)$, $\phi(\cdot)$, and $\varepsilon(\cdot)$ are
equal to zero. Correspondingly, optimal controls can be obtained by
solving the resulting Euler-Lagrange equations. Equivalently, optimal
controls can be derived through application of the Pontryagin maximum
principle (PMP) \cite{Jurdjevic1997, DAlessandro2001a,
  Bryson1975}. Satisfaction of the first-order conditions following
from the PMP is a necessary but not sufficient condition for
optimality of a control $\varepsilon(\cdot)$. So-called Legendre
conditions on the Hessian, which depend on the type of cost, are also
required for optimality \cite{Bryson1975, Stengel1994}.

An important issue is the existence of optimal control fields (i.e.,
maxima of the objective functional) for realistic situations that
involve practical constraints on the applied laser fields. It is
important to distinguish between the existence of an optimal control
field and controllability; in the former case, a field is designed,
subject to particular constraints, that guides the evolution of the
system towards a specified target until a maximum of the objective
functional is reached, while in the latter case, the exact coincidence
between the attained evolution operator (or state) and the target
evolution operator (or state) is sought. The existence of optimal
controls for quantum systems was analyzed in a number of works.
Peirce~\emph{et~al.}~\cite{Peirce1988} proved the existence of optimal
solutions for state control in a spatially bounded quantum system that
necessarily has spatially localized states and a discrete
spectrum. Zhao and Rice~\cite{ZhaoRice1991} extended this analysis to
a system with both discrete and continuous states and proved the
existence of optimal controls over the evolution in the subspace of
discrete states. Demiralp and Rabitz \cite{DemiralpRabitz1993} showed
that, in general, there is a denumerable infinity of solutions to a
particular class of well-posed quantum control problems; the solutions
can be ordered in quality according to the achieved optimal value of
the objective functional. The existence of multiple control solutions
has important practical consequences, suggesting that there may be
broad latitude in the laboratory, even under strict experimental
restrictions, for finding successful controls for well-posed quantum
objectives. The existence and properties of critical points (including
global extrema) of objective functionals for various types of quantum
control problems were further explored using the analysis of control
landscapes \cite{RabitzHsiehRosenthal2005PRA, HsiehRabitz2008PRA,
  HoDominyRabitz2009PRA, RabitzHsiehRosenthal2006JCP,
  ShenHsiehRabitz2006JCP,  WuRabitzHsieh2008JPA, HsiehWuRabitz2009JCP,
  RabitzHsiehRosenthal2004, RabitzHoHsieh2006PRA,
  WuPechenRabitz2008JMP} (see section~\ref{sec:Landscapes}). 

A number of optimization algorithms were adapted or specially
developed for use in QOCT, including the conjugate gradient search
method \cite{Kosloff1989}, the Krotov method \cite{PalaoKosloff2003,
  TannorKazakovOrlov1992, SomloiKazakovTannor1993}, monotonically
convergent algorithms \cite{ZhuBotinaRabitz1998, ZhuRabitz1998,
  Maday2003, Ohtsuki2004, Ohtsuki2007, BorziSalomon2008,
  DitzBorzi2008}, non-iterative algorithms \cite{ZhuRabitz1999}, the
gradient ascent pulse engineering (GRAPE) algorithm
\cite{KhanejaReiss2005}, a hybrid local/global algorithm
\cite{BeyversSaalfrank2008}, and homotopy-based methods
\cite{Hillermeier2001, RothmanHoRabitz2005JCP,
  RothmanHoRabitz2006PRA}. Faster convergence of iterative QOCT
algorithms was demonstrated using ``mixing'' strategies
\cite{CastroGross2009PRE}. Also, the employment of propagation
toolkits \cite{Yip2003, BalintKurtiManby2005, HsiehRabitz2008PRE}
greatly increases the efficiency of numerical optimizations and allows
for fast combinatorial optimization~\cite{StroheckerRabitz2010JCC}.
Detailed discussions of the QOCT formalism and algorithms are
available in the literature \cite{RiceZhao2000, RabitzZhu2000,
  WerschnikGross2007, BalintKurti2008}.

\subsection{An example of QOCT applied to a molecular system}
\label{sec:OCT-examples}

In order to illustrate optimal control of molecules, we consider an
instructive example. In one of the pioneering QOCT studies, Kosloff
\emph{et al.} \cite{Kosloff1989} considered two electronic states
(ground and excited) of a model molecular system, with the wave
function (in the coordinate representation) of the form\footnote{For
  the sake of notation consistency, the control problem is presented
  here slightly differently than in the original work
  \cite{Kosloff1989}.}
\begin{equation}
\label{eq:Kosloff-1}
\psi(\mathbf{r}, t) = \langle \mathbf{r} | \psi (t) \rangle 
= \left( \begin{array}{c}
\psi_{\mathrm{e}} (\mathbf{r}, t) \\
\psi_{\mathrm{g}} (\mathbf{r}, t)
    \end{array} \right) ,
\end{equation}
where $\psi_{\mathrm{g}}$ and $\psi_{\mathrm{e}}$ are the projections
of the wave function on the ground and excited state,
respectively. The time evolution of the wave function is determined by
the Schr\"{o}dinger equation:
\begin{equation}
\label{eq:Kosloff-2}
\rmi \hbar \frac{\partial}{\partial t} 
\left( \begin{array}{c}
\psi_{\mathrm{e}} (\mathbf{r}, t) \\
\psi_{\mathrm{g}} (\mathbf{r}, t)
    \end{array} \right)
= \left( \begin{array}{cc}
H_{\mathrm{e}} (\mathbf{r}) & H_{\mathrm{g e}} (\mathbf{r}, t) \\
H^{\dagger}_{\mathrm{g e}} (\mathbf{r}, t) & H_{\mathrm{g}} (\mathbf{r})
    \end{array} \right)
\left( \begin{array}{c}
\psi_{\mathrm{e}} (\mathbf{r}, t) \\
\psi_{\mathrm{g}} (\mathbf{r}, t)
    \end{array} \right) ,
\end{equation}
where $H_i (\mathbf{r}) = \mathbf{p}^2/(2 m) + V_i (\mathbf{r})$ ($i =
\mathrm{g}, \mathrm{e}$), $\mathbf{p}$ is the momentum operator,
$V_{\mathrm{g}} (\mathbf{r})$ and $V_{\mathrm{e}} (\mathbf{r})$ are
the adiabatic potential energy surfaces for the ground and excited
state, respectively. The off-diagonal term $H_{\mathrm{g e}}
(\mathbf{r}, t)$ represents the field-induced coupling between the
molecular states:
\begin{equation}
H_{\mathrm{g e}} (\mathbf{r}, t) = - \mu_{\mathrm{g e}}(\mathbf{r}) 
\varepsilon(t) ,
\end{equation}
where $\mu_{\mathrm{g e}}(\mathbf{r})$ is the electric dipole operator
and $\varepsilon(t)$ is the time-dependent electric field of the
control laser pulse applied to the molecule. 

The goal is to control a dissociation reaction in the presence of two
distinct exit channels on the ground potential energy surface. The
corresponding objective functional (including the dynamical
constraint) is given by
\begin{equation}
\label{eq:Kosloff-3}
\tilde{J} = \langle\psi(T)| P |\psi(T)\rangle 
- \lambda \int_0^T \varepsilon^2 (t) \, \rmd t
- 2 \Re \int_0^T \langle\chi(t)| \left(
\frac{\partial}{\partial t} + \frac{\rmi}{\hbar} H \right) 
|\psi(t)\rangle \, \rmd t ,
\end{equation}
The first term in (\ref{eq:Kosloff-3}) represents the main control
goal, where $P$ is the projection operator on the state corresponding
to the target exit channel (i.e., the part of the wave function which
is beyond the target saddle point on the ground-state surface and is
characterized by the outgoing momentum); the second term is used to
manage the fluence of the control field, with $\lambda$ being a scalar
weight factor; the third term includes an auxiliary state
$|\chi(t)\rangle$ that is a Lagrange multiplier employed to enforce
satisfaction of the Schr\"{o}dinger equation ($H$ is the $2 \times 2$
Hamiltonian matrix defined by (\ref{eq:Kosloff-2})). In order to find
the control field that maximizes the objective, the first-order
functional derivatives of $\tilde{J}$ with respect to $\chi(\cdot)$,
$\psi(\cdot)$, and $\varepsilon(\cdot)$ are set to zero, producing the
following Euler-Lagrange equations:
\begin{align}
\label{eq:Kosloff-4a}
& \rmi \hbar \frac{\partial}{\partial t} |\psi(t)\rangle 
= H |\psi(t)\rangle ,
\ \ \ |\psi(0) \rangle = |\psi_0 \rangle , \\
\label{eq:Kosloff-4b}
& \rmi \hbar \frac{\partial}{\partial t} |\chi(t)\rangle 
= H |\chi(t)\rangle ,
\ \ \ |\chi(T) \rangle = P |\psi(T) \rangle , \\
\label{eq:Kosloff-4c}
& \varepsilon(t) = - \frac{1}{\hbar \lambda} 
\Im \left\{ \langle\chi_{\mathrm{g}}(t)| \mu_{\mathrm{g e}} 
|\psi_{\mathrm{e}}(t)\rangle +
\langle\chi_{\mathrm{e}}(t)| \mu_{\mathrm{g e}} 
|\psi_{\mathrm{g}}(t)\rangle \right\} .
\end{align}
An initial guess is selected for the control field (e.g., a pair of
transform-limited pulses with a time delay between them, as in
pump-dump control), and equations
(\ref{eq:Kosloff-4a})--(\ref{eq:Kosloff-4c}) are solved using an
appropriate algorithm, as discussed in section \ref{sec:OCT-formalism}
above. This optimization procedure identifies a shaped control field
$\varepsilon_{\mathrm{opt}}(\cdot)$ that maximizes photoinduced
molecular dissociation into the target channel. Successful application
of QOCT to this model molecular system \cite{Kosloff1989} demonstrated
the benefits of optimally tailoring the time-dependent laser field to
achieve the desired dynamic outcome.

\subsection{Applications of QOCT}
\label{sec:OCT-applications}

Originally, QOCT was developed to design optimal fields for
manipulation of molecular systems~\cite{Shi1988, Peirce1988,
  ShiRabitz1989, Kosloff1989, Jakubetz1990, ShiRabitz1990a,
  ShiRabitz1990b, Dahleh1990, ShiRabitz1991, Gross1991, Kaluza1994,
  Sugawara1994} and has been applied to a myriad of problems (e.g.,
rotational, vibrational, electronic, reactive, and other
processes)~\cite{RiceZhao2000, RabitzZhu2000, BalintKurti2008}. Some
recent applications include, for example, control of molecular
isomerization~\cite{ArtamonovHo2004CP, ArtamonovHo2006CP,
  ArtamonovHo2006JCP, KurosakiArtamonov2009}, control of electron ring
currents in chiral aromatic molecules~\cite{KannoHoki2007}, control of
current flow patterns through molecular wires~\cite{LiWelack2008}, and
control of heterogeneous electron transfer from surface attached
molecules into semiconductor band states~\cite{WangMay2009}. Beyond
molecules, QOCT has been applied to various physical objectives
including, for example, control of electron states in semiconductor
quantum structures~\cite{Kosionis2007, RasanenCastro2007,
  RasanenCastro2008}, control of atom transport in optical
lattices~\cite{ChiaraCalarco2008}, control of Bose-Einstein condensate
transport in magnetic microtraps~\cite{HohenesterRekdal2007}, control
of a transition of ultracold atoms from the superfluid phase to a Mott
insulator state \cite{DoriaCalarcoMontangero2010}, control of coherent
population transfer in superconducting quantum interference
devices~\cite{JirariHekking2009}, and control of the local
electromagnetic response of nanostructured
materials~\cite{GrigorenkoRabitz2009}. Recent interest has rapidly
grown in applications of QOCT to the field of quantum information
sciences, including optimal protection of quantum systems against
decoherence~\cite{JirariPotz2005, HohenesterStadler2004,
  SklarzTannorKhaneja2004, Grigorenko2005a, JirariPotz2006,
  WeninPotz2006, Potz2007JCE, Pelzer2007, Pelzer2008,
  PalaoKosloffKoch2008, CuiXiPan2008PRA, Jirari2009, Gordon2009},
optimal operation of quantum gates in closed systems
\cite{SandersKimHolton1999, DAlessandro2001a, PalaoKosloff2002,
  PalaoKosloff2003, SklarzTannor2004, SklarzTannor2006,
  SchulteSporl2005, SporlSchulte2007, VivieRiedleTroppmann2007,
  WuRaj2008, DominyRabitz2008JPA, SchroderBrown2009JCP, Nebendahl2009,
  Schirmer2009JMO, NigmatullinSchirmer2009, FisherHelmerGlaser2010,
  GollubKowalewski2008, SchroderBrown2009NJP, LiGaitan2010} and in
open systems (i.e., in the presence of decoherence)
\cite{GraceDominy2010NJP, Grigorenko2005b, SchulteSporlKhaneja2006,
  Hohenester2006, MontangeroCalarcoFazio2007PRL, GraceBrif2007JPB,
  GraceBrif2007JMO, WeninPotz2008PRA, WeninPotz2008PRB,
  Rebentrost2009PRL, RebentrostWilhelm2009, MotzoiGambetta2009PRL,
  SafaeiMontangero2009, RoloffPotz2009PRB, WeninRoloffPotz2009,
  RoloffWeninPotz2009JCE, RoloffWeninPotz2009JCTN}, optimal generation
of entanglement \cite{MishimaYamashita2009a, MishimaYamashita2009b,
  RoloffWeninPotz2009JCE, GalveLutz2009, FisherYuan2009,
  WangSchirmer2009PRA}, and optimal (i.e., maximum-rate) transfer of
quantum information \cite{MurphyMontangero2010}. In a recent
experiment with trapped ion qubits, shaped pulses designed using QOCT
were applied to enact single-qubit gates with enhanced robustness to
noise in the control field \cite{TimoneyElman2008}. Optimal control
methods were also applied to the problem of storage and retrieval of
photonic states in atomic media, including both theoretical
optimization \cite{NunnWalmsley2007, GorshkovAndreLukin2007,
  GorshkovCalarcoLukin2008} and experimental tests
\cite{NovikovaGorshkovPhillips2007, NovikovaPhillipsGorshkov2008,
  PhillipsGorshkovNovikova2008}.

\subsection{Advantages and limitations of QOCT}
\label{sec:OCT-limitations}

An advantage of QOCT relative to the laboratory method of AFC (to be
discussed in detail in section~\ref{sec:AFC}) is that the former can
be used to optimize a well defined objective functional of virtually
any form, while the latter relies on information obtained from
measurements and thus is best suited to optimize expectation values of
directly measurable observables. In numerical optimizations, there is
practically no difference in effort between computing the expectation
value of an observable, the density matrix, or the evolution
operator. In the laboratory, however, it is much more difficult to
estimate a quantum state or, even more so, the evolution operator,
than to measure the expectation value of an observable. Moreover,
state estimation error increases rapidly with the Hilbert-space
dimension \cite{Buzek1998,Chakrabarti2009}. The very large number of
measurements required for accurate quantum state/process tomography
\cite{DarianoPresti2001, KosutWalmsleyRabitz2004, Branderhorst2008JPB,
  Mohseni2008PRA} renders (at least, presently) the use of adaptive
laboratory methods for state/evolution-operator control rather
impractical (although not impossible). 

QOCT is often used to explore new quantum phenomena in relatively
simple models to gain physical insights. The realization of quantum
control is ultimately performed in the laboratory. In this context
QOCT fits into what is called open-loop control. Generally, in
open-loop control, a theoretical control design (e.g., obtained by
using QOCT or another theoretical method) is implemented in the
laboratory with the actual system.  Unfortunately, there are not many
problems for which theoretical control designs are directly applicable
in the laboratory. QOCT is most useful when detailed knowledge of the
system's Hamiltonian is available. Moreover, for open quantum systems,
it is essential to know the details of the system-environment
interaction. Therefore, the practical applicability of QOCT in the
context of open-loop control is limited to very simple systems, i.e.,
mostly to cases when a small number of degrees of freedom can be
controlled separately from the remainder of the system. This may be
possible when the controlled subsystem has characteristic frequencies
well separated from those of other transitions, and/or evolves on a
time scale which is very different from that of the rest of the larger
system. A well known example of such a separately controllable
subsystem is nuclear spins in a molecule, which can be very well
controlled using RF fields without disturbing rotational, vibrational,
and electronic degrees of freedom. Another example is a subset of
several discrete levels in an atom or diatomic molecule, the
transitions between which can be controlled in a very precise way
without any significant leakage of population to other
states. However, for a majority of interesting physical and chemical
phenomena, controlled systems are too complex and/or too strongly
coupled to other degrees of freedom. For such complex systems, the
accuracy of control designs obtained using model-based QOCT is usually
inadequate, and hence laboratory AFC is generally the preferred
strategy. In these situations, QOCT may be more useful for feasibility
analysis and exploration of control mechanisms, as basic features of
the controlled dynamics can be identified in many cases even using
relatively rough models. 


\section{Quantum control landscapes}
\label{sec:Landscapes}

An important practical goal of quantum control is the discovery of
optimal solutions for manipulating quantum phenomena. Early studies
\cite{Peirce1988, DemiralpRabitz1993, ZhaoRice1991} described
conditions under which optimal solutions exist, but did not explore
the complexity of finding them. Underlying the search for optimal
controls is the landscape which specifies the physical objective as a
function of the control variables. Analysis of quantum control
landscapes \cite{ChakrabartiRabitz2007review} can not only establish
the existence of optimal control solutions and determine their types
(e.g., global versus local maxima and true maxima versus saddle
points), but also deal with establishing necessary conditions for
convergence of optimization algorithms to global maxima along with
bounds on the scaling of convergence effort. Surprisingly, these
properties are independent of details of a particular Hamiltonian
(provided that the system is controllable), which makes the results of
landscape analysis applicable across a wide range of controlled
quantum phenomena. 

\subsection{Control landscape definition and critical points}
\label{sec:L-formalism}

Properties of the search space associated with Mayer-type cost
functionals play a fundamental role in the ability to identify optimal
controls. To characterize these properties, it is convenient to
express the cost functional in a form where the dynamical constraints
are implicitly satisfied. Consider a control problem with a fixed
target time $T$ for a closed quantum system with unitary
evolution. Denote by $V_T: \varepsilon(\cdot) \mapsto U(T)$ the
endpoint map from the space of control functions to the space of
unitary evolution operators, induced by the Schr\"{o}dinger equation
(\ref{schrodinger}), so that $U(T) = V_T(\varepsilon(\cdot))$. A
Mayer-type cost functional $F(U(T))$ itself describes a map $F$ from
the space of evolution operators to the space of real-valued
costs. Thus the composition of these maps, $J = F \circ V_T
:~\mathbb{K} \rightarrow \mathbb{R}$, is a map from the space of
control functions to the space of real-valued costs. This map
generates the functional $J[\varepsilon(\cdot)] =
F(V_T(\varepsilon(\cdot)))$. We will refer to the functional
$J[\varepsilon(\cdot)]$ as the \emph{control landscape}. The optimal
control problem may then be expressed as the unconstrained search for
\begin{equation}
\label{reducedcost}
J_{\mathrm{opt}} = \max_{\varepsilon(\cdot)} J [\varepsilon(\cdot)] .
\end{equation}
The topology of the control landscape (i.e., the character of its
critical points, including local and global extrema) determines
whether local search algorithms will converge to globally optimal
solutions to the control problem \cite{HoRab2006a}. Studies of quantum
control landscape topology are presently an active research area
\cite{ChakrabartiRabitz2007review, RabitzHsiehRosenthal2005PRA,
  HsiehRabitz2008PRA, HoDominyRabitz2009PRA,
  RabitzHsiehRosenthal2006JCP, WuRabitzHsieh2008JPA,
  HsiehWuRabitz2009JCP, RabitzHsiehRosenthal2004,
  RabitzHoHsieh2006PRA, GirardeauKoch1998, Glaser1998}.

The critical points (extrema) of the landscape are controls, at which
the first-order functional derivative of $J[\varepsilon(\cdot)]$ with
respect to the control field is zero for all time, i.e., 
\begin{equation}
\label{eq:c-point}
\frac{\delta J[\varepsilon(\cdot)]}{\delta \varepsilon(t)} = 0 ,
\ \ \ \forall t \in [0,T] .
\end{equation}
The critical manifold $\mathcal{M}$ of the control landscape is the
collection of all critical points:
\begin{equation}
\label{c-manifold}
\mathcal{M} = \left\{\varepsilon(\cdot) 
~\left| ~ \delta J / \delta \varepsilon(t) = 0 , 
\right. ~\forall t \in [0,T] \right\} .
\end{equation}
A central concept in landscape topology is the classification of a
critical point as regular or singular \cite{BonnardChyba2003,
  WuDominyHo2009}. Most generally, a critical point of
$J[\varepsilon(\cdot)]$ is regular if the map $V_T$ is locally
surjective in its vicinity, i.e., if for any local increment $\delta
U(T)$ in the evolution operator there exists an increment $\delta
\varepsilon(\cdot)$ in the control function such that
$V_T(\varepsilon(\cdot) + \delta \varepsilon(\cdot)) =
V_T(\varepsilon(\cdot)) + \delta U(T)$. This condition is equivalent
to requiring that the elements $\mu_{i j}(t)$ of the time-dependent
dipole-operator matrix (in the Heisenberg picture) form a set of $N^2$
linearly independent functions of time \cite{HoDominyRabitz2009PRA}.
In its turn, this condition is satisfied for all non-constant
admissible controls if and only if the quantum system is
evolution-operator controllable \cite{Ramakrishna1995,
  HoDominyRabitz2009PRA}. Note that for landscapes of some particular
physical objectives the conditions for regularity of the critical
points can be less stringent. For example, in the important special
case of state-transition control, a critical point is regular if the
matrix elements $\mu_{i j}(t)$ contain a set of just $2 N - 1$
linearly independent functions of time. This condition is satisfied
for all non-constant admissible controls if and only if the quantum
system is pure-state controllable (which is a weaker condition than
evolution-operator controllability, as discussed in section
\ref{sec:OCT-controllability}).

A critical point of $J[\varepsilon(\cdot)]$ is singular if the map
$V_T$ is not locally surjective in the point's vicinity.  Using the
chain rule, one obtains:
\begin{equation}
\label{jchain}
\frac{\delta J}{\delta \varepsilon(t)}  = \left\langle \nabla F(U(T)),
\frac{\delta U(T)}{\delta \varepsilon(t)} \right\rangle ,
\end{equation}
where $\nabla F(U(T))$ is the gradient of $F$ at $U(T)$, $\delta U(T)
/ \delta \varepsilon(t)$ is the first-order functional derivative of
$U(T)$ with respect to the control field, and $\langle A , B \rangle =
\Tr (A^{\dag} B)$ is the Hilbert-Schmidt inner product. From
(\ref{jchain}), if a critical point of $J$ is regular, $\nabla
F(U(T))$ must be zero. A critical point is called kinematic if $\nabla
F(U(T)) = 0$ and non-kinematic if $\nabla F(U(T)) \neq 0$. Thus, all
regular critical points are kinematic. A singular critical point may
be either kinematic or non-kinematic; in the latter case, $\delta J /
\delta \varepsilon(t) = 0$ whereas $\nabla F(U(T)) \neq 0$
\cite{WuDominyHo2009}. On quantum control landscapes, the measure of
regular critical points appears to be much greater than that of
singular ones \cite{WuDominyHo2009}. Therefore attention has been
focused on the characterization of regular critical points, and
several important results have been obtained
\cite{ChakrabartiRabitz2007review}. Nevertheless, singular critical
points on quantum control landscapes have been recently studied
theoretically \cite{WuDominyHo2009} and demonstrated experimentally
\cite{LapertZhangBraun2010}.

The condition for kinematic critical points, $\nabla F(U(T)) = 0$, can
be cast in an explicit form for various types of quantum control
problems. For evolution-operator control with the objective functional
$J = F_1(U(T))$ of (\ref{gatefn}), this condition becomes
\cite{RabitzHsiehRosenthal2005PRA, HsiehRabitz2008PRA}
\begin{equation}
\label{gatecrit}
W^{\dag} U(T) = U^{\dag}(T) W ,
\end{equation}
i.e., $W^{\dag} U(T)$ is required to be a Hermitian operator. It was
shown \cite{RabitzHsiehRosenthal2005PRA, HsiehRabitz2008PRA} that this
condition implies $W^{\dag} U(T) = Y^{\dag} (-I_m \oplus I_{N-m}) Y$,
where $Y$ is an arbitrary unitary transformation and $m = 0, 1, \ldots
, N$. There are $N+1$ distinct critical submanifolds labeled by $m$,
with corresponding critical values of $J$ given by $J_m = 1 - (2 m /
N)$. The global optima corresponding to $m = 0$ and $m = N$ (with $J_0
= 1$ and $J_N = -1$, respectively) are isolated points, while local
extrema corresponding to $m = 1, 2, \ldots , N-1$ are smooth, compact,
Grassmannian submanifolds embedded in $\mathrm{U}(N)$. It can be shown
that all regular local extrema are saddle-point regions
\cite{HsiehRabitz2008PRA}.

For observable control with the objective functional $J = F_3(U(T))$
of (\ref{obsfn}), the condition for a kinematic critical point becomes
\cite{RabitzHsiehRosenthal2006JCP, RabitzHsiehRosenthal2004,
  GirardeauKoch1998}
\begin{equation}
\label{obscrit}
[U(T) \rho_0 U^{\dag}(T), \Theta] = 0 ,
\end{equation}
i.e., the density matrix at the final time is required to commute with
the target observable operator. This condition was studied in the
context of optimization of Lagrange-type cost functionals with an
endpoint constraint \cite{Glaser1998, Brockett1991, VonNeumann1937a}
as well as in the context of regular critical points for Mayer-type
cost functionals \cite{RabitzHsiehRosenthal2006JCP,
  GirardeauKoch1998}. Let $R$ and $S$ denote unitary matrices that
diagonalize $\rho_0$ and $\Theta$, respectively, and define
$\tilde{U}(T) = S^{\dag} U(T) R$. The condition (\ref{obscrit}) that
$\rho(T)$ and $\Theta$ commute is equivalent to the condition that the
matrix $\tilde{U}(T)$ is in the double coset $\mathcal{M}_{\pi}$ of
some permutation matrix $P_{\pi}$ \cite{WuRabitzHsieh2008JPA}:  
\begin{equation}
\label{quotient}
\tilde{U}(T) \in \mathcal{M}_{\pi} 
= \mathrm{U}(\textbf{n}) P_{\pi} \mathrm{U}(\textbf{m}) .
\end{equation}
Here, $\mathrm{U}(\textbf{n})$ is the product group $\mathrm{U}(n_1)
\times \cdots \times \mathrm{U}(n_r)$, where $\mathrm{U}(n_{l})$
corresponds to the $l$th eigenvalue of $\rho_0$ with $n_l$-fold
degeneracy, and $\mathrm{U}(\textbf{m})$ is the product group
$\mathrm{U}(m_{1}) \times \cdots \times \mathrm{U}(m_{s})$, where
$\mathrm{U}(m_{l})$ corresponds to the $l$th eigenvalue of $\Theta$
with $m_{l}$-fold degeneracy. Thus, each critical submanifold
$\mathcal{M}_{\pi}$ corresponds to a particular choice of the
permutation $\pi$. All permutations on $N$ indices form the symmetric
group $\mathfrak{S}_N$, and the entire critical manifold $\mathcal{M}$
is given by $\mathcal{M} = \bigcup_{\pi \in \mathfrak{S}_N}
\mathcal{M}_{\pi}$. The structure of $\mathcal{M}$ depends on any
degeneracies in the spectra of $\rho_0$ and $\Theta$. When both
$\rho_0$ and $\Theta$ are fully nondegenerate, then
$\mathrm{U}(\textbf{n}) = \mathrm{U}(\textbf{m}) = [\mathrm{U}(1)]^N$,
and $\mathcal{M}$ consists of $N!$ disjoint $N$-dimensional tori,
labeled by the permutation matrices. The occurrence of degeneracies in
the spectra of $\rho_0$ and $\Theta$ will merge two or more tori
together, thereby reducing the number of disjoint critical
submanifolds and increasing their dimensions
\cite{WuRabitzHsieh2008JPA}.

\subsection{Optimality of control solutions}

Satisfaction of the condition (\ref{eq:c-point}) for a critical point
is a necessary but not sufficient condition for optimality of a
control \cite{Stengel1994, Jurdjevic1997}. For Mayer-type cost
functionals, a sufficient condition for optimality is negative
semidefiniteness of the Hessian of $J$, which is defined as
\begin{equation}
\mathcal{H}(t,t') \colonequals 
\frac{\delta^2 J}{\delta \varepsilon(t') \delta \varepsilon(t)} .
\end{equation}
The characteristics of critical points (in particular, the presence or
absence of local optima) are important for the convergence properties
of search algorithms \cite{ChakrabartiRabitz2007review}. To classify
critical points as global maxima and minima, local maxima and minima,
and saddle points, one examines the second-order variation in $J$ for
an arbitrary control variation $\delta \varepsilon(\cdot)$, which for
Mayer-type functionals can be written as
\begin{equation}
\label{second-order-1}
\delta^2 J = \mathcal{Q}_F(\delta U(T), \delta U(T)) + \langle \nabla
F(U(T)) , \delta^2 U(T) \rangle ,
\end{equation}
where $\delta U(T)$ and $\delta^2 U(T)$ are the first- and
second-order variations, respectively, of $U(T)$ caused by a control
variation $\delta \varepsilon(\cdot)$, and $\mathcal{Q}_F$ is the
Hessian quadratic form of $F(U)$. Assuming that the critical point
$\varepsilon(\cdot)$ is regular, one obtains:
\begin{equation}
\label{second-order-2}
\delta^2 J = \mathcal{Q}_F(\delta U(T), \delta U(T)) .
\end{equation}
Explicit expressions for the Hessian and/or Hessian quadratic form
were obtained for evolution-operator control
\cite{RabitzHsiehRosenthal2005PRA, HsiehRabitz2008PRA,
  HoDominyRabitz2009PRA} and observable control
\cite{ShenHsiehRabitz2006JCP, HoRab2006a}.

The optimality of regular critical points can be determined by
inspecting the number of positive, negative and null eigenvalues of
the Hessian (or, equivalently, the coefficients of the Hessian
quadratic form when written in a diagonal basis). An issue of special
interest is to determine whether any of the regular critical points
are local maxima (frequently referred to as local traps due to their
ability to halt searches guided by gradient algorithms before reaching
the global maximum). Detailed analyses for evolution-operator control
and observable control reveal \cite{ChakrabartiRabitz2007review,
  RabitzHsiehRosenthal2005PRA, HsiehRabitz2008PRA,
  HoDominyRabitz2009PRA, ShenHsiehRabitz2006JCP, HoRab2006a} that all
regular optima are global and the remainder of regular critical points
(i.e., except for the global maximum and global minimum) are
saddles. This discovery means that no local traps exist in the control
landscapes of controllable closed quantum systems. The same result was
also obtained for observable-control landscapes of controllable open
quantum systems with Kraus-map dynamics \cite{WuPechenRabitz2008JMP}.
Due attention still needs to be given to consideration of singular
critical points, although numerical evidence suggests that their
effect on optimization is likely insignificant \cite{WuDominyHo2009}.

\subsection{Pareto optimality for multi-objective control}
\label{sec:L-Pareto}

Many practical quantum control problems seek to optimize multiple,
often competing, objectives. In such situations the usual notion of
optimality is replaced by that of Pareto optimality. The \emph{Pareto
  front} of a multi-objective control problem is the set of all
controls such that all other controls have a lower value for at least
one of the objectives \cite{ChankongHaimes1983, Steuer1986,
  Miettinen1998}. The analysis of the Pareto front reveals the nature
of conflicts and tradeoffs between different control objectives. The
structure of the landscape for multi-observable control is of interest
and follows directly from that of single-observable control
\cite{RajWu2008a}. Of particular relevance to many chemical and
physical applications is the problem of simultaneous maximization of
the expectation values of multiple observables. Such simultaneous
maximization is possible if the intersection $\bigcap_k
\mathcal{M}_k^{(\max)}$ (where $\mathcal{M}_k^{(\max)}$ is the maximum
submanifold for the $k$th observable) is nonempty and a point in the
intersection can be reached under some control $\varepsilon(\cdot)$;
in this regard, the dimension of the intersection manifold $\bigcap_k
\mathcal{M}_k^{(\max)}$ has been analyzed \cite{RajWu2008b}. It has
been shown that the common QOCT technique of running many independent
maximizations of a cost functional like (\ref{multiobsfn}) (using
different weight coefficients $\{\alpha_k\}$) is incapable of sampling
many regions of the Pareto front \cite{RajWu2008b}. Alternative
methods for Pareto front sampling are discussed further below. 

\subsection{Landscape exploration via homotopy trajectory control} 
\label{sec:L-trajectory}

The absence of local traps in landscapes for observable control and
evolution-operator control with Mayer-type cost functionals has
important implications for the design of optimization algorithms. Many
practical applications require algorithms capable of searching quantum
control landscapes for optimal solutions that satisfy additional
criteria, such as minimization of the field fluence or maximization of
the robustness to laser noise. So-called \emph{homotopy trajectory
  control} algorithms (in particular, diffeomorphic modulation under
observable-response-preserving homotopy, or D-MORPH)
\cite{Hillermeier2001, RothmanHoRabitz2005JCP, RothmanHoRabitz2006PRA}
can follow paths to the global maximum of a Mayer-type cost
functional, exploiting the trap-free nature of the control landscape,
while locally optimizing auxiliary costs. The essential prerequisite
for successful use of these algorithms is the existence of a connected
path between the initial and target controls. Homotopy trajectory
control is closely related to the notion of a level set which is
defined as the collection of controls that all produce the same value
of the cost functional $J$. Theoretical analysis
\cite{ChakrabartiRabitz2007review, RothmanHoRabitz2005JCP,
  RothmanHoRabitz2006PRA} predicts that for controllable quantum
systems each level set is a continuous manifold. A homotopy trajectory
algorithm is able to move on such a manifold exploring different
control solutions that result in the same value of the cost
functional, but may differ in other properties (e.g., the field
fluence or robustness). A version of the D-MORPH algorithm was also
developed for evolution-operator control of closed quantum systems; it
was able to identify optimal controls generating a target unitary
transformation up to machine precision \cite{DominyRabitz2008JPA}.

Homotopy trajectory algorithms are also very useful for exploring
quantum control landscapes for multiple objectives. For example, in
order to track paths in the space of expectation values of multiple
observables while locally minimizing a Lagrange-type cost,
multi-observable trajectory control algorithms were developed
\cite{RajWu2008a}. Such algorithms are generally applicable to the
treatment of multi-objective quantum control problems (Pareto quantum
optimal control) \cite{RajWu2008b}. They can traverse the Pareto front
to identify admissible tradeoffs in optimization of multiple control
objectives (e.g., maximization of multiple observable expectation
values). This method can continuously sample the Pareto front during
the course of one optimization run \cite{RajWu2008b} and thus can be
more efficient than the use of standard QOCT with cost functionals of
the form (\ref{multiobsfn}). Also, the D-MORPH algorithm was recently
extended to handle optimal control problems involving multiple quantum
systems and multiple objectives \cite{BeltraniGhosh2009}.

\subsection{Practical importance of control landscape analysis}
\label{sec:L-importance}

The absence of local traps in control landscapes of controllable
quantum systems has very important implications for the feasibility of
AFC experiments (see section~\ref{sec:AFC}). The relationship between
the quantum control landscape structure and optimization complexity of
algorithms used in AFC has been the subject of recent theoretical
analyses \cite{ChakrabartiRabitz2007review, RajWu2007b,
  MooreHsiehRabitz2008JCP, OzaPechen2009JPA}. Results of these studies
support the vast empirical evidence \cite{HoRab2006a} indicating that
the favorable landscape topology strongly correlates with fast mean
convergence times to the global optimum. The trap-free control
landscape topology also ensures convergence of gradient-based
optimization algorithms to the global maximum. These algorithms can be
used to search for optimal solutions to a variety of quantum control
problems. In addition to theoretical studies (mostly using QOCT),
gradient algorithms are also applicable in quantum control experiments
\cite{RoslundRabitz2009}, provided that measurement of the gradient is
sufficiently robust to laser and detection noise. The use of
deterministic algorithms in AFC experiments is discussed in more
detail in section~\ref{sec:AFC-algorithms}. 

\subsection{Experimental observation of quantum control landscapes}
\label{sec:L-observation}

Significant efforts have been recently devoted to experimental
observation of quantum control landscapes, aiming both at testing the
predictions of the theoretical analysis and at obtaining a better
understanding of control mechanisms. Roslund \emph{et al.}
\cite{RoslundRoth2006} observed quantum control level sets for
maximization of non-resonant two-photon absorption in a molecule and
second harmonic generation (SHG) in a nonlinear crystal and found them
to be continuous manifolds (closed surfaces) in the control
landscape. A diverse family of control mechanisms was encountered, as
each of the multiple control fields forming a level set preserves the
observable value by exciting a distinct pattern of constructive and
destructive quantum interferences. 

Wollenhaupt, Baumert, and co-workers \cite{WollenhauptPrakelt2005JMO,
  Bayer2008} used parameterized pulse shapes to reduce the
dimensionality of the optimization problem (maximization of the
Autler-Townes contrast in strong-field ionization of potassium atoms)
and observed the corresponding two-dimensional quantum control
landscape. In order to better understand the performance of AFC, the
evolution of different optimization procedures was visualized by means
of trajectories on the surface of the measured control
landscape. Marquetand \emph{et al.}  \cite{MarquetandNuernberger2007}
observed a two-dimensional quantum control landscape (for maximization
of the retinal photoisomerization yield in bacteriorhodopsin) and used
it to elucidate the properties of molecular wave-packet evolution on
an excited potential energy surface.

The theoretical analysis of control landscape topology has been
carried out with no constraints placed on the controls (see
section~\ref{sec:L-formalism}). A main conclusion from these studies
is the inherent lack of local traps on quantum control landscapes
under normal circumstances. Recently, Roslund and Rabitz
\cite{RoslundRabitz2009b} experimentally demonstrated the trap-free
monotonic character of control landscapes for optimization of
frequency unfiltered and filtered SHG. For unfiltered SHG, the
landscape was randomly sampled and interpolation of the data was found
to be devoid of traps up to the level of data noise. In the case of
narrow-band-filtered SHG, trajectories taken on the landscape revealed
the absence of traps, although a rich local structure was observed on
the landscape in this case. Despite the inherent trap-free nature of
the landscapes, significant constraints on the controls can distort
and/or isolate portions of the erstwhile trap-free landscape to
produce apparent (i.e., false) traps \cite{RoslundRabitz2009b}. Such
artificial structure arising from the forced sampling of the landscape
has been seen in some experimental studies
\cite{WollenhauptPrakelt2005JMO, Bayer2008,
  MarquetandNuernberger2007}, in which the number of control variables
was purposely reduced.


\section{Adaptive feedback control in the laboratory}
\label{sec:AFC}

There are important differences between quantum control theory and its
experimental implementation. Control solutions obtained in theoretical
studies strongly depend on the employed model Hamiltonian. However,
for real systems controlled in the laboratory, the Hamiltonians
usually are not known well (except for the simplest cases), and the
Hamiltonians for the system-environment coupling are known to an even
lesser degree. An additional difficulty is the computational
complexity of accurately solving the optimal control equations for
realistic polyatomic molecules. Another important difference between
control theory and experiment arises from the difficulty of reliably
implementing theoretical control designs in the laboratory, due to
instrumental noise and other limitations. As a result, optimal
theoretical control designs generally will not be optimal in the
laboratory.  Notwithstanding these comments, control simulations
continue to be very valuable, and they even set forth the logic
leading to practical laboratory control as explained below.

A crucial step towards selective laser control of physical and
chemical  phenomena on the quantum scale was the introduction of AFC
(also referred to as closed-loop laboratory control or learning
control). AFC was proposed and theoretically grounded by Judson and
Rabitz in their paper ``Teaching lasers to control molecules'' in 1992
\cite{JudsonRabitz1992}. In AFC, a loop is closed in the laboratory,
with results of measurements on the quantum system used to evaluate
the success of the applied control and to refine it, until the control
objective is reached as best as possible. At each cycle of the loop,
the external control (e.g., a shaped laser pulse) is applied to the
system (e.g., an ensemble of molecules). The signal (e.g., the yield
of a particular reaction product or population in a target state) is
detected and fed back to the learning algorithm (e.g., a genetic
algorithm). The algorithm evaluates each control based on its measured
outcome with respect to a predefined control goal, and searches
through the space of available controls to move towards an optimal
solution. 

While AFC can be simulated on the computer \cite{JudsonRabitz1992,
  GeremiaZhu2000, OmenettoLuceTaylor1999, BrixnerAbajo2005PRL,
  BrixnerAbajo2006PRB, BrixnerAbajo2006APB, HertzRouzee2007,
  Voronine2006JCP, Voronine2007JCP, Tuchscherer2009, ZhuRabitz2003,
  GraceBrif2006, GollubVivieRiedle2008PRA, GollubVivieRiedle2009}, the
important advantage of this approach lies in its ability to be
directly implemented in the laboratory. Most importantly, the
optimization is performed in the laboratory with the actual system,
and thus is independent of any model. As a result, the AFC method
works remarkably well for systems even of high complexity, including,
for example, large polyatomic molecules in the liquid phase, for which
only very rough models are available. Second, there is no need to
measure the laser field in AFC, because any systematic
characterization of the control ``knobs'' (such as pulse shaper
parameters) is sufficient. This set of control ``knobs'' determined by
the experimental apparatus defines the parameter space searched by the
learning algorithm for an optimal laser shape. This procedure
naturally incorporates any laboratory constraints on the control laser
fields. Third, optimal controls identified in AFC are characterized by
a natural degree of robustness to instrumental noise, since non-robust
solutions will be rejected by the algorithm. Fourth, in AFC, it is
possible to operate at a high-duty cycle of hundreds or even thousands
of experiments per second, by exploiting (i) the conceptual advantage
of the evolving quantum system solving its own Schr\"{o}dinger
equation in the fastest possible fashion and (ii) the technological
advantage of high-repetition-rate pulsed laser systems under full
automation. Fifth, in AFC, a new quantum ensemble (e.g., a new
molecular sample) is used in each cycle of the loop, which completely
avoids the issue of back action exerted by the measurement process on
a quantum system. Thus AFC is technologically distinct from
measurement-based RTFC \cite{Belavkin1983, WisemanMilburn1993PRL,
  Wiseman1994PRA, DohertyHabib2000, DohertyDoyle2000,
  WisemanMilburn2010}, in which the same quantum system is manipulated
until the final target objective is reached and for which measurement
back action is an important effect that needs to be taken into account
(see section~\ref{sec:RTFC}).

\subsection{Femtosecond pulse-shaping technology}

The majority of current AFC experiments employ shaped ultrafast laser
pulses. In such experiments, one usually starts with a random or
nearly random selection of trial shaped pulses of length $\sim
10^{-13}$~s or less. The pulses are shaped by modulating the phases
and/or amplitudes of the spatially resolved spectral components, for
example, by means of a liquid crystal modulator (LCM), acousto-optic
modulator (AOM), or a micromechanical mirror array (MMA). The
experiments employ fully automated computer control of the pulse
shapes guided by a learning algorithm. The shaped laser pulses
produced by this method can be viewed as ``photonic reagents,'' which
interact with matter at the atomic or molecular scale to facilitate
desired controlled outcomes of various physical and chemical
phenomena.

Significant femtosecond pulse-shaping capabilities were already
available in the early 1990s, with the development of a programmable
multi-element liquid-crystal phase modulator that operated on a
millisecond time scale \cite{WeinerLeaird1990}. Devices with two LCMs
made possible simultaneous and independent phase and amplitude
modulation of spectral components \cite{WefersNelson1993,
  WefersNelson1995}. Similar capabilities are also available with
AOM-based pulse shapers \cite{HillegasWarren1994}. These and other
developments have been reviewed \cite{Kawashima1995, Weiner1995,
  Weiner2000, Goswami2003}. During the last decade, physical and
chemical applications of AFC motivated further advances in femtosecond
pulse-shaping technology, including arbitrary amplitude and phase
modulation in an acousto-optic programmable dispersive filter
\cite{VerluiseLaude2000}, enhanced resolution of LCMs
\cite{Stobrawa2001, Monmayrant2004}, compact and robust pulse-shaping
\cite{PrakeltWollenhaupt2003}, pulse-shape modulation at nanosecond
time scales using an electro-optical gallium arsenide array with
controlled waveguides \cite{FrumkerTal2005}, and spectral line-by-line
pulse shaping \cite{JiangWeiner2007a, JiangWeiner2007b}. The
development of polarization pulse shaping \cite{BrixnerGerber2001,
  BrixnerKrampert2002APB} brought an additional dimension to control
of quantum phenomena, which is particularly important in some
applications (e.g., for increasing the yield of multiphoton ionization
in molecules); recent improvements in this area also include full
control of the spectral polarization of ultrashort laser pulses
\cite{PolachekOron2006}, simultaneous phase, amplitude, and
polarization shaping \cite{PlewickiWeise2006, NinckGaller2007,
  MasihzadehSchlup2007, PlewickiWeber2007, WeiseLindinger2009}, and a
simplified ultrafast polarization shaper using a birefringent prism
\cite{KupkaSchlup2009}. Most recently, the shaping of ultraviolet (UV)
femtosecond pulses has been demonstrated, including phase modulation
\cite{NuernbergerVogtSelle2007}, simultaneous phase and amplitude
modulation \cite{ParkerNunn2009}, and polarization shaping
\cite{SelleNuernberger2008, NuernbergerSelle2009}.

\subsection{Optical applications of AFC}

The AFC approach can be used to produce optical fields with prescribed
properties, which, in turn, can be applied to control physical and
chemical phenomena (e.g., in atoms, molecules, and semiconductor
structures). In particular, some of the earliest AFC experiments aimed
at the maximal compression of femtosecond laser pulses
\cite{BaumertBrixner1997, Yelin1997, BrixnerStrehleGerber1999,
  Zeek1999, Zeek2000, ZeidlerHornung2000}. In these experiments, the
light produced through SHG of the shaped pulse in a thin nonlinear
crystal served as the feedback signal. The SHG yield is directly
proportional to the intensity of the incident light pulse, and, for
pulses with a fixed energy, the most intense pulse is the shortest
one. AFC-optimized compression of broadband laser pulses was also
demonstrated using a feedback signal derived from two-photon
absorption in semiconductors \cite{SiegnerHaimlKunde2002}. Application
of AFC makes it possible to generate maximally compressed laser pulses
in a simple and effective way, without requiring knowledge of the
input pulse's shape. Such adaptive pulse compressors (with AOM-based
pulse shapers and SHG-based feedback signal) are now employed as
built-in components in some commercially available femtosecond
amplification systems. However, since the amplification process is
nearly linear, the full-scale application of AFC is usually not
necessary, as pulses can be compressed in a single feedback step using
spectral interferometry. The resulting transform-limited pulses can be
used as a starting point for the study and control of various
photophysical and photochemical processes (e.g., they can be used to
excite and track localized fine-structure and Rydberg wave packets in
atoms and vibrational wave packets in molecules). In many AFC
experiments, transform-limited pulses are used as a reference, to
separate off the intensity dependence which is ubiquitous in nonlinear
processes. 

Another optical application of AFC is optimal amplification of chirped
femtosecond laser pulses \cite{Efimov1998, Efimov2000}. The AFC method
is used to minimize the higher-order phase dispersion that is inherent
in the amplification process. Furthermore, AFC was applied to generate
almost arbitrary target temporal shapes starting with uncharacterized
input pulses \cite{MeshulachYelinSilberberg1998,
  BrixnerOehrlein2000}. These experiments used a cross-correlation
measurement \cite{MeshulachYelinSilberberg1998} and electric field
characterization via frequency-resolved optical gating (FROG)
\cite{BrixnerOehrlein2000} of the output pulses as the feedback
signal. As polarization shaping technology for femtosecond laser
pulses developed \cite{BrixnerGerber2001, BrixnerKrampert2002APB}, AFC
was used to generate pulses with target polarization profiles
\cite{BrixnerDamrauerKrampert2003, Suzuki2004ApplOpt}. One experiment
\cite{BrixnerDamrauerKrampert2003} used the SHG feedback signal to
compensate for material dispersion and time-dependent modulation of
the polarization state. Another experiment \cite{Suzuki2004ApplOpt}
employed a sophisticated feedback signal based on dual-channel
spectral interferometry to generate shaped femtosecond pulses whose
ellipticity increased at a constant rate. In a further development, a
recent experiment \cite{Aeschlimann2007} used polarization-shaped
laser pulses and AFC to manipulate the optical near field on a
nanometer scale.

\subsection{AFC of high-harmonic generation}
\label{sec:AFC-hh}

Among important physical applications of AFC is coherent manipulation
of soft X-rays produced via high-harmonic generation. In a pioneering
experiment, Murnane, Kapteyn, and co-workers \cite{Bartels2000} used
shaped ultrashort, intense laser pulses (with 6-8 optical cycles) for
AFC of high-harmonic generation in atomic gases. Their results
demonstrate that optimally shaped laser pulses identified by the
learning algorithm can improve the efficiency of X-ray generation by
an order of magnitude, manipulate the spectral characteristics of the
emitted radiation, and ``channel'' the interaction between nonlinear
processes of different orders. All these effects result from complex
interferences between the quantum amplitudes of the atomic states,
created by the external laser field. The learning algorithm guides the
pulse shaper to tailor the laser field to produce the optimal
interference pattern. Several consequent AFC experiments
\cite{Bartels2001, Bartels2004, Reitze2004} explored various aspects
of optimal high-harmonic generation in atomic gases, including the
analysis of the control mechanism via a comparison of experimental
data with predictions of theoretical models. Further experimental
studies used AFC for optimal spatial control of high-harmonic
generation in hollow fibers \cite{PfeiferKemmer2005,
  WalterPfeifer2006}, optimal control of the brilliance of
high-harmonic generation in gas jet and capillary setups
\cite{SpitzenpfeilEyring2009}, and optimal control of the spectral
shape of coherent soft X-rays \cite{PfeiferWalter2005}. The latter
work \cite{PfeiferWalter2005} has been a precursor to a more recent
development, in which spectrally shaped femtosecond X-ray fields were
themselves used to adaptively control photofragmentation yields of
SF$_6$
\cite{PfeiferSpitzenpfeil2007}. Advances in optimal control of
high-harmonic generation (including related AFC experiments) have been
recently reviewed \cite{PfeiferSpielmann2006, Winterfeldt2008}.

Beyond the physical interest in achieving control over high-harmonic
generation, these experiments also demonstrated a dramatic degree of
inherent robustness to laser-field noise in strongly non-linear
control. This behavior can be understood in terms of an extensive null
space of the Hessian at the top of the control landscape, implying a
very gentle slope near the global maximum
\cite{ShenHsiehRabitz2006JCP, RabitzHoHsieh2006PRA}. This
characteristic of the quantum control landscape makes it possible to
tolerate much of the laser noise while maintaining a high control
yield. Such robustness is expected to be a key attractive feature of
observable control across virtually all quantum phenomena.

\subsection{AFC of multiphoton transitions in atoms}

Control of bound-to-bound multiphoton transitions in atoms with
optimally shaped femtosecond laser pulses provides a vivid
illustration of the control mechanism based on multi-pathway quantum
interference. Non-resonant multiphoton transitions involve many routes
through a continuum of virtual levels. The interference pattern
excited by the multiple frequency components of the control pulse can
enhance or diminish the total transition probability. The interference
effect depends on the spectral phase distribution of the laser
pulse. A number of experiments \cite{MeshulachSilberberg1998,
  MeshulachSilberberg1999, HornungMeierZeidler2000} used AFC to
identify pulse shapes that are optimal for enhancing or cancelling the
probability of making a transition. In particular, it was possible to
tailor \emph{dark pulses} that do not excite the atom at all due to
destructive quantum interference. On the other hand, AFC was able to
find shaped pulses that induce transitions as effectively as
transform-limited pulses, even though their peak intensities are much
lower. Due to the relative simplicity of the atomic systems studied,
it was possible to compare the results of the AFC experiments with
theoretical predictions and verify the control mechanism based on
quantum interference of multiple laser-driven transition amplitudes.
In a related experiment \cite{Dudovich2001}, AFC was helpful for
demonstrating that transform-limited pulses are not optimal for
inducing resonant multiphoton transitions. It was shown that optimally
shaped pulses enhance resonant multiphoton transitions significantly
beyond the level achieved by maximizing the pulse's peak intensity. A
recent experiment \cite{TralleroHerreroCohen2006} considered
non-resonant multiphoton absorption in atomic sodium in the
strong-field limit. It was demonstrated that in this regime the
stimulated emission induced by the dynamic Stark shift becomes
important, which makes transform-limited pulses not optimal for
strong-field non-resonant multiphoton transitions. AFC was used to
discover strong-field shaped laser pulses that optimally counteract
the dynamic Stark shift-induced stimulated emission and thus maximize
the absorption probability.

A more complex problem in atomic physics is control of multiphoton
ionization. In one experiment \cite{Papastathopoulos2005}, AFC was
applied to optimize multiphoton ionization of atomic calcium by shaped
femtosecond laser pulses. The feedback signals were measured using ion
and electron spectroscopy, and the optimization results were used to
elucidate the intermediate resonances involved in the photoionization
process. Another experiment \cite{WollenhauptPrakelt2005JOptB} studied
photoionization of potassium atoms controlled by phase-locked pairs of
intense femtosecond laser pulses. Measurements of the Autler-Townes
doublet in the photoelectron spectra enabled analysis of the induced
transient processes. The AFC experimental results were helpful for
exploring the control mechanism based on the selective population of
dressed states.

\subsection{AFC of Rydberg wave packets in atoms}

In one of the first applications of AFC, in 1999, Bucksbaum and
co-workers \cite{Weinacht1999a} manipulated the shape of an atomic
radial wave function (a so-called Rydberg wave packet). Non-stationary
Rydberg wave packets were created by irradiating cesium atoms with
shaped ultrafast laser pulses. A variation of the quantum holography
method \cite{Leichtle1998} was used to measure the atomic radial wave
function generated by the laser pulse. In order to reconstruct the
wave function, the amplitude of each energy eigenstate in the total
wave packet was measured independently via state selective field
ionization. The distance between the measured and target wave packet
provided the feedback signal. In the weak-field limit, a simple linear
relationship exists between the amplitudes of the energy eigenstates
in the wave packet decomposition and the amplitudes of the
corresponding spectral components of the control laser field. Based on
this relationship, a simple gradient-type algorithm was employed to
adjust the spectral phase distribution of the control field. AFC
equipped with this algorithm was able to change the shape of the
Rydberg wave packet to match the target within two iterations of the
feedback control loop. If the wave packet is created in the
strong-field regime, then a more sophisticated learning algorithm is
generally required to implement AFC.

\subsection{AFC of electronic excitations in molecules}

The first AFC experiment was reported in 1997 by Wilson and co-workers
\cite{BardeenYakovlevWilson1997}. Femtosecond laser pulses shaped by a
computer-controlled AOM were used to excite an electronic transition
in molecules (laser dye IR125 in methanol solution). The measured
fluorescence served as the feedback signal in AFC to optimize the
population transfer from the ground to first excited molecular
electronic state. Both excitation efficiency (the ratio of the excited
state population to the laser energy) and effectiveness (the total
excited-state population) were optimized. Similar AFC experiments were
later performed with different molecules in the liquid phase: laser
dyes LDS750 in acetonitrile and ethanol solutions \cite{Nahmias2005},
DCM in methanol solution \cite{LeeJungSungHongNam2002}, rhodamine 101
in methanol solution \cite{Prokhorenko2005}, coumarin 515 in ethanol
solution \cite{ZhangSun2005}, coumarin 6 in a range of non-polar
solvents (linear and cyclic alkanes) \cite{vanderWalleHerek2009}, the
charge-transfer coordination complex [Ru(dpb)$_3$](PF$_6$)$_2$ (where
dpb is 4,$4'$-diphenyl-2,$2'$-bipyridine) in methanol solution
\cite{BrixnerDamrauer2003} and acetonitrile solution
\cite{MontgomeryMeglen2006, MontgomeryMeglen2007,
  MontgomeryDamrauer2007}, a donor-acceptor macromolecule (a phenylene
ethynylene dendrimer tethered to perylene) in dichloromethane solution
\cite{KurodaKleiman2009}, and perylene in chloroform solution
\cite{OtakeKanoWada2006}. Two-photon electronic excitations in flavin
mononucleotide in aqueous solution were controlled using
multi-objective optimization (a genetic algorithm was employed to
simultaneously maximize the fluorescence intensity and the ratio of
fluorescence and SHG intensities) \cite{BonacinaWolf2007}. Molecular
electronic excitations were also optimized in AFC experiments in the
solid state, with a crystal of $\alpha$-perylene
\cite{OtakeKanoWada2006, Okada2004}. 

Since these AFC experiments are performed in the condensed phase, an
important issue is the degree of coherence of the controlled
dynamics. This issue has been recently explored in a series of AFC
experiments \cite{vanderWalleHerek2009}, in which the level of
attained control was investigated by systematically varying properties
of the environment. Specifically, AFC was applied to optimize the
stimulated emission from coumarin 6 (a laser dye molecule) dissolved
in cyclohexane, and the recorded optimal pulse shape (characterized by
a significantly nonlinear negative chirp) was used with several other
solvents (linear and cyclic alkanes). In these experiments, the
molecule was excited in the linear absorption regime in order to
exclude the trivial intensity dependence characteristic of
multiphoton processes. The results revealed an inverse correlation
between the obtained degree of control (as measured by the enhancement
of stimulated emission relative to that achieved by excitation with
the transform-limited pulse) and the viscosity of the solvent. This
study indicates that the control mechanism involves a coherent process
(i.e., based on quantum interference of coherent pathways) and that
environmentally-induced decoherence limits the leverage of control on
the particular molecular system. Also, in a recent AFC experiment
\cite{KurodaKleiman2009}, the photoemission yield of a donor-acceptor
macromolecule was maximized, and a coherent control mechanism was
identified by analyzing the pulse optimization process and optimal
pulse features. Another AFC experiment \cite{RothGuyonRoslund2009}
controlled flavin mononucleotide and riboflavin dissolved in water,
through a vibronic transition driven by a shaped 400~nm control pulse
followed by a delayed 800~nm pulse that produced irreversible further
excitation. Fluorescence depletion served as a feedback signal. The
experiment only functioned when the delay between the two pulses was
less than $\sim$~1~ps, indicating that the underlying control
mechanism employs coherent dynamics (see further discussion of this
experiment in section~\ref{sec:AFC-ODD}).

\subsection{AFC of photodissociation reactions in molecules}
\label{sec:AFC-fragment}

A long-standing goal of photochemistry is selective control of
molecular fragmentation. During the last decade, AFC employing shaped
femtosecond laser pulses has been applied to achieve significant
successes towards meeting this goal \cite{LevisRabitz2002,
  BrixnerGerber2003, Nuernberger2007}.  Selective quantum control of
photodissociation reactions in molecules using AFC was first
demonstrated by Gerber and co-workers in 1998 \cite{Assion1998}. They
studied photodissociation of the organometallic complex CpFe(CO)$_2$Cl
(where Cp = C$_5$H$_5$) that contains particular types of iron-ligand
bonds and exhibits different fragmentation channels upon excitation
with shaped femtosecond laser pulses. The branching ratio
[CpFe(CO)Cl]$^+$/[FeCl]$^+$ was maximized and minimized in AFC
experiments employing an evolutionary algorithm. The experiment was
performed in a molecular beam, and the feedback signal was obtained
from measurements of the ionized photofragments in a time-of-flight
mass spectrometer. Using AFC, it was possible to change the branching
ratio between 5:1 and 1:1. 

The success of the AFC experiment described above triggered an ongoing
wave of research activity in this area. In particular, Gerber's group
explored various aspects of AFC of photodissociation and
photoionization reactions in molecules. In one AFC experiment
\cite{BergtBrixner1999}, the relative yields of photodissociation and
photoionization of iron pentacarbonyl, Fe(CO)$_5$, were controlled in
the gas phase, using femtosecond laser pulses with carrier wavelengths
at 800~nm and 400~nm (the latter produced via SHG of the former). The
AFC-based optimization (both maximization and minimization) of the
branching ratio [Fe(CO)$_5$]$^+$/Fe$^+$ demonstrated that the control
mechanism is not simply intensity-dependent, but rather employs the
spectral phase distribution of the shaped laser pulse to steer the
dynamics of the excited molecular vibrational wave packet towards the
target reaction channel. Another gas-phase AFC experiment
\cite{BrixnerKiefer2001} also analyzed the relative importance of
intensity-dependent and coherent effects in control of photochemical
reactions that involve nonlinear (multiphoton) optical excitations.
The control goals were the direct photoionization of CpFe(CO)$_2$Cl
(i.e., maximization of the [CpFe(CO)$_2$Cl]$^+$ yield) and selective
photofragmentation (i.e., maximization of the branching ratio
[CpFe(CO)Cl]$^+$/[FeCl]$^+$). For each pulse shape during the
AFC-based optimization, the target reaction yield and SHG efficiency
(which is directly proportional to the pulse intensity) were
recorded. In the case of direct ionization control, a clear
correlation between the [CpFe(CO)$_2$Cl]$^+$ yield and SHG efficiency
was observed, which implies that the photoionization control mechanism
is mainly intensity-dependent. However, for fragmentation control, no
correlation between the [CpFe(CO)Cl]$^+$/[FeCl]$^+$ ratio and SHG
efficiency was found. Moreover, for different pulses with the same SHG
intensity, a large range of different [CpFe(CO)Cl]$^+$/[FeCl]$^+$
values was obtained, depending on the specific pulse shape. These
results indicate that while photofragmentation involves multiphoton
excitation, it is not regulated by the pulse intensity alone. Rather,
the control mechanism for a particular photofragmentation reaction
requires a specially tailored laser pulse to guide the complex wave
packet dynamics towards the desired outcome. Results of a similar AFC
experiment \cite{Damrauer2002} that analyzed photofragmentation of
CH$_2$ClBr in the gas phase (including maximization and minimization
of the [CH$_2$Br]$^+$/[CH$_2$Cl]$^+$ ratio) also indicate that the
control mechanism involves manipulation of the wave packet dynamics on
neutral dissociative surfaces rather than purely intensity-dependent
effects. Experiments that demonstrated AFC of photofragmentation in
the molecules CpFe(CO)$_2$Cl and CpFe(CO)$_2$Br
\cite{BergtBrixnerDietl2002} will be discussed later in the context of
optimal dynamic discrimination of similar quantum systems.

In 2001, Levis and co-workers \cite{Levis2001} used AFC with shaped,
strong-field laser pulses to demonstrate selective cleavage and
rearrangement of chemical bonds in polyatomic organic molecules (in
the gas phase), including (CH$_3$)$_2$CO (acetone), CH$_3$COCF$_3$
(trifluoroacetone), and C$_6$H$_5$COCH$_3$ (acetophenone). Control
over the formation of CH$_3$CO from (CH$_3$)$_2$CO, CF$_3$ or CH$_3$
from CH$_3$COCF$_3$, and C$_6$H$_5$CH$_3$ (toluene) from
C$_6$H$_5$COCH$_3$ was achieved with high selectivity. The use of
strong laser fields (with intensities of about $10^{13}$~W/cm$^2$)
helps to effectively increase the available bandwidth, as transitions
to excited molecular states are facilitated by the dynamic Stark
shift. This effect opens up many reaction pathways which are
inaccessible in the weak-field (perturbative) regime due to resonant
spectral restrictions \cite{LevisRabitz2002}.  While theoretical
treatment of the complex strong-field molecular dynamics is extremely
difficult, this complexity in no way affects employment of AFC in the
laboratory, where the molecule solves its own Schr\"{o}dinger equation
on a femtosecond time scale. By operating at a high-duty control
cycle, a learning algorithm is typically able to identify optimal
laser pulses in a matter of minutes.

Significant attention has been devoted to the analysis of quantum
dynamical processes involved in molecular photofragmentation control
achieved in gas-phase AFC experiments with shaped femtosecond laser
pulses.  W\"{o}ste and co-workers \cite{Daniel2001, VajdaRosendo2001,
  Daniel2003} studied mechanisms of photofragmentation control for
CpMn(CO)$_3$ optimizing the branching ratios
[CpMn(CO)]$^+$/[CpMn(CO)$_3$]$^+$ and
[CpMn(CO)$_2$]$^+$/[CpMn(CO)$_3$]$^+$. Weinacht and co-workers
\cite{Cardoza2005JCP, Cardoza2005CPL, Cardoza2004,
  LanghojerCardoza2005} analyzed mechanisms underlying control of
photofragmentation in a series of AFC experiments with similar
molecules: CH$_3$COCF$_3$ (trifluoroacetone), CH$_3$COCCl$_3$
(trichloroacetone), and CH$_3$COCD$_3$ (tri-deuterated acetone). The
yield of the [CX$_3$]$^+$ fragment and the [CX$_3$]$^+$/[CH$_3$]$^+$
ratio (where X is F, Cl, and D for trifluoroacetone, trichloroacetone,
and tri-deuterated acetone, respectively) were optimized in these AFC
experiments using intense shaped laser pulses. AFC was also used to
optimize the branching ratios Br$^+$/[CH$_2$Br]$^+$ and
[CH$_2$I]$^+$/[CH$_2$Br]$^+$ in photofragmentation of CH$_2$BrI
(bromoiodomethane) \cite{LanghojerCardoza2005, Cardoza2005a}.  In a
number of works \cite{Daniel2001, Daniel2003, Cardoza2005JCP,
  Cardoza2005CPL}, AFC experiments were supplemented by theoretical
\emph{ab initio} quantum calculations to help clarify
photofragmentation control mechanisms. In several other studies
\cite{Cardoza2004, LanghojerCardoza2005, Cardoza2005a}, a change in
the basis of the control variables made it possible to reduce the
dimension of the search space and thus elucidate control mechanisms of
selective molecular photofragmentation. In another work
\cite{CardozaPearson2006}, pump-probe spectroscopy was utilized to
explore the control mechanism of photofragmentation of
CHBr$_2$COCF$_3$ (1,1-3,3,3 dibromo-trifluoroacetone) in AFC
experiments with intense shaped laser pulses. In particular,
optimization of the [CF$_3$]$^+$/[CHBr$_2$]$^+$ ratio revealed a
charge-transfer-based control mechanism. 

Selective control of molecular fragmentation in the gas phase was
demonstrated in several other AFC experiments with shaped femtosecond
laser pulses. W\"{o}ste and co-workers \cite{VajdaBartelt2001,
  BarteltMinemoto2001EPJD, LindingerLupulescu2003SAB,
  BarteltLindinger2004PCCP} controlled photoionization and
photofragmentation dynamics of alkali clusters. Jones and co-workers
\cite{WellsBetsch2005} optimized the S$_N^+$/S$_M^+$ ratios for
various $N$ and $M$ values in strong-field photofragmentation of S$_8$
molecules. It was found that optimally shaped pulses dramatically
outperform the transform-limited pulses. Wells \emph{et al.}
\cite{WellsMcKenna2010JPB} controlled the vibrational population
distribution in the transient CO$^{2+}$ to manipulate the branching
ratio of the CO$^{2+}$ and $\mathrm{C}^{+} + \mathrm{O}^{+}$
products. Hill and co-workers \cite{ChenWangHill2009} controlled the
amplitude of the bending vibrational mode in highly ionized CO$_2$
(during strong-field Coulomb explosion) to enhance the symmetric
six-electron fragmentation channel, $\mathrm{CO}_2^{6+} \rightarrow
\mathrm{O}^{2+} + \mathrm{C}^{2+} + \mathrm{O}^{2+}$. They constrained
the search space by expressing the spectral phase of the laser pulse
as a Taylor series, in order to elucidate the controlled
photodissociation dynamics. Laarmann \emph{et al.}
\cite{LaarmannShchatsinin2007JCP, LaarmannShchatsinin2008JPB} achieved
selective cleavage of strong backbone bonds in amino acid complexes
(in particular, a peptide bond in Ac-Phe-NHMe and Ac-Ala-NHMe), while
keeping weaker bonds intact. Based on these results, they suggested
the possibility of employing AFC with optimally tailored laser pulses
as an analytical tool in mass spectrometry of complex polyatomic
systems (with potential applicability, e.g., to protein sequencing of
large biopolymers). In a recent AFC experiment, Levis and co-workers
\cite{Palliyaguru2008} used intense laser pulses to manipulate
branching ratios of various photofragmentation products of dimethyl
methylphosphonate (DMMP), a simulant for the nerve agent sarin. The
optimization in this experiment was performed in the presence of a
high background of a hydrocarbon and water in the extraction region of
a time-of-flight mass spectrometer. The ability to achieve highly
selective control under these conditions demonstrates that AFC may
provide the means to identify complex airborne molecules. As mentioned
in section~\ref{sec:AFC-hh}, photofragmentation of SF$_6$ was
controlled (including optimization of the ratio
[SF$_5$]$^+$/[SF$_3$]$^+$) in an AFC experiment
\cite{PfeiferSpitzenpfeil2007} that used spectrally shaped femtosecond
X-ray fields produced from intense shaped laser pulses via
high-harmonic generation.

Recently, Dantus and co-workers \cite{LozovoyZhuDantus2008,
  ZhuGunaratneLozovoy2009} reported a study of molecular fragmentation
using intense femtosecond pulses, in which they did not use
algorithm-guided AFC, but rather evaluated a large set of
predetermined pulse shapes. They concluded that the yields of
photofragmentation products are mainly controlled by the pulse
intensity, while the details of the pulse shape (e.g., the spectral
phase distribution) are not important. The contradiction between this
conclusion and the results of numerous AFC experiments discussed above
appears to be explained by the fact that Dantus and co-workers
performed their measurements \cite{LozovoyZhuDantus2008} at laser
intensities above the saturation threshold for ionization. As pointed
out by Levis \cite{Levis2009}, the experimental conditions employed by
the Dantus group obscure connections to coherent processes which occur
below the saturation threshold. Since these coherent processes play
the main role in control mechanisms underlying reaction selectivity
achieved by optimally shaped laser pulses, the proper experimental
conditions must be satisfied for coherent control of photoinduced
molecular fragmentation. Moreover, the search over a set of
predetermined pulse shapes \cite{LozovoyZhuDantus2008} cannot
guarantee discovery of optimal controls, even if the employed set is
very large. In a parameter space of the size characteristically
available from a typical pulse shaper, only a dedicated optimization
algorithm is capable of consistently identifying optimal control
fields which commonly lie in a null space of the full search
space. This situation is fully consistent with classical engineering
control practice where optimization is the design tool employed almost
without exception to meet complex operational objectives.

\subsection{AFC of multiphoton ionization in molecules}

The use of polarization-shaped femtosecond laser pulses can
significantly enhance the level of control over multiphoton ionization
in molecules. In 2004,
Brixner~\emph{et~al.}~\cite{BrixnerKrampert2004} demonstrated that a
suitably polarization-shaped laser pulse increased the photoionization
yield in K$_2$
beyond that obtained with an optimally shaped linearly polarized laser
pulse. This effect is explained by the existence of different
multiphoton ionization pathways in the molecule involving dipole
transitions which are preferably excited by different polarization
directions of the laser field. Suzuki \emph{et al.}
\cite{Suzuki2004PRL} applied AFC with polarization-shaped laser pulses
to multiphoton ionization of I$_2$ molecules and optimized the
production of oddly charged (I$_2^{+}$ and I$_2^{3 +}$) and evenly
charged (I$_2^{2 +}$) molecular ions. Weber \emph{et al.}
\cite{Weber2008} performed AFC experiments with polarization-shaped
laser pulses to optimize the photoionization yield in NaK
molecules. Free optimization of the pulse phase, amplitude, and
polarization resulted in a higher ionization yield than parameterized
optimization with a train of two pulses.

W\"{o}ste and co-workers \cite{LupulescuLindingerPlewicki2004CP,
  WeberLindinger2004CP, SchaferBungMitric2004, LindingerWeber2005,
  BarteltFeurer2005, LindingerWeberMerli2006} investigated AFC of
multiphoton ionization in K$_2$ and NaK using femtosecond laser pulses
with phase and amplitude modulation, but without polarization
shaping. In particular, good agreement between the optimal shapes of
laser pulses obtained theoretically (via QOCT) and experimentally (via
AFC) was reported \cite{SchaferBungMitric2004}. Significant attention
was also devoted to studying properties of the controlled ionization
dynamics and revealing underlying control mechanisms. In one series of
AFC experiments \cite{LindingerWeber2005}, the photoionization
dynamics was explored using control pulse cleaning (CPC), which is a
process of removing extraneous control field features by applying
pressure in the optimization algorithm with an appropriate cost
function on the spectral components of the pulse
\cite{GeremiaZhu2000}. Weak pressure applied in the case of
isotope-selective ionization of K$_2$ was sufficient to remove
unnecessary pulse components and thus expose the participating
vibronic transitions. For ionization of NaK, strong pressure was
applied and multi-objective optimization was performed. The resultant
Pareto-optimal curve revealed the correlation of the two conflicting
objectives of maximizing the ionization yield versus cleaning the
control pulse. The optimal ionization pathway depends on the CPC
strength, which helps to identify the important electronic transitions
to particular vibrational states. These results demonstrate that the
spectra of optimal pulses obtained with CPC contain important
information about the control mechanism. In another series of AFC
experiments \cite{BarteltFeurer2005}, the control mechanism of
multiphoton ionization in NaK was analyzed by systematically reducing
the complexity of the search space. The spectral phase function of the
control pulse was expressed as a truncated Fourier series, whose
parameters were examined with respect to the ionization yield and the
obtained optimal field. By progressively reducing the number of phase
modulation parameters, it was possible to generate optimized pulses
that allowed for a simple mechanistic interpretation of the controlled
dynamics. In an earlier study, Leone and co-workers
\cite{BallardStauffer2002} applied AFC to optimize the weak-field
pump-probe photoionization signal in Li$_2$ and used first-order
time-dependent perturbation theory to investigate the dynamics of a
rotational wave packet excited by the pump pulse and explain the
corresponding control mechanism.

W\"{o}ste and co-workers \cite{WeberLindinger2004CP,
  LindingerVetter2004CPL, Lindinger2004, LindingerLupulescu2005} also
demonstrated that AFC is capable of achieving isotope-selective
ionization of diatomic molecules such as K$_2$ and NaK. They showed
that optimally tailored control pulses can increase the divergence
between the dynamics of excited vibrational wave packets in distinct
isotopomers (these studies will be discussed in more detail in
section~\ref{sec:AFC-ODD} below).

\subsection{AFC of molecular alignment}

The controlled alignment of molecules has attracted considerable
attention as it can provide a well defined sample for subsequent
additional control experiments. At high laser intensities ($\sim
10^{13}$--$10^{14}$~W/cm$^2$), dynamical variations of the molecular
polarization can have a significant effect on alignment. By shaping
the temporal profile of such an intense femtosecond laser pulse, it is
possible to achieve control over molecular alignment. Quantum dynamics
of laser-induced molecular alignment is amenable to theoretical
treatment and optimization \cite{Shir2007, SiedschlagShir2006OC,
  ShirBeltrani2008JPB, RouzeeGijsbertsen2009NJP,
  LeibscherAverbukhRabitz2003, LeibscherAverbukhRabitz2004}, and can
be successfully controlled using simple ultrafast laser pulses
\cite{StapelfeldtSeideman2003, BisgaardPoulsen2004, RenardHertz2004,
  RenardHertz2005, LeeVilleneuve2006}. AFC provides a very effective
general laboratory tool for alignment manipulation, making the best
use of the laser resources. AFC of molecular alignment with intense
shaped laser pulses was successfully demonstrated at room temperature
for N$_2$ \cite{HornWollenhaupt2006, deNaldaHorn2007} and CO
\cite{PinkhamMooneyJones2007}.

\subsection{Applications of AFC in nonlinear molecular spectroscopy} 

Shaped femtosecond laser pulses were successfully used to enhance
resolution and improve detection in several areas of nonlinear
spectroscopy and microscopy \cite{Silberberg2009}. Of particular
interest are experiments that employ AFC to identify optimal pulse
shapes. One area of nonlinear molecular spectroscopy is control of
vibrational modes via stimulated Raman scattering (SRS). In the gas
phase, a number of AFC experiments \cite{HornungMeierMotzkus2000,
  WeinachtBartels2001CPL, BartelsWeinacht2002PRL} manipulated
molecular vibrations excited via SRS by intense ultrafast laser pulses
in the impulsive regime (i.e., when the duration of the control laser
pulse is shorter than the vibrational period). In one of these
experiments \cite{HornungMeierMotzkus2000}, control of the vibrational
dynamics of K$_2$ was achieved via impulsive SRS in a degenerate
four-wave-mixing optical setup. Different parameterizations of shaped
femtosecond laser pulses in the frequency and time domains were
employed to decipher the physical mechanism responsible for the
achieved control. In other gas-phase experiments, mode suppression and
enhancement in sulfur hexafluoride \cite{WeinachtBartels2001CPL},
mode-selective excitation in carbon dioxide
\cite{WeinachtBartels2001CPL}, and creation of shaped multimode
vibrational wave packets with overtone and combination mode excitation
in CCl$_4$ \cite{BartelsWeinacht2002PRL} were demonstrated using
impulsive SRS at room temperature and high pressures. In the liquid
phase, AFC was applied to control relative intensities of the peaks in
the Raman spectrum, corresponding to the symmetric and antisymmetric
C--H stretch modes of methanol \cite{Weinacht1999b, Pearson2001,
  WhitePearson2004, PearsonBucksbaum2004PRL}. The modes were excited
via SRS in the non-impulsive regime (i.e., the duration of the control
laser pulse exceeded the vibrational period). The control pulse was
shaped and the forward scattered Raman spectrum was measured to obtain
the feedback signal, with the goal of achieving selective control of
the vibrational modes. However, it was argued
\cite{SpannerBrumer2006a, SpannerBrumer2006b} that in non-impulsive
SRS the relative peak heights in the Raman spectrum do not reflect the
relative populations of the vibrational modes and that control of the
spectral features demonstrated in the experiments \cite{Weinacht1999b,
  Pearson2001, WhitePearson2004, PearsonBucksbaum2004PRL} does not
involve quantum interference of vibrational excitations, but rather is
based on classical nonlinear optical effects.

Another important area of nonlinear molecular spectroscopy is control
of molecular vibrational modes via CARS. Materny and co-workers
\cite{ZeidlerFrey2002, KonradiSingh2005, KonradiScaria2007,
  KonradiSingh2006JRS, KonradiSingh2006JPPA, ScariaKonradi2008},
Zhang~\emph{et~al.}~\cite{ZhangZhang2007} and von Vacano \emph{et
  al.}~\cite{VacanoWohllebenMotzkus2006} used AFC in a CARS setup to
optimally control vibrational dynamics in complex molecules. The
Stokes pulse was shaped and the feedback signal was derived from the
intensities observed in the CARS spectrum. In an AFC experiment with
polymers (fluorobenzene sulfonate diacetylenes), selective excitation
of one vibrational ground-state mode and suppression of all other
modes was achieved, and the decay times of different modes were
modified \cite{ZeidlerFrey2002}. In liquid-phase AFC experiments,
selective enhancement or suppression of one or more vibrational modes
was demonstrated for toluene \cite{KonradiSingh2005,
  KonradiScaria2007}, benzene \cite{ZhangZhang2007}, and
$\beta$-carotene in hexane solution \cite{KonradiSingh2006JRS}. In a
related experiment \cite{KonradiSingh2006JPPA}, AFC was used to obtain
molecule-specific CARS spectra from a mixture of benzene and
chloroform. Molecule-specific enhancement or suppression of the CARS
spectral lines achieved in this AFC experiment is an example of
optimal dynamic discrimination that will be discussed in more detail
in section~\ref{sec:AFC-ODD} below. Another related study
\cite{ScariaKonradi2008} compared selective excitation of molecular
vibrational modes (achieved in AFC experiments with shaped femtosecond
pulses) in the gas phase with carbon disulfide and the liquid phase
with toluene. Interestingly, it was found that the relative
intensities of the CARS spectral lines could be changed more
effectively in the liquid phase than in the gas phase. An experiment
reported by von Vacano \emph{et al.}
\cite{VacanoWohllebenMotzkus2006} employed the method of single-beam
CARS spectroscopy with shaping of broadband pulses from a photonic
crystal fiber. Each broadband pulse provides numerous pairs of pump
and Stokes frequencies, and the spectral phase of the pulse was
optimized with AFC to produce the desired interference pattern of the
molecular vibrational modes. The optimally compressed and shaped
pulses enabled the unambiguous assignment of the participating
vibrational modes of toluene between 500 and 1000 cm$^{-1}$ in a
blue-shifted CARS signal.

Control of molecular vibrational dynamics is possible not only through
Raman-type processes, but also directly in the infra-red (IR)
regime. Zanni and co-workers \cite{StrasfeldShim2007} demonstrated
selective control of vibrational excitations on the ground electronic
state of W(CO)$_6$, using shaped femtosecond mid-IR (5.2~$\mu$m,
1923~cm$^{-1}$) pulses. The spectral phase distribution of the pulse
was optimized using AFC to achieve selective population of the excited
vibrational levels of the $T_{1 u}$ CO-stretching mode. Systematic
truncation of optimal pulses was employed to analyze the control
mechanism. In a related AFC experiment \cite{StrasfeldMiddleton2009},
polarization-shaped mid-IR pulses were used to selectively control
vibrational excitations of the two carbonyl stretching modes in
Mn(CO)$_5$Br.

\subsection{Applications of AFC in multiphoton microscopy} 

An important application of AFC with shaped femtosecond laser pulses
is in multiphoton excited fluorescence (MPEF) microscopy. For a given
pulse energy, the transform-limited pulse has the maximum peak
intensity, which helps to increase the fluorescence signal intensity,
but unfortunately also increases the rate of photobleaching of the
molecules (which is especially undesirable with samples of live
cells). Use of optimally shaped pulses instead of a transform-limited
pulse can reduce the bleaching rate, enhance spatial resolution, and
increase contrast in biological fluorescence imaging. In a series of
AFC experiments with shaped laser pulses, Midorikawa and co-workers
\cite{Kawano2003, ChenKawano2004, TadaKono2007, IsobeSudaTanaka2009a}
optimally controlled MPEF microscopy in different fluorescent
biomolecules. Attenuation of photobleaching by a factor of four
(without decreasing the fluorescence signal intensity) was
demonstrated in two-photon excitation fluorescence (TPEF) from a green
fluorescent protein \cite{Kawano2003}. Another AFC experiment
\cite{ChenKawano2004} achieved selective control of two-photon and
three-photon fluorescence in a mixture of two biosamples. The use of
optimally tailored pulses helped to minimize the harmful three-photon
fluorescence from the amino acid L-Tryptophan, without a significant
loss of useful two-photon fluorescence from a green fluorescence
protein. Optimally shaped super-continuum pulses from a microstructure
fiber were used in TPEF microscopy in another experimental study
\cite{TadaKono2007}. The pulse was shaped prior to propagation through
the fiber, and AFC maximized the fluorescence signal contrast between
two fluorescent proteins. A novel phase modulation technique for
ultra-broadband laser pulses was developed for selective excitation of
multiple fluorophores in TPEF microscopy \cite{IsobeSudaTanaka2009a}.
This technique was applied to dual-color imaging of cells containing
two types of fluorescent proteins, and AFC was employed to find the
phase modulation that maximizes or minimizes the individual TPEF
intensity from one of the fluorophores.

\subsection{Applications of AFC for optimal dynamic discrimination}
\label{sec:AFC-ODD}

Discrimination of similar systems is important for many practical
problems in science and engineering. In particular, selective
identification of target molecules in a mixture of structurally and
spectroscopically similar compounds is a challenge in such areas as
selective excitation of multiple fluorescent proteins in microscopy of
live samples, targeted component excitation in solid-state arrays, and
selective transformation of chemically similar molecules. Theoretical
studies \cite{LiTuriniciRamakhrishna2002, TuriniciRamakhrishnaLi2004,
  LiRabitzWolf2005, LiZhuRabitz2006, BeltraniGhosh2009} indicate that
quantum systems differing even very slightly in structure may be
distinguished by means of their dynamics when acted upon by a suitably
tailored ultrafast control field. Such optimal dynamic discrimination
(ODD) can in principle achieve dramatic levels of control, and hence
also provides a valuable test of the fundamental selectivity limits of
quantum control despite noise and constrained laser resources. AFC
provides a very effective laboratory means for practical
implementation of ODD.

In 2001, Gerber and co-workers \cite{BrixnerDamrauer2001}
experimentally demonstrated selective multiphoton excitation of two
complex molecules, a laser dye DCM and [Ru(dpb)$_3$](PF$_6$)$_2$, in
methanol solution. The goal was to electronically excite DCM while
simultaneously suppressing electronic excitation of [Ru(dpb)$_3$]$^{2
  +}$. While these two molecules are electronically and structurally
distinct, the DCM/[Ru(dpb)$_3$]$^{2 +}$ emission ratio is practically
unaffected by variations in single control parameters such as
wavelength, intensity, and linear chirp. Nevertheless, selective
excitation was successfully achieved using AFC with shaped femtosecond
laser pulses. The DCM/[Ru(dpb)$_3$]$^{2 +}$ emission ratio was used as
the feedback signal, and the evolutionary algorithm identified
optimally shaped control pulses that improved the signal by
approximately 50\%. These results obtained in the presence of complex
solvent/solute interactions aroused significant interest to ODD. As
mentioned in section~\ref{sec:AFC-fragment}, Gerber and co-workers
\cite{BergtBrixnerDietl2002} demonstrated AFC of photoproduct
branching ratios in CpFe(CO)$_2$Cl and CpFe(CO)$_2$Br. Despite the
chemical similarity of these two molecules, AFC was sensitive enough
to detect differences due to the electronic metal-halogen bonding
properties. This finding suggests the possibility of performing ODD of
individual compounds in mixtures of chemically similar
molecules. Other examples of ODD include molecule-specific
manipulation of CARS spectra from a mixture of benzene and chloroform
\cite{KonradiSingh2006JPPA}, selective excitation of multiple
fluorophores in TPEF microscopy \cite{TadaKono2007,
  IsobeSudaTanaka2009a}, and quantitative differentiation of dyes with
overlapping one-photon spectra \cite{Tkaczyk2010JLumin}.

A recent experimental demonstration of ODD by
Roth~\emph{et~al.}~\cite{RothGuyonRoslund2009} achieved distinguishing
excitations of two nearly identical flavin molecules in aqueous~phase.
The absorption spectra for flavin mononucleotide (FMN) and riboflavin
(RBF)~are practically indistinguishable throughout the entire visible
and far UV. This~implementation of ODD used a shaped UV pulse centered
at 400~nm and a time-delayed unshaped IR pulse centered at 800~nm. The
first pulse creates a coherent vibrational wave packet on an excited
electronic state, and the second pulse disrupts the wave packet motion
and results in additional excitation to a higher state and
consequential depletion of the recorded fluorescence signal. The
effect of slight differences in the vibronic structure of the two
molecules upon the dynamics of the excited wave packets is amplified
by tailoring the spectral phase of the UV pulse. Since further
excitation produced by the second pulse depends on the precise
structure, position, and coherence of the tailored wave packet, it is
possible to dynamically interrogate the two statically nearly
identical systems and thereby produce a discriminating difference in
their respective depleted fluorescence signals. In contrast, if the UV
pulse is transform-limited, then the fluorescence depletion signals
from the flavins are indistinguishable. UV pulse shapes that optimally
discriminate between FMN and RBF were discovered using AFC. The
optimized depletion ratio $D_{\mathrm{FMN}}/D_{\mathrm{RBF}}$ could be
changed by $\sim \pm 28$\%, despite the initially indistinguishable
linear and nonlinear optical spectra. Although the laser resources
consisted of a modest $\sim$~3.5~nm of UV bandwidth and $\sim$~10~nm
of IR bandwidth, significant selectivity was achieved with optimal UV
pulses working in concert with the time-delayed unshaped IR pulse.
Although the static spectra appear nearly identical, subtle
differences in the molecular structure are nonetheless made profound
during the tailored evolution of wave packets generated by optimal
controls. System complexity (e.g., high vibrational state density,
thermal population, solvent-induced line broadening) effectively
enhances the ODD capabilities of the control field and compensates for
the limited bandwidth constraint, thus making dramatic levels of
control possible even in the weak-field limit.

Another example of ODD is isotope-selective ionization of molecules
demonstrated in a number of AFC experiments by W\"{o}ste and
co-workers \cite{WeberLindinger2004CP, LindingerVetter2004CPL,
  Lindinger2004, LindingerLupulescu2005}. In particular, in an
illustrative study \cite{Lindinger2004, LindingerLupulescu2005}, they
applied shaped femtosecond laser pulses to the $^{39,39}$K$_2$ and
$^{39,41}$K$_2$ isotopomers and used AFC to maximize and minimize the
isotope ion ratio $R = I( ^{39,39}\mathrm{K}_2 )/ I(
^{39,41}\mathrm{K}_2 )$. K$_2$ molecules can be ionized in a
three-photon process at relatively low pulse energies within the
available wavelength range (810--833~nm). Differences between the
evolutions of vibrational wave packets on an excited electronic state
in the two isotopomers can be amplified by optimized control fields.
Operation in this fashion made it possible to achieve significant
selectivity of isotope ionization, with a variation by a factor of
$R_{\mathrm{max}}/R_{\mathrm{min}} \sim 140$ between the maximal and
minimal values of the isotope ion ratio. Information about the
dynamics of the controlled vibrational wave packets was extracted from
the optimal pulse shapes to help reveal ionization pathways
\cite{LindingerWeber2005, Lindinger2004, LindingerLupulescu2005}.

\subsection{AFC of energy flow in biomolecular complexes}

Applications of quantum control to increasingly complex molecular
systems have been considered. In particular, Motzkus and co-workers
used AFC with shaped femtosecond laser pulses to control and analyze
the energy flow pathways in the light-harvesting antenna complex LH2
of \emph{Rhodopseudomonas acidophila} (a photosynthetic purple
bacterium) \cite{Herek2002, WohllebenBuckup2003} and $\beta$-carotene
\cite{BuckupLebold2006}. They demonstrated that by shaping the
spectral phase distribution of the control pulse, it is possible to
manipulate the branching ratio of energy transfer between intra- and
inter-molecular channels in the donor-acceptor system of the LH2
complex \cite{Herek2002}. Analysis of the transient absorption data
was used to decipher the control mechanism and identify the molecular
states participating in energy transfer within LH2
\cite{WohllebenBuckup2003} and $\beta$-carotene
\cite{BuckupLebold2006}. These results indicate that coherent quantum
control is possible even in very complex molecular systems in a
condensed-phase environment.

\subsection{AFC of photoinduced electron transfer}

AFC has been applied to quantum control of inter-molecular
photoinduced electron transfer. Yartsev and co-workers
\cite{BruggemannYartsev2006} reported an AFC experiment that maximized
the yield of ultrafast electron injection from the sensitizer to
TiO$_2$ nanocrystals in the core part of a dye-sensitized solar
cell. The electron transfer process was monitored using the transient
absorption signal. The impulsive structure of the optimal laser pulse
was observed to correlate with the coherent nuclear motion of the
photoexcited dye. The pulse shape and the transient absorption
kinetics were explained by an impulsive stimulated (anti-Stokes) Raman
scattering process, followed by electronic excitation.

\subsection{AFC of photoisomerization in complex molecules}

The control of molecular structure transformations is a coveted goal
in chemistry. In particular, control of \emph{cis-trans} isomerization
has attracted much attention due to the importance of this process in
chemistry and biology (e.g., it is a primary step of vision). AFC of
\emph{cis-trans} photoisomerization in cyanines (in the liquid phase)
with shaped femtosecond laser pulses was first reported by Gerber and
co-workers \cite{VogtKrampert2005}.  This experiment demonstrated that
by using optimally shaped laser pulses it is possible to enhance or
reduce isomerization efficiencies. The mechanism underlying
isomerization control in this experiment was discussed in a number of
theoretical works \cite{HokiBrumer2005, HuntRobb2005,
  ImprotaSantoro2005}. In particular, Hoki and Brumer
\cite{HokiBrumer2005} suggested that isomerization control involves an
incoherent pump-dump process and that the role of quantum coherence
effects in the evolution of the excited vibrational wave packet is
negligible due to strong environmentally-induced decoherence. On the
other hand, Hunt and Robb \cite{HuntRobb2005} and Improta and Santoro
\cite{ImprotaSantoro2005} used a more sophisticated model and argued
that control of isomerization does rely on quantum coherence of the
photoexcited vibrational wave packet that moves on the
multidimensional potential-energy surface of an excited electronic
state. Moreover, Hunt and Robb \cite{HuntRobb2005} showed that the
generation of the \emph{trans} versus \emph{cis} product is affected
by the presence of an extended conical intersection seam on the
potential-energy surface, and that photoisomerization can therefore be
coherently controlled by tuning the distribution of momentum
components in the photoexcited vibrational wave packet. The validity
of this coherent control mechanism was corroborated in a further AFC
experiment by Yartsev and co-workers \cite{DietzekYartsev2006}. They
demonstrated that optimally shaped laser pulses can be used to modify
the momentum composition of the photoexcited wave packet and thus
achieve significant control of the absolute yield of isomerization
(i.e., the photoisomer concentration versus the laser pulse
energy). The coherent character of liquid-phase control of
\emph{cis-trans} photoisomerization in cyanines was further studied in
another AFC experiment by Yartsev and co-workers
\cite{DietzekYartsev2007}. They used a control scheme with an unshaped
pump pulse and a time-delayed shaped dump pulse (an unshaped probe
pulse was also applied to measure the effect of control). By using the
optimally shaped dump pulse, they achieved control of
photoisomerization closer to the decisive points of the reaction. This
approach made it possible to explore the effect of the wave-packet's
momentum composition at different time scales and assign the dynamics
to distinct parts of the excited-state potential. 

AFC of the retinal molecule in bacteriorhodopsin (from the
all-\emph{trans} to the 13-\emph{cis} state) was demonstrated by
Miller and co-workers \cite{ProkhorenkoNagy2006}. This experiment
employed both phase and amplitude modulation of femtosecond laser
pulses and operated in the weak-field regime (with pulse energies of
16--17~nJ). By using optimally shaped pulses, it was possible to
enhance or suppress the quantity of molecules in the 13-\emph{cis}
state by about 20\%, relative to the yield observed using a
transform-limited pulse with the same energy. They further explored
the mechanism of coherent control of retinal photoisomerization in
bacteriorhodopsin using time- and frequency-resolved pump-probe
measurements \cite{Prokhorenko2007}.  Experimental data together with
a theoretical analysis suggest that the isomerization yield depends on
the coherent evolution of the photoexcited vibrational wave packet on
an excited-state potential-energy surface in the presence of a conical
intersection. According to this analysis, control of retinal
photoisomerization is dominated by amplitude modulation of the
spectral components of the excitation pulse. Gerber and co-workers
\cite{VogtNuernberger2006} also demonstrated AFC of retinal
isomerization in bacteriorhodopsin. In this experiment, they pioneered
the control scheme with unshaped-pump and time-delayed shaped-dump
femtosecond laser pulses. As mentioned above, the use of the optimally
shaped dump pulse makes it possible to control the molecule in a
region of the potential-energy surface where the decisive reaction
step occurs. Moreover, by changing the time delay between the pulses,
it is possible to obtain information on the wave packet evolution.

The role of quantum coherence effects in control of retinal
isomerization in bacteriorhodopsin is still not fully clear, as a
recent experiment by Bucksbaum and co-workers
\cite{FloreanCardoza2009} found no dependence of the isomerization
yield on the control pulse shape at pulse energies below 30~nJ. In the
high-intensity regime (at pulse energies above 30 nJ), they found that
the yield of the 13-\emph{cis} isomer is maximized by a
transform-limited pulse, which could indicate that the yield depends
only on the pulse intensity, with quantum coherence not playing a
significant role. These findings (especially for lower intensities)
apparently contradict the optimization results obtained by Miller and
co-workers \cite{ProkhorenkoNagy2006}. It is possible that these
discrepancies could be explained by differences in experimental
setups. In particular, Bucksbaum and co-workers
\cite{FloreanCardoza2009} used only phase modulation of the control
pulse, whereas Miller and co-workers \cite{ProkhorenkoNagy2006,
  Prokhorenko2007} argued that control is mainly achieved by amplitude
modulation. Additional experimental and theoretical work will be
needed to fully explore the mechanisms underlying condensed-phase
control of photoisomerization in complex molecular systems and clarify
the role of quantum coherence in the controlled dynamics. For example,
recent theoretical studies \cite{HokiBrumer2009CPL,
  KatzRatnerKosloff2010NJP} suggest that coherent control of
photoisomerization and other branching reactions in an excited state
may be affected by and, moreover, take advantage of
environmentally-induced relaxation effects. Such cooperation between
coherent control and environmentally-induced decoherence may be
important in various quantum phenomena \cite{ShuangRabitz2006JCP} and
hence its optimal exploitation deserve further investigation.

Other examples of structural transformations in complex molecules
include ring opening in cyclohexadiene along with isomerization as
well as cyclization reactions in \emph{cis}-stilbene. Carroll \emph{et
  al.} \cite{CarrollPearson2006, CarrollWhite2008} demonstrated AFC of
the photoinduced ring-opening reaction of 1,3-cyclohexadiene (CHD) to
form 1,3,5-cis-hexatriene (Z-HT). The feedback signal was obtained
from measurements of the UV absorption spectrum. The learning
algorithm was able to identify optimal pulse shapes that increased the
formation of Z-HT by a factor of two. For a different control
objective, the AFC optimization produced pulse shapes that decreased
solvent fragmentation while leaving the formation of Z-HT essentially
unaffected. Kotur \emph{et al.} \cite{KoturWeinacht2009} used AFC with
shaped ultrafast laser pulses in the deep UV to control the ring
opening reaction of CHD to form 1,3,5-hexatriene (HT). The feedback
signal was obtained from measurements of fragmentation products
following strong-field ionization with a time-delayed IR laser
pulse. The learning algorithm discovered shaped UV pulses that
increased the HT yield by $\sim$~37\% relative to an unshaped (nearly
transform-limited) pulse of the same energy.  Greenfield \emph{et al.}
\cite{GreenfieldMcGrane2009} demonstrated AFC of the
photoisomerization and cyclization reactions in \emph{cis}-stilbene
dissolved in \emph{n}-hexane. This experiment employed phase-modulated
266~nm femtosecond pulses to maximize or minimize the yields of
\emph{cis}- to \emph{trans}-stilbene isomerization as well as
\emph{cis}-stilbene to 4$a$,4$b$-dihydrophenanthrene cyclization. The
yields of both isomerization and cyclization were minimized by
transform-limited pulses that enhanced competing multiphoton
processes, while the yields were maximized by complex pulse shapes
that helped to avoid multiphoton effects.

\subsection{AFC of nuclear motion in fullerenes}

Fullerenes are a class of molecules of considerable interest in many
areas of science. Laarmann \emph{et al.} \cite{Laarmann2007} employed
AFC-optimized femtosecond laser pulses to coherently excite
large-amplitude oscillations in C$_{60}$ fullerene. The structure of
the optimal pulses in combination with complementary two-color
pump-probe data and time-dependent density functional theory
calculations provided information on the underlying control
mechanism. It was found that the strong laser field excites many
electrons in C$_{60}$, and the nuclear motion is excited, in turn, due
to coupling of the electron cloud to a radially symmetric breathing
mode. Despite the complexity of this multi-particle system with
various electronic and nuclear degrees of freedom, the optimal control
fields generated essentially one-dimensional oscillatory motion for up
to six cycles with an amplitude of $\sim 130$\% of the molecular
diameter.

\subsection{Applications of AFC in semiconductors}

Quantum control has found applications beyond atomic and molecular
phenomena. In particular, it is possible to use optimal control
methods to manipulate various processes in semiconductors. Kunde
\emph{et al.} \cite{Kunde2000, Kunde2001} demonstrated AFC of
semiconductor nonlinearities using phase-modulated femtosecond laser
pulses, with the purpose of creating an ultrafast all-optical switch.
The feedback signal was obtained by measuring the differential
transmission (DT) in a control-probe setup. Optimizations were
performed on the spectrally integrated DT as well as DT in narrow
spectral windows. The learning algorithm was able to identify optimal
pulse shapes that enhanced ultrafast semiconductor nonlinearities by
almost a factor of four. Chung and Weiner \cite{ChungWeiner2006} used
AFC with phase-modulated femtosecond laser pulses to coherently
control two-photon-induced photocurrents in two different
semiconductor diodes. Because of their spectrally distinct two-photon
absorption responses, the diodes generated noticeably different
photocurrent yields depending on the pulse shape. An evolutionary
algorithm guided the AFC experiment to discover pulse shapes that
maximize or minimize the photocurrent yield ratio for the two diodes.

\subsection{AFC of decoherence}
\label{sec:AFC-decoherence}

Manipulation of quantum interference requires that the system under
control remains coherent, avoiding (or at least postponing) the
randomization induced by coupling to an uncontrolled
environment. Therefore, the ability to manage environmentally-induced
decoherence would bring substantial advantages to control of many
physical and chemical phenomena. In particular, decoherence is a
fundamental obstacle to quantum information processing
\cite{NielsenChuang2000}, and therefore the ability to protect quantum
information systems against decoherence is indispensable.

The possibility of using AFC for optimal suppression of decoherence
was first proposed by Brif \emph{et al.} \cite{Brif2001}, and
numerical simulations in a model system were performed by Zhu and
Rabitz \cite{ZhuRabitz2003}. Walmsley and co-workers
\cite{Branderhorst2008} have recently used AFC to achieve coherent
control of decoherence of molecular vibrational wave packets in the
laboratory. The concept underlying this experiment is the use of
coherent preparation of the quantum system to alter non-unitary
decoherent dynamics induced by an uncontrolled environment. In this
experiment, a gas-phase ensemble of K$_2$ at $400\,^{\circ}\mathrm{C}$
was irradiated by a shaped femtosecond laser pulse, inducing a
vibrational wave packet in the lowest excited electronic state $A^1
\Sigma_{\mathrm{u}}^{+}$ of the molecules. This wave packet undergoes
dephasing (a form of decoherence that does not involve dissipation of
energy). Dephasing is caused by coupling of the vibrational mode to
the thermalized rotational quasi-bath. The amplitude of quantum beats
in the fluorescence signal (measured at a chosen delay time after the
excitation pulse) served as the feedback signal. This amplitude
provides an estimate of the degree of wave packet localization in the
phase space and therefore can be used as a coherence surrogate. The
optimal pulse identified by AFC increased the quantum-beat visibility
from zero to more than four times the noise level and prolonged the
coherence lifetime by a factor of $\sim 2$ relative to the beats
produced by the transform-limited pulse. The main characteristic of
the optimal pulse is a high degree of linear negative chirp. This
indicates that the mechanism of decoherence control is based on
exciting a wave packet which is initially amplitude-squeezed in the
phase-space representation. A theoretical analysis confirmed that the
coherence lifetime is extended if the initial state is
amplitude-squeezed (corresponding to the initial orientation in the
phase space along the classical trajectory).

A well-known strategy for suppressing decoherence in quantum systems
is through application of pulse sequences designed to dynamically
decouple the system from the environment \cite{Haeberlen1976,
  ViolaLloyd1998, ViolaKnillLloyd1999, Zanardi1999, VitaliTombesi1999,
  VitaliTombesi2001, ByrdLidar2003, KhodjastehLidar2005,
  FacchiTasaki2005, ViolaKnill2005, Uhrig2007, PasiniUhrig2010,
  UhrigPasini2010NJP}.  Experiments \cite{FravalSellars2005,
  MortonTyryshkin2006, MortonTyryshkin2008,
  DamodarakurupLucamarini2009, SagiAlmogDavidson2009,
  SagiAlmogDavidson2010} have employed theoretically designed pulse
sequences based on particular environment models (i.e., an example of
open-loop control). However, the actual noise power spectra for
realistic environments can significantly differ from the models. To
overcome this difficulty, Bollinger and co-workers
\cite{BiercukBollinger2009} recently used AFC to tailor the dynamical
decoupling pulse sequence to an actual experimental noise
environment. In this experiment, the system was a quantum memory
realized in an array of $\sim 1000$ $^9$Be$^+$ ions (cooled to a
temperature of $\sim 1$~mK) in a Penning ion trap, with qubit states
realized using a ground-state electron-spin-flip transition. These
qubit states are highly susceptible to magnetic field fluctuations,
making such noise a significant source of decoherence. The qubits were
coherently controlled by directly driving the $\sim 124$~GHz
transition in a microwave setup similar to optical ones. A sequence of
microwave $\pi$ pulses used for qubit control in this laboratory
configuration is technologically quite different from shaped
femtosecond optical laser pulses typically employed in molecular
control experiments; however, the fundamental concept of AFC is still
fully applicable. The feedback signal was obtained from fluorescence
detection on a cycling transition (with decoherence-induced errors
manifested as non-zero fluorescence). The Nelder-Mead simplex method
was utilized to search for optimal pulse positions in a fixed-length
sequence of $n$ pulses. Optimal pulse sequences discovered in the AFC
experiment, without \emph{a priori} knowledge of the noise
environment, suppressed the qubit error rate by a factor of five to
ten relative to benchmark model-based sequences.

\subsection{Algorithmic advances for laboratory AFC}
\label{sec:AFC-algorithms}

The learning algorithm is an important component of laboratory AFC.
The majority of AFC experiments employ stochastic algorithms such as
evolutionary strategies \cite{Schwefel1995} and genetic algorithms
\cite{Goldberg2007}. Historically, genetic algorithms were
characterized by the use of recombination, while evolutionary
strategies primarily relied on mutation; however, modern algorithms
guiding AFC experiments and simulations typically incorporate both
types of genetic operations and are variably called genetic algorithms
or evolutionary algorithms. These algorithms are very well suited to
laboratory optimizations as they naturally match the discrete
structure of control ``knobs'' (e.g., the pixels of a pulse shaper)
and are robust to noise. Moreover, robustness to noise in AFC
experiments can be enhanced by incorporating the signal-to-noise ratio
into the control objective functional \cite{GeremiaZhu2000,
  BarteltRoth2005}. Various aspects of evolutionary algorithms and
their application to AFC of quantum phenomena were
assessed~\cite{ZeidlerFreyKompa2001, ShirSiedschlag2006}.
Evolutionary algorithms can be also used in multi-objective
optimization \cite{FonsecaFleming1995, Deb1999}, and the application
of this approach to quantum control problems was studied theoretically
\cite{Shir2007, RajWu2008a, RajWu2008b, Gollub2009NJP} and
demonstrated in AFC experiments \cite{LindingerWeber2005,
  BonacinaWolf2007}.

Other types of stochastic algorithms include, for example, simulated
annealing \cite{Kirkpatrick1983} and ant colony optimization
\cite{Dorigo1996, Bonabeau2000}. Simulated annealing was utilized in
some AFC experiments \cite{MeshulachYelinSilberberg1998,
  IsobeSudaTanaka2009a}, and it seems best suited to situations where
just a few experimental parameters are optimized \cite{Feurer1999,
  GlassRozgonyi2000}. Ant colony optimization recently has been used
in an AFC simulation \cite{GollubVivieRiedle2009}, but it is yet to be
tested in the laboratory.

As mentioned in section~\ref{sec:Landscapes}, the absence of local
traps in control landscapes for controllable quantum systems has
important practical consequences for the optimization complexity of
AFC experiments. In particular, deterministic search algorithms can be
used to reach a globally optimal solution. Deterministic algorithms
(in particular, the downhill simplex method) were successfully
implemented in several AFC experiments \cite{BrixnerOehrlein2000,
  GreenfieldMcGrane2009, BiercukBollinger2009}. Recently, Roslund and
Rabitz \cite{RoslundRabitz2009} demonstrated the efficiency of a
gradient algorithm in laboratory AFC of quantum phenomena. They
implemented a robust statistical method for obtaining the gradient on
a general quantum control landscape in the presence of noise. The
experimentally measured gradient was utilized to climb along
steepest-ascent trajectories on the landscapes of three quantum
control problems: spectrally filtered SHG, integrated SHG, and
excitation of atomic rubidium. The optimization with the gradient
algorithm was very efficient, as it required approximately three times
fewer experiments than needed by a standard genetic algorithm in these
cases. High algorithmic efficiency is especially important for AFC of
laser-driven processes in live biological samples, as damage (e.g.,
due to photobleaching) can be reduced by decreasing the number of
trial laser pulses. Still, evolutionary or other stochastic algorithms
may be preferable over deterministic algorithms in many AFC
experiments due to their inherent robustness to noise. Hybrid
stochastic-deterministic algorithms (e.g., derandomized evolution
strategies) seem to offer the most flexibility and efficiency
\cite{RoslundShir2009}.


\section{The role of theoretical quantum control designs in the laboratory}
\label{sec:Designs}

A very significant portion of theoretical research in the area of
quantum control is devoted to model-based computations which employ
QOCT (or other similar methods) to design optimal control fields for
various physical and chemical problems. Such computations are
widespread in theoretical studies of molecular applications of quantum
control and are becoming increasingly popular in considering control
of quantum information systems, including optimal operation of quantum
gates and optimal generation of entanglement (see
section~\ref{sec:OCT-applications}). Notwithstanding these extensive
QOCT-based control field designs, experiments seeking optimal control
of molecular processes overwhelmingly employ AFC methods as described
in section~\ref{sec:AFC}. Such experiments in most cases work
remarkably well with random initial trials, and thus exhibit no
evident need to operate or possibly start with theoretical control
designs. This raises important questions about the practical
usefulness of open-loop control and role of theoretical methods such
as QOCT in control experiments \cite{Rabitz2003}. In considering this
matter it is important to keep in mind that the AFC procedure grew out
of observations from QOCT simulations and associated analyses.

\subsection{The effect of system complexity}

As discussed in section~\ref{sec:OCT-limitations}, the practical
laboratory relevance of theoretical designs depends on the complexity
of the system of interest, with simpler cases yielding theoretical
models closer to reality. For example, in numerous NMR experiments
employing RF fields to manipulate nuclear spins \cite{Abragam1983,
  Ernst1990, Freeman1998, Levitt2008, Slichter2010}, including NMR
realizations of quantum gates \cite{Cory2000FP, Jones2000FP,
  Jones2001PNMRS, VandersypenChuang2005}, theoretically designed
sequences of pulses (some of which were developed using QOCT
\cite{RyanNegrevergne2008, KhanejaReiss2005}) can function quite
well. The model of a collection of spins (with empirical coupling and
decay constants) interacting with classical fields is often
sufficiently accurate for NMR-based applications, allowing for
successful employment of open-loop control. A QOCT-based design was
also successfully applied experimentally to enhance robustness of
single-qubit gate operations in a system of trapped ions
\cite{TimoneyElman2008}.

At the other extreme of system complexity are electronic and
vibrational processes in polyatomic molecules whose dynamics cannot be
accurately modelled at the present time. An objective assessment is
that models used for polyatomic molecules in control computations are
currently too simplified and computational techniques inadequate for
the true levels of complexity, resulting in theoretical designs that
are not directly applicable to control experiments which work with
\emph{real} systems. There are several aspects of laser control of
molecules, which make the difference between theoretical models and
actual systems very important. First, optimal control is generally
based on creating interference of multiple quantum pathways, which can
be very sensitive to the detailed properties of the system (e.g.,
evolution of a laser-induced vibrational wave packet can be strongly
influenced by small variations of a molecular potential energy
surface, as evident from experimental ODD results
\cite{RothGuyonRoslund2009}). Therefore, even small inaccuracies of
theoretical models or associated numerical procedures may result in
control designs that are not optimal for the actual systems. Second,
the wavelength of the laser field is typically much larger than the
molecule, which makes it impossible to focus the field on a particular
group of atoms. In many applications, the goal is to attain control
(e.g., generate an excitation or break a bond) in a localized part of
a molecule. However, the external control field interacts with the
molecule as a whole, possibly exciting a multitude of accessible
transitions within the bandwidth of the laser (including multiphoton
processes). The ``global'' effect of photonic reagents becomes ever
more important with stronger fields for which resonance-based
``localization'' is not valid. An \emph{ab initio} quantitative
theoretical account of laser-driven molecular dynamics is presently
not feasible, unless the studies are limited to cases of very simple
molecules and weak fields. Third, it is difficult to calibrate the
laser and pulse-shaping apparatus to reliably reproduce theoretical
control designs in the laboratory. In many cases, directly using AFC
optimization is much easier and much more effective, than calibrating
the laser and pulse-shaper for generation of theoretically computed
control fields with the required accuracy.

These considerations lead to the conclusion that open-loop control
experiments employing theoretical designs may be useful for some
systems and impractical for others, depending on how well the system
is known and which computational capabilities are available,
consequently determining how accurately the controlled dynamics can be
modelled. Thus, the boundary between the systems for which modelling
is sufficiently reliable and the systems for which it is not, depends
on available Hamiltonian data, numerical algorithms and computational
power. Of course, with time, better modelling will become available
for more complex systems, although the exponential increase of the
Hilbert-space dimension with the system complexity is a fundamental
feature of quantum mechanics, which significantly hinders the
effectiveness of numerical control designs for practical laboratory
implementation. 

Consider, for example, quantum information processing systems which
are typically modelled as collections of two-level particles (qubits)
with controlled interactions between them \cite{NielsenChuang2000}.
\emph{Prima facie}, such a system appears to be quite simple, so that
controls for all desired transformations can be theoretically designed
(e.g., using QOCT). However, the difficulty of accurately modelling
the actual environmental noise is significant even for simple
few-qubit systems. As was recently demonstrated with trapped-ion
qubits, dynamical decoupling pulse sequences obtained via AFC
significantly outperformed the best available theoretical designs
\cite{BiercukBollinger2009}. Moreover, as the Hilbert-space dimension
increases exponentially with the number of qubits, the unwanted
effects of uncontrolled couplings between the qubits in multi-particle
systems will be very difficult to model and, consequently, to manage
via open-loop control. Therefore, the effectiveness of theoretical
control designs for realistic quantum computers will depend on the
ability to engineer systems in which couplings between small blocks of
qubits can be made arbitrarily small. Hopefully, further technological
advances will make such systems available, which in turn will make
open-loop control with QOCT-based designs useful for practical quantum
computing.

\subsection{Importance of theoretical control designs for feasibility
  analysis} 

In molecules, the interactions between the atoms are inherently strong
in order to hold the atoms together. Therefore, in the foreseeable
future, for optimal manipulation of electronic and vibrational
processes in molecules with four and more atoms, AFC will continue to
be much more effective than employing theoretical control
designs. Notwithstanding this assessment, theoretical control studies
should continue to have high significance; however, for complex
systems the value of theoretical studies is not in generating specific
control designs for immediate laboratory use. Control solutions
obtained via theoretical model-based computations (in particular,
those employing QOCT) should play an important role by advancing the
general understanding of the character of controlled dynamics and
control mechanisms. One practically important issue is that while each
cycle of a typical AFC experiment is very fast (from microseconds to
milliseconds) and cheap, the initial setup of the experiment is
usually quite difficult and expensive, since advanced methods of pulse
shaping and control-yield measurement need to be incorporated together
as well as adjusted to the particular nature of a physical or chemical
system. Therefore, theory can be especially important in exploring the
feasibility of various control outcomes for quantum dynamics of a
system of interest; even semiquantitative modelling may be successful
for such purposes in many applications. Theoretical control
simulations can provide important guidance for the selection of the
experimental configuration, thereby helping to make AFC a more
effective practical tool. Additional such high value utilizations of
theory and simulations can be expected in the future.

\subsection{Open-loop quantum control experiments with non-optimal
  designs}

The open-loop control procedure is not limited to the use of optimal
theoretical designs generated via QOCT and similar methods. Moreover,
optimality in not always required in quantum control. In the
conceptually allied field of synthetic chemistry, progress has often
been achieved via intuition-guided trials leading to a gradual
increase of reaction yields. Following this venerable tradition, some
recent open-loop control experiments seek improvement by employing
ultrafast shaped laser pulses with so-called ``rational'' or
``judicious'' control designs obtained using a combination of
intuition and arguments based on some knowledge of system properties
(e.g., spectral information or symmetry). This approach is popular in
nonlinear spectroscopy and microscopy \cite{Silberberg2009,
  DudovichOron2002Nature, OronDudovichYelin2002PRL,
  OronDudovich2002PRL, OronDudovichYelin2002PRA, DudovichOron2003JCP,
  OronDudovich2003PRL, GershgorenBartels2003, PastirkDantus2003,
  LimCasterLeone2005, OgilvieDebarre2006, VacanoMotzkus2007,
  PestovWang2008, Postma2008, IsobeSudaTanaka2009b} as well as some
other atomic \cite{DudovichOron2004, WollenhauptKrug2009,
  PrakeltWollenhaupt2004, BarrosLozano2005OL, BarrosFerraz2006PRA,
  WollenhauptPrakeltBaumert2006PRA, DudovichPolack2005, Amitay2008,
  ZhdanovichShapiroHepburn2009}, molecular
\cite{StapelfeldtSeideman2003, BisgaardPoulsen2004, RenardHertz2004,
  RenardHertz2005, LeeVilleneuve2006, ViftrupKumarappan2009,
  IbrahimSchwentner2009}, and solid-state
\cite{NakamuraPashkinTsai1999, FeurerVaughanNelson2003,
  FanciulliWeiner2005PRB, GolanFradkin2009} applications. While in
many situations such ``rationally'' designed control pulses enhance
the spectroscopic resolution or increase the control yield as compared
to results obtained with transform-limited pulses, in general they are
not optimal. Experience gained from numerous quantum control
experiments indicates that intuition generally fails to discover the
most effective controls (except for the simplest systems), and
therefore in most cases the degree of control can be increased via
closed-loop optimization employing a suitable learning algorithm. 
In some situations, intuition-driven control may be effective for
providing a guide to initial fields for subsequent optimization under
AFC.


\section{Concepts and applications of real-time feedback control}
\label{sec:RTFC}

Feedback is very important in classical engineering where it is
routinely used for control of complex systems in the presence of
uncertainties. In quantum control, two important approaches based on
the concept of feedback have been introduced for similar reasons. One
is AFC, which was extensively discussed in section~\ref{sec:AFC}. A
fundamental characteristic of AFC is that in each control cycle a
fresh quantum ensemble is used (either a new sample is prepared or the
system is reset to its initial state before each run), which makes
measurement back action irrelevant. The other approach is real-time
feedback control (RTFC) \cite{Belavkin1983, WisemanMilburn1993PRL,
  Wiseman1994PRA, DohertyHabib2000, DohertyDoyle2000,
  WisemanMilburn2010, Lloyd2000}, in which the same quantum system is
followed in real time around the feedback loop, and the measurement
(or interaction with a quantum ``controller'') has a significant
effect on the system's evolution.

There are two distinct types of RTFC, which differ by the nature of
the controller. In one approach to RTFC, measurements are employed to
probe the quantum system, and the gathered information is processed
classically off-line in real time to assess the best, next control
action \cite{Belavkin1983, WisemanMilburn1993PRL, Wiseman1994PRA,
  DohertyHabib2000, DohertyDoyle2000, WisemanMilburn2010}. The
evolution of the controlled quantum system is governed by two effects:
coherent (unitary) action exerted by the classical
controller\footnote{The free evolution of the system can be included
  while the off-line modelling is performed.}  and incoherent
(non-unitary) back action exerted by the measurements. A generalized
description of measurement-based RTFC employs quantum filtering theory
\cite{Belavkin1999, BoutenHandelJames2007, BoutenHandelJames2009}.
Recently, another type of RTFC --- referred to as coherent feedback
control\footnote{We will also use the term coherent RTFC. The choice
  of terminology is standard in the field and should not be confused
  with the notion of coherent control employed in AFC and QOCT.} ---
has drawn much attention \cite{Lloyd2000, NelsonWeinsteinCory2000,
  YanagisawaKimura2003I, YanagisawaKimura2003II, WisemanMilburn1994,
  DHelonJames2006, JamesNurdinPetersen2008TAC, Mabuchi2008PRA}. In
this approach, no measurements with a classical output signal are
performed; instead, an ancillary quantum system serves as the
controller. The controller influences the evolution of the system of
interest via a direct interaction between them. Additionally, external
classical forces also can be used to act upon the system, the
controller, or both. The system of interest together with the
controller are characterized by the entirely quantum nature of the
information flow --- coherence is not destroyed by measurements, which
is the source of the name given to this type of control
\cite{Lloyd2000, NelsonWeinsteinCory2000}.  Coherent feedback control
can be viewed as a quantum analog of Watt's flyball governor --- an
automatic self-regulating quantum machine \cite{KallushKosloff2006}.

It was recently shown \cite{PechenBrif2010} that the evolution of a
quantum system undergoing any type of RTFC (or a combination thereof),
including effects of measurements, feedback actions and interactions
with auxiliary quantum systems, can be generally represented by a
Kraus map. This result has an important consequence for the
optimization complexity of RTFC, since observable-control landscapes
of open quantum systems with Kraus-map dynamics are free from local
traps (under the controllability assumption)
\cite{WuPechenRabitz2008JMP}.  Remarkably, this trap-free landscape
topology unifies virtually all types of control, including
measurement-based RTFC, coherent RTFC, AFC, and open-loop control (the
Kraus-map description of open-system dynamics is also generally valid
for the latter two types when the controlled system is coupled to an
environment). The possibility of employing this general unifying
feature of controlled quantum dynamics for development of hybrid
control schemes is discussed in section \ref{sec:Future-hybrid}.

Although the history of AFC in the quantum realm is short (it was
proposed in 1992 and first experimentally implemented in 1997), it has
become a well established and popular laboratory tool successfully
employed in growing numbers of experimental studies (see
section~\ref{sec:AFC}). In contrast, while RTFC of quantum systems was
first proposed in 1983~\cite{Belavkin1983}, its laboratory
implementation~\cite{NelsonWeinsteinCory2000, Mabuchi2008PRA,
  BushevRotterWilson2006, BerglundMcHaleMabuchi2007,
  GillettDaltonLanyon2010} thus far has been much less
extensive. Implementing RTFC on the atomic or molecular scale
encounters significant technical difficulties. One especially
important obstacle is that many interesting quantum phenomena occur on
a time scale which is too short to allow for processing of the
measurement data in classical controllers based on conventional
electronics (i.e., the issue of loop latency). Coherent RTFC can
overcome the latency issue, but in this case the controller itself may
require precise engineering to assure quality control performance of
the quantum system. Nevertheless, interest in potential applications
of RTFC for manipulation of quantum systems is high and the
theoretical activity in this field is growing \cite{DohertyJacobs2001,
  ThomsenManciniWiseman2002PRA, ThomsenManciniWiseman2002JPB,
  StocktonGeremiaDoherty2004PRA, BerglundMabuchi2004APB,
  MabuchiKhaneja2005, WisemanDoherty2005, GoughBelavkinSmolyanov2005,
  BelavkinNegrettiMolmer2009, JacobsShabani2008,
  NielsenHopkinsMabuchi2009, Jacobs2009}. Of particular importance
(especially for applications in quantum information sciences) is the
ability of RTFC to stabilize the dynamics of quantum systems in the
presence of noise \cite{DHelonJames2006, Mabuchi2008PRA,
  MirrahimiHandel2007, JacobsLund2007, YamamotoTsumuraHara2007,
  DotsenkoMirrahimi2009} and achieve robust control performance in the
presence of uncertainties in the system Hamiltonian
\cite{DongPetersen2009NJP}. Other interesting possibilities include
the use of RTFC for quantum error correction
\cite{AhnDohertyLandahl2002, AhnWisemanMilburn2003,
  AhnWisemanJacobs2004, SarovarAhnJacobs2004, ChaseLandahlGeremia2008,
  KerckhoffNurdin2009}, generation and protection of entanglement
\cite{YamamotoTsumuraHara2007, StocktonHandelMabuchi2004,
  HandelStocktonMabuchi2005JOB, Mancini2006, ManciniWiseman2007,
  YamamotoNurdinJames2008, Nielsen2010PRA}, cooling of quantum systems
\cite{BushevRotterWilson2006, HopkinsJacobs2003PRB,
  SteckJacobsMabuchi2004, SteckJacobsMabuchi2006,
  WilsonCarvalhoHope2007}, and quantum state purification
\cite{Jacobs2003PRA, CombesJacobs2006, HandelStocktonMabuchi2005,
  WisemanRalph2006NJP, WisemanBouten2008, ChiruvelliJacobs2008,
  ShabaniJacobs2008PRL}. In the related field of quantum metrology,
real-time feedback was employed to approach fundamental quantum limits
of measurement accuracy \cite{ArmenAuStockton2002PRL,
  CookMartinGeremia2007N}. Both theoretical and experimental aspects
of RTFC should continue to draw significant attention in the future.


\section{Future directions of quantum control}
\label{sec:Future}

Common sense dictates that the future is notoriously difficult to
predict, but it is also the nature of science to try and anticipate
new directions that will expand current knowledge. The evident paths
followed in the development of the quantum control field during the
last two decades provide a basis for projection, with due caution, on
how current trends may evolve upon going forward. Below we try to
identify some critical theoretical and technological issues, where
breakthroughs are required to significantly increase the capability of
controlling quantum phenomena. 

\subsection{Input-output maps for quantum control simulations}
\label{sec:Future-IOM}

Except for the special situation of measurement-based RTFC (where
measurement back action is a distinctively non-classical feature), one
may naively conclude that there are no fundamental differences between
designing controls for quantum and classical systems. The distinctions
seem to lie in solving classical equations of motion in one case and
the Schr\"{o}dinger equation in the other, but otherwise the method of
finding the optimal control solution is basically the same. However,
from a practical perspective, the difference between solving classical
and quantum equations of motion is fundamental due to the exponential
increase of the Hilbert space dimension characteristic of quantum
systems. This is the reason why simulating controlled quantum dynamics
of multi-particle systems is so difficult.

A qualitative breakthrough in open-loop control of complex quantum
systems would be possible, if a way could be found to replace the
laborious calculations of quantum dynamics with ``black-box'' models
that essentially capture the main features of the processes leading to
a particular control objective (e.g., breaking of a specific molecular
bond). The goal is to perform a modest number of simulations and use
the information to generate an input-output map from the applied
control field to the resultant change in the control objective. In
this fashion, the input-output map aims to capture the relationship
between the control and the system's reaction to it. This approach is
commonly used in classical control problems in many areas of
engineering; however, at the present time, we do not know how to
effectively determine these input-output maps for complex quantum
systems, such as molecules.

An example of a method proposed for identifying nonlinear input-output
maps for quantum control studies is the high-dimensional model
representation (HDMR) technique \cite{ShorterIpRabitz1999,
  RabitzAlis1999, RabitzAlisShorter1999, AlisRabitz2001,
  LiRosenthalRabitz2001}. The total number of points in the search
space for a quantum control optimization problem (and for many other
problems in science and technology) grows exponentially with the
number of input variables (this situation is sometimes called the
``curse of dimensionality''). In HDMR, the input-output map is
characterized by a hierarchy of contributions from the input variables
acting independently, in pairs, triples, etc. For many important
problems, with an appropriate choice of the variables, only low order
input variable cooperativity is significant. This property can be used
to dramatically reduce the effort required to explore the
map. Approaches such as HDMR are designed for systems with a large
number of input variables with the aim of learning the input-output
map using a number of simulations that grows relatively slowly (e.g.,
polynomially) with the number of input variables
\cite{LiRosenthalRabitz2001}. Specifically, the use of nonlinear
functional HDMR-type maps for quantum control problems was discussed
by Geremia~\emph{et~al.}~\cite{GeremiaWeissRabitz2001}. Such
input-output maps would be of value as well when generated from
experimental data, as they would constitute the control
landscape. Although much is now understood about the topology of
quantum control landscapes, there is little information about
non-critical point structural features.

Another approach popular for constructing nonlinear input-output maps
employs neural networks \cite{HertzKroghPalmer1991,
  FreemanSkapura1991}. Recently, neural networks were used to model
ultrafast laser control of SHG, molecular fluorescence yield, and
photoelectron spectra from resonant strong-field ionization of
potassium atoms \cite{SelleVogt2007, SelleBrixner2008}. However, it
was found that the amount of data required for the training of a
neural network significantly increases with the complexity of the
correlations which are to be modelled. While reproduction of the
training data worked very well, extrapolation to regions of the
parameter space which were not covered by the training data was a
challenge. 

Development of efficient and accurate input-output maps for control of
complex quantum phenomena remains an important objective. Ideally,
after a modest effort at learning a map, it could then be used in a
highly efficient manner to seek out new controls and dynamical regimes
with favorable characteristics.

\subsection{Analysis of quantum control landscapes}
\label{sec:Future-L}

The introduction of quantum control landscapes in the last few years
is an important theoretical advance in the field. The nature of the
control landscape topology has direct implications for the ease of
finding effective controls in the laboratory. The analysis of the
control landscape topology and other structural features can provide
the basis for investigating the complexity of optimizing different
types of control objectives. 
In turn, this understanding can help identify the most suitable
optimization algorithms for various theoretical and experimental
applications of quantum control (see sections~\ref{sec:L-trajectory},
\ref{sec:L-importance}, and \ref{sec:AFC-algorithms}). In addition,
the landscape analysis may be extended to the study of quantum control
problems involving simultaneous optimization of multiple objectives
(see sections~\ref{sec:L-Pareto} and \ref{sec:L-trajectory}). This
research area is still rapidly developing with much remaining for
investigation. In particular, an open issue that deserves significant
attention is the effect of field constraints (e.g., due to limited
laboratory resources) upon the accessible regions of quantum control
landscapes.

There are several additional research directions for which the
analysis of the control landscape features may provide important
insights. One ubiquitous problem with wide-ranging implications is
evaluation of the robustness of control solutions to noise and
imperfections, which depends on the degree of flatness of the control
landscape around an optimal solution. 
Another interesting issue is related to a phenomenon discovered for
observable control of an open quantum system prepared in a mixed state
and coupled to a thermal environment \cite{WuPechenRabitz2008JMP}.
Specifically, the range of the control landscape (i.e., the difference
between the maximum and minimum expectation values of the target
observable) decreases when the temperature of the environment
raises. Therefore, an important application of control landscape
analysis would be determination of the fundamental thermodynamic
limits on the control yield for open quantum systems.

\subsection{Future applications of AFC}
\label{sec:Future-AFC}

As discussed in section~\ref{sec:AFC}, AFC has proved to have broad
practical success as a means for achieving optimal control of quantum
phenomena in the laboratory. Particularly impressive is the breadth of
applications, ranging from optical systems, to atoms, to semiconductor
structures, to biologically relevant photochemical processes in
complex molecules, etc. 
One clear trend is the extension of AFC applications towards the
manipulation of increasingly more complex systems and phenomena. Along
this avenue, implementation of AFC could bring significant benefits to
such areas as near and even remote detection of chemical compounds
(first steps in this direction have been recently made
\cite{Palliyaguru2008, McGraneScharff2009}), optimal control of
molecular electronics devices, and optimal control of photochemical
phenomena in live biological samples (including nonlinear microscopy
and ODD, as indicated by several recent experiments \cite{Kawano2003,
  ChenKawano2004, TadaKono2007, IsobeSudaTanaka2009a}). 

We can also envision increasing use of AFC for optimal quantum control
of photophysical phenomena. One important area is coherent
manipulation of quantum processes in solid-state systems, especially
in semiconductor quantum structures \cite{RossiKuhn2002RMP,
  Hanson2007RMP}. Another potential application is optimal storage and
retrieval of photonic states in atomic-vapor and solid-state quantum
memories \cite{NovikovaPhillipsGorshkov2008,
  PhillipsGorshkovNovikova2008, AppelFigueroaLvovsky2008,
  ChoiDengLauratKimble2008, StaudtGisin2007, ReimNunnLorenz2009,
  LvovskySandersTittel2009NP, SimonAfzeliusAppel2010}. The AFC
methodology may be also applicable to physical problems where, instead
of laser pulses, other means (e.g., voltages applied to an array of
electrodes) are used to implement the control. Examples could include
optimal control of coherent electron transport in semiconductors by
means of adaptively shaped electrostatic potentials
\cite{SolasAshton2009}, coherent control of charge qubits in
superconducting quantum devices by gate voltages
\cite{MakhlinSchonShnirman2001RMP, ClarkeWilhelm2008N}, and coherent
control of photonic qubits in integrated optical circuits via the
thermo-optic effect \cite{Walmsley2009privatecommun}. Several types of
quantum systems (e.g., flux qubits in superconducting quantum devices,
hyperfine-level qubits in trapped neutral atoms and ions, electron
spins of donor atoms in silicon, etc.) can be controlled by pulses of
microwave radiation (e.g., AFC-optimized dynamical decoupling
\cite{BiercukBollinger2009} of trapped-ion qubits by a sequence of
microwave $\pi$ pulses was discussed in
section~\ref{sec:AFC-decoherence}). Many possible applications of AFC
could have significant implications for the progress in the field of
quantum information sciences. A new domain of quantum control involves
manipulation of relativistic quantum dynamics with extremely intense
laser fields \cite{LiuKohlerKeitel2009} for accelerating particles and
even intervening in nuclear processes in analogy with what is
happening in atomic-scale control. It is reasonable to forecast that
AFC methods could become useful for optimal control of such
laser-driven high-energy phenomena.

Despite significant advances achieved in the field of quantum control
during the last decade, the experimental capabilities are limited by
currently available laser resources. It is likely that existing
practical limitations, in particular, the relatively narrow bandwidth
of femtosecond lasers, restrict the achievable yields in some AFC
experiments. One might expect that many new applications would open
up, if reliable sources of coherent laser radiation with a much wider
bandwidth became available. Such resources could make possible the
simultaneous manipulation of rotational, vibrational, and electronic
processes in molecules in a more effective fashion, thereby achieving
hitherto unattainable levels of control. Moreover, if pulse-shaping
technology can be extended to coherent radiation in the attosecond
regime as well as in the range of MeV photon energies, a multitude of
new applications in X-ray spectroscopy, medical physics, and control
of nuclear dynamics could arise.

\subsection{Hybrid methods of quantum feedback control}
\label{sec:Future-hybrid}

Despite the significant technological difficulties on the path to
routine practical application of RTFC, the potential benefits are
alluring (see section~\ref{sec:RTFC}). An interesting question is
whether AFC, whose practical utility has already been well
established, can be used to aid in the implementation of RTFC
(measurement-based, coherent, or both). Due to the apparent
technological differences between AFC and RTFC, thus far they have
been considered as separate branches of quantum control. However, it
has been recently shown \cite{PechenBrif2010} that AFC and RTFC share
a common fundamental landscape topology characterized by the absence
of local traps (i.e., all sub-optimal extrema are saddles provided
that the controllability condition is satisfied). Since the control
landscape topology strongly influences the optimization complexity,
this finding may have immediate practical importance. Moreover, the
unification of the seemingly different AFC and RTFC approaches at a
fundamental level suggests the possibility of developing new
laboratory realizations that combine these currently distinct
techniques of quantum feedback control in a synergistic way. For
example, some form of AFC might be used to optimize the design or
construction of quantum controllers employed in coherent RTFC.
Development of ``hybrid'' quantum control schemes incorporating both
AFC and RTFC (in particular, for control and stabilization of quantum
computing systems) could provide significantly enhanced flexibility in
the laboratory.

\subsection{Material control}
\label{sec:Future-material}

In addition to the manipulation of quantum dynamics via application of
optimal external fields, there is the prospect of performing material
control through alteration of the internal system properties (i.e.,
the spatial structure or matrix elements of the system Hamiltonian),
with the aim of identifying optimal materials and system
designs. Analogous to the circumstance of a particular quantum system
acted upon by a family of homologous external control fields, we can
consider the controlled response of a family of homologous quantum
systems to a particular field. In the former case, a control level set
consists of all homologous control fields that produce the same
expectation value of the target observable when applied to a
particular quantum system. This level set can be explored, for
example, by homotopy trajectory methods (e.g., D-MORPH), in order to
identify control solutions with desired properties (see
section~\ref{sec:L-trajectory}). In the latter case of material
control, a level set consists of all dynamically homologous quantum
systems that produce the same expectation value of the target
observable when controlled by a particular field. For example, each
quantum system may be specified by a point in a hypercube whose edges
are labeled by Hamiltonian matrix elements. A variation of the D-MORPH
method can be used to explore a system level set by continuously
warping the corresponding Hamiltonian
\cite{BeltraniDominyHoRabitz2007}. At this juncture little is known
about either homologous control fields or homologous quantum
systems. Exploration of these topics could reveal the systematic
aspects of control over quantum phenomena.

Morphing through Hamiltonian structure in the laboratory can be
physically realized in many different ways, with broad and yet largely
unexplored possibilities. For example, the properties of
light-sensitive materials could be varied using families of
structurally similar chemical compounds, characteristics of doped
semiconductors can be varied by changing the concentration of dopant
atoms and the depth of implanting, etc. Material control is
potentially applicable to a wide set of problems in various areas of
science and technology. Possible applications include, for example,
development of photodetectors with higher efficiency and faster
response time, molecular switches with increased sensitivity and
durability, quantum computing systems with enhanced immunity to
environmentally-induced decoherence and improved robustness to
instrumental noise, etc. Exploiting the accessible variations in
Hamiltonian structure as a means for achieving optimal quantum control
is a potentially important area of future research, including
exploration of the associated control landscapes, development of
adaptive and open-loop techniques, investigation of effective methods
of combining material and electromagnetic control, and adaptation of
the theoretical concepts to various practical applications. 

\subsection{Scientific and engineering goals of optimal quantum
  control} 
\label{sec:Future-goals}

The general goal of science is to understand nature, including the
structure of physical systems and characteristics of the system
dynamics, while the goal of engineering is to design and implement a
system that will function in a prescribed manner in the best possible
way. Optimal quantum control draws together science and engineering to
incorporate both goals: (1) to understand the dynamical behavior of
quantum systems and the mechanisms by which these processes can be
managed, and (2) to require optimal functional performance through the
achievement of prescribed control objectives in the best possible
way. 

An important feature evident in the prior development of quantum
control is the impact of progress in one aspect of the subject on
advancing another. We expect that this trend will continue in the
future, as a better understanding of the underlying physical
processes would aid in choosing better control tools and thereby
achieving a higher degree of performance. 
In turn, the ability to steer system evolution in an optimal fashion
should facilitate the acquisition of knowledge about the underlying
control mechanisms and other properties of the system. For example, in
many AFC experiments, the characteristics of the resultant optimal
control fields were used (often in combination with additional
measurements and/or simulations) to decipher physical mechanisms
responsible for the achieved control \cite{MontgomeryMeglen2006,
  MontgomeryMeglen2007, MontgomeryDamrauer2007, KurodaKleiman2009,
  OtakeKanoWada2006, Okada2004, BergtBrixner1999, BrixnerKiefer2001,
  Daniel2001, Daniel2003, Cardoza2005JCP, Cardoza2005CPL, Cardoza2004,
  LanghojerCardoza2005, Cardoza2005a, CardozaPearson2006,
  BarteltFeurer2005, deNaldaHorn2007, HornungMeierMotzkus2000,
  WeinachtBartels2001CPL, KonradiSingh2006JPPA, ZhangZhang2007,
  StrasfeldShim2007, DietzekYartsev2006, DietzekYartsev2007,
  Prokhorenko2007, CarrollWhite2008, KoturWeinacht2009,
  GreenfieldMcGrane2009}. Also, a method for analysis of quantum
control mechanisms through Hamiltonian encoding was recently developed
\cite{MitraRabitz2003, MitraSolaRabitz2003, MitraRabitz2004,
  SharpRabitz2004, MitraRabitz2006, MitraRabitz2008,
  MitraSolaRabitz2008, SharpMitraRabitz2008} and experimentally tested
\cite{ReydeCastroRabitz2010}. However, much additional theoretical and
experimental work is still needed to better understand the controlled
dynamics of complex systems, especially those in the condensed phases.

Recent experiments \cite{EngelFleming2007Nat, LeeChengFleming2007Sci,
  MercerElTahaKajumba2009PRL, ColliniWongWilk2010, Engel2010}
discovered manifestations of long-lived electronic quantum coherence
in energy transfer processes in light-harvesting complexes of
photosynthetic systems. Evidence of long-lived electronic and
vibrational quantum coherence was also found in intrachain energy
transfer in a conjugated polymer \cite{ColliniScholes2009Sci,
  ColliniScholes2009JPCA}. These findings raise an important question
about the role of coherent quantum effects in the dynamics of energy
transfer and other photoinduced processes in complex chemical and
biological systems. This issue already attracted significant attention
\cite{ChengFleming2009ARPC, BeljonneCurutchetScholes2009,
  AbramaviciusPalmieriMukamel2009, ArndtJuffmannVedral2009}, and a
number of theoretical models \cite{ChengFleming2008JPCA,
  IshizakiFleming2009PNAS, IshizakiFleming2009a, IshizakiFleming2009b,
  OlayaCastro2008PRB, YuBerdingWang2008, JangCheng2008JCP,
  Jang2009JCP, MohseniRebentrost2008JCP, RebentrostMohseni2009NJP,
  RebentrostMohseni2009JPCB, RebentrostChakraborty2009JCP,
  PlenioHuelga2008, CarusoChinPlenio2009, PalmieriAbramavicius2009,
  ThorwartEckel2009, Nazir2009PRL, CarusoChinDatta2009,
  BradlerWilde2009, PerdomoVogt2010, FassioliOlayaCastro2010} were
developed to explain the existence of quantum transport in the
presence of strong coupling to a thermal environment. Further advances
in this area may lead to a better understanding and more effective
control of photophysical and photochemical quantum phenomena in the
condensed phase. 

Hamiltonian identification is a potentially important application of
quantum control aimed at revealing detailed information about physical
systems. Extraction of the Hamiltonian from measured data is an
inverse problem that generally suffers from being ill-posed (i.e., the
Hamiltonian information is unstable against small changes of the
data), which arises because the data used for inversion are inevitably
incomplete. Recent attempts to address this challenging problem
include Hamiltonian identification via inversion of time-dependent
data (instead of traditional use of time-independent spectroscopic
data) \cite{LuRabitz1995PRA, LuRabitz1995JPC, ZhuRabitz1999b,
  ZhuRabitz1999c, BrifRabitz2000, KurtzRabitz2002} and application of
global, nonlinear, map-facilitated inversion procedures
\cite{GeremiaRabitz2001JCP, GeremiaRabitz2004PRA}. In this context, it
appears that suitable controls can be used to significantly increase
the information content of the measured data. For example, it may be
possible to control the motion of a molecular wave packet to gain more
information on interatomic forces in selected regions of a potential
energy surface \cite{RabitzZhu2000}. This concept has seen some
nascent development by Geremia and Rabitz \cite{GeremiaRabitz2002PRL,
  GeremiaRabitz2003JCP} who proposed the notion of optimal Hamiltonian
identification (OHI). OHI aims to employ coherent control of quantum
dynamics to minimize the uncertainty in the extracted Hamiltonian
despite data limitations such as finite resolution and noise. The
proposed OHI implementation operates in a manner similar to an AFC
experiment, using closed-loop optimization guided by a learning
algorithm to discover controls that minimize the dispersion of the
distribution of Hamiltonians consistent with the measured
data. Numerical simulations indicate that an optimal experiment can
act as a tailored filter to prevent laboratory noise from
significantly propagating into the extracted Hamiltonian
\cite{GeremiaRabitz2003JCP}. Ideally, upon each cycle of the
experiment more information about the Hamiltonian will be extracted,
which in turn would be used to guide the next cycle, etc. A critical
component of OHI is the need for real-time numerical simulations of
the quantum system's dynamics on the fly. 
OHI will require further development \cite{KosutWalmsleyRabitz2004} of
inversion algorithms, computational capabilities (e.g., possibly
including input-output maps discussed in
section~\ref{sec:Future-IOM}), and experimental techniques to become
practical. 


\section{Concluding remarks}
\label{sec:Concl}

It would be impossible to cover in a paper of any reasonable length
all of the advances that have been made in the last two decades in the
field of quantum control. Some areas that did not receive full
attention here were considered in more detail in other review articles
and books (in particular, those cited in section~\ref{sec:Intro}), to
which we refer the interested reader. For example, thematic reviews
are available on control via two-pathway quantum interference
\cite{BrumerShapiro1992, ShapiroBrumer1997, BrumerShapiro2003,
  ShapiroBrumer2003}, pump-dump control \cite{GordonRice1997,
  RiceZhao2000}, control via STIRAP \cite{Bergmann1998, Vitanov2001},
control via WPI \cite{Ohmori2009}, the formalism and applications of
QOCT \cite{RabitzZhu2000, DAlessandro2007, WerschnikGross2007,
  BalintKurti2008}, controllability of quantum systems
\cite{DAlessandro2007}, the formalism of quantum control landscape
theory \cite{ChakrabartiRabitz2007review}, femtosecond pulse-shaping
technology \cite{Kawashima1995, Weiner1995, Weiner2000, Goswami2003},
femtosecond laser control of X-ray generation
\cite{PfeiferSpielmann2006, Winterfeldt2008}, quantum control
experiments with ``rational'' control designs
\cite{MarcosLozovoy2004}, quantum control applications in nonlinear
spectroscopy and microscopy \cite{Wohlleben2005, Silberberg2009}, and
control of quantum dynamics on the attosecond time scale
\cite{KrauszIvanov2009}. While we tried to provide a comprehensive
account of laboratory AFC of quantum phenomena, more detailed
discussions of some important AFC experiments are available in earlier
reviews \cite{Brixner2001, LevisRabitz2002, WeinachtBucksbaum2002,
  Brixner2003, BrixnerGerber2003, BrixnerPfeifer2005, Wohlleben2005,
  PfeiferSpielmann2006, Nuernberger2007, Winterfeldt2008}. New papers,
often containing significant theoretical and experimental results in
quantum control, appear now almost on a daily basis.

In this paper, our goal was to give a perspective and prospective on
the field highlighting the evolution of important trends in quantum
control. A look into the past together with a review of current,
cutting-edge research were used to cautiously forecast topics of
future interest. We also attempted to emphasize the synergistic
connection between the theoretical and experimental advances, which
has been immensely beneficial for the development of the field. We
believe that sustaining this productive interplay between theory and
experiment will be critical for future progress. This paper aimed to
provide the basis to better understand which aspects of theoretical
research are having a high impact on laboratory control of quantum
phenomena. At the same time, a complementary goal of this work was to
point out the experimental aspects of quantum control that have
special significance and relation to theory. 
Although the scope of experimental and theoretical research in quantum
control is vast, we hope that this work provides a valuable bridge for
the community involved as well as for those outside who are interested
in understanding the reasons for the fervor in the field. 

\ack This work was supported by DOE, NSF, ARO, and Lockheed Martin.

\section*{References}
\bibliography{Brif_QuantumControlReview_NJP}

\providecommand{\newblock}{}
\begin{thebibliography}{100}
\expandafter\ifx\csname url\endcsname\relax
  \def\url#1{{\tt #1}}\fi
\expandafter\ifx\csname urlprefix\endcsname\relax\def\urlprefix{URL }\fi
\providecommand{\eprint}[2][]{\url{#2}}

\bibitem{BrumerShapiro1992}
Brumer P and Shapiro M 1992 {\em Ann. Rev. Phys. Chem.\/} {\bf 43} 257--282

\bibitem{WarrenRabitzDahleh1993}
Warren W~S, Rabitz H and Dahleh M 1993 {\em Science\/} {\bf 259} 1581--1589

\bibitem{KohlerKrause1995}
Kohler B, Krause J~L, Raksi F, Wilson K~R, Yakovlev V~V, Whitnell R~M and Yan Y
  1995 {\em Acc. Chem. Res.\/} {\bf 28} 133--140

\bibitem{Kawashima1995}
Kawashima H, Wefers M~M and Nelson K~A 1995 {\em Ann. Rev. Phys. Chem.\/} {\bf
  46} 627--656

\bibitem{Weiner1995}
Weiner A~M 1995 {\em Prog. Quantum Electron.\/} {\bf 19} 161--237

\bibitem{Manz1996}
Manz J 1996 Molecular wavepacket dynamics: Theory for experiments 1926--1996
  {\em Femtochemistry and Femtobiology: Ultrafast Reaction Dynamics at
  Atomic-Scale Resolution\/} ed Sundstr\"{o}m V (London, UK: Imperial College
  Press) chap~3, pp 80--318

\bibitem{GordonRice1997}
Gordon R~J and Rice S~A 1997 {\em Ann. Rev. Phys. Chem.\/} {\bf 48} 601--641

\bibitem{ShapiroBrumer1997}
Shapiro M and Brumer P 1997 {\em J. Chem. Soc., Faraday Trans.\/} {\bf 93}
  1263--1277

\bibitem{GaspardBurghardt1997}
Gaspard P and Burghardt I (eds) 1997 {\em Chemical Reactions and Their Control
  on the Femtosecond Time Scale\/} ({\em Adv. Chem. Phys.\/} vol 101) (New
  York: Wiley)

\bibitem{Bergmann1998}
Bergmann K, Theuer H and Shore B~W 1998 {\em Rev. Mod. Phys.\/} {\bf 70}
  1003--1025

\bibitem{RiceZhao2000}
Rice S~A and Zhao M 2000 {\em Optical Control of Molecular Dynamics\/} (New
  York: Wiley)

\bibitem{Weiner2000}
Weiner A~M 2000 {\em Rev. Sci. Instr.\/} {\bf 71} 1929--1960

\bibitem{Rabitz2000}
Rabitz H, de~Vivie-Riedle R, Motzkus M and Kompa K 2000 {\em Science\/} {\bf
  288} 824--828

\bibitem{RabitzZhu2000}
Rabitz H and Zhu W~S 2000 {\em Acc. Chem. Res.\/} {\bf 33} 572--578

\bibitem{Brixner2001}
Brixner T, Damrauer N and Gerber G 2001 Femtosecond quantum control {\em Adv.
  At. Mol. Opt. Phys.\/} vol~46 ed Bederson B and Walther H (San Diego:
  Academic Press) pp 1--54

\bibitem{Vitanov2001}
Vitanov N~V, Halfmann T, Shore B~W and Bergmann K 2001 {\em Ann. Rev. Phys.
  Chem.\/} {\bf 52} 763--809

\bibitem{BrownRabitz2002}
Brown E and Rabitz H 2002 {\em J. Math. Chem.\/} {\bf 31} 17--63

\bibitem{LevisRabitz2002}
Levis R~J and Rabitz H~A 2002 {\em J. Phys. Chem.\/} A {\bf 106} 6427--6444

\bibitem{WeinachtBucksbaum2002}
Weinacht T~C and Bucksbaum P~H 2002 {\em J. Opt. B: Quantum Semiclass. Opt.\/}
  {\bf 4} R35--R52

\bibitem{Bandrauk2002}
Bandrauk A~D, Fujimura Y and Gordon R~J (eds) 2002 {\em Laser Control and
  Manipulation of Molecules\/} ({\em Symposium Series\/} vol 821) (Washington,
  DC: ACS Publications)

\bibitem{BrumerShapiro2003}
Brumer P and Shapiro M 2003 {\em Principles of the Quantum Control of Molecular
  Processes\/} (Hoboken, NJ: Wiley-Interscience)

\bibitem{BrifRabitz2003}
Brif C and Rabitz H 2003 Optimal control of molecular scale phenomena {\em
  Fundamentals of Chemistry\/} ({\em Encyclopedia of Life Support Systems\/}
  vol~6) ed Carra S (Oxford, UK: EOLSS Publishers)

\bibitem{Rabitz2003}
Rabitz H 2003 {\em Theor. Chem. Acc.\/} {\bf 109} 64--70

\bibitem{ShapiroBrumer2003}
Shapiro M and Brumer P 2003 {\em Rep. Prog. Phys.\/} {\bf 66} 859--942

\bibitem{Brixner2003}
Brixner T, Damrauer N~H, Krampert G, Niklaus P and Gerber G 2003 {\em J. Mod.
  Opt.\/} {\bf 50} 539--560

\bibitem{BrixnerGerber2003}
Brixner T and Gerber G 2003 {\em ChemPhysChem\/} {\bf 4} 418--438

\bibitem{Goswami2003}
Goswami D 2003 {\em Phys. Rep.\/} {\bf 374} 385--481

\bibitem{WalmsleyRabitz2003}
Walmsley I and Rabitz H 2003 {\em Phys. Today\/} {\bf 56}(8) 43--49

\bibitem{BrixnerGerber2004}
Brixner T and Gerber G 2004 {\em Phys. Scr.\/} {\bf T110} 101--107

\bibitem{MarcosLozovoy2004}
Dantus M and Lozovoy V~V 2004 {\em Chem. Rev.\/} {\bf 104} 1813--1860

\bibitem{BrixnerPfeifer2005}
Brixner T, Pfeifer T, Gerber G, Wollenhaupt M and Baumert T 2005 Optimal
  control of atomic, molecular and electron dynamics with tailored femtosecond
  laser pulses {\em Femtosecond Laser Spectroscopy\/} ed Hannaford P (New York:
  Springer) chap~9

\bibitem{Carley2005}
Carley R~E, Heesel E and Fielding H~H 2005 {\em Chem. Soc. Rev.\/} {\bf 34}
  949--969

\bibitem{Wollenhaupt2005}
Wollenhaupt M, Engel V and Baumert T 2005 {\em Ann. Rev. Phys. Chem.\/} {\bf
  56} 25--56

\bibitem{Wohlleben2005}
Wohlleben W, Buckup T, Herek J~L and Motzkus M 2005 {\em ChemPhysChem\/} {\bf
  6} 850--857

\bibitem{PfeiferSpielmann2006}
Pfeifer T, Spielmann C and Gerber G 2006 {\em Rep. Prog. Phys.\/} {\bf 69}
  443--505

\bibitem{DAlessandro2007}
D'Alessandro D 2007 {\em Introduction to Quantum Control and Dynamics\/} (Boca
  Raton, FL: Chapman \& Hall/CRC)

\bibitem{Nuernberger2007}
Nuernberger P, Vogt G, Brixner T and Gerber G 2007 {\em Phys. Chem. Chem.
  Phys.\/} {\bf 9} 2470--2497

\bibitem{WerschnikGross2007}
Werschnik J and Gross E~K~U 2007 {\em J. Phys. B: At. Mol. Opt. Phys.\/} {\bf
  40} R175--R211

\bibitem{ChakrabartiRabitz2007review}
Chakrabarti R and Rabitz H 2007 {\em Int. Rev. Phys. Chem.\/} {\bf 26} 671--735

\bibitem{BalintKurti2008}
Balint-Kurti G~G, Zou S and Brown A 2008 Optimal control theory for
  manipulating molecular processes {\em Adv. Chem. Phys.\/} vol 138 ed Rice S~A
  (New York: Wiley) pp 43--94

\bibitem{Winterfeldt2008}
Winterfeldt C, Spielmann C and Gerber G 2008 {\em Rev. Mod. Phys.\/} {\bf 80}
  117--140

\bibitem{Ohmori2009}
Ohmori K 2009 {\em Ann. Rev. Phys. Chem.\/} {\bf 60} 487--511

\bibitem{Rego2009}
Rego L~G~C, Santos L~F and Batista V~S 2009 {\em Ann. Rev. Phys. Chem.\/} {\bf
  60} 293--320

\bibitem{Silberberg2009}
Silberberg Y 2009 {\em Ann. Rev. Phys. Chem.\/} {\bf 60} 277--292

\bibitem{KrauszIvanov2009}
Krausz F and Ivanov M 2009 {\em Rev. Mod. Phys.\/} {\bf 81} 163--234

\bibitem{Letokhov1977}
Letokhov V~S 1977 {\em Phys. Today\/} {\bf 30}(5) 23--32

\bibitem{BloemYabl1978}
Bloembergen N and Yablonovitch E 1978 {\em Phys. Today\/} {\bf 31}(5) 23--30

\bibitem{Zewail1980}
Zewail A~H 1980 {\em Phys. Today\/} {\bf 33}(11) 25--33

\bibitem{BloemZewail1984}
Bloembergen N and Zewail A~H 1984 {\em J. Phys. Chem.\/} {\bf 88} 5459--5465

\bibitem{ElsaesserKaiser1991}
Elsaesser T and Kaiser W 1991 {\em Ann. Rev. Phys. Chem.\/} {\bf 42} 83--107

\bibitem{Zewail1996}
Zewail A~H 1996 {\em J. Phys. Chem.\/} {\bf 100} 12701--12724

\bibitem{BrumerShapiro1986a}
Brumer P and Shapiro M 1986 {\em Chem. Phys. Lett.\/} {\bf 126} 541--546

\bibitem{BrumerShapiro1986b}
Brumer P and Shapiro M 1986 {\em Faraday Discuss. Chem. Soc.\/} {\bf 82}
  177--185

\bibitem{Shapiro1988}
Shapiro M, Hepburn J~W and Brumer P 1988 {\em Chem. Phys. Lett.\/} {\bf 149}
  451--454

\bibitem{BrumerShapiro1989}
Brumer P and Shapiro M 1989 {\em Acc. Chem. Res.\/} {\bf 22} 407--413

\bibitem{ChanBrumerShapiro1991}
Chan C~K, Brumer P and Shapiro M 1991 {\em J. Chem. Phys.\/} {\bf 94}
  2688--2696

\bibitem{ChenBrumerShapiro1993}
Chen Z, Brumer P and Shapiro M 1993 {\em J. Chem. Phys.\/} {\bf 98} 6843--6852

\bibitem{Lee1998}
Lee S 1998 {\em J. Chem. Phys.\/} {\bf 108} 3903--3908

\bibitem{ChenYinElliott1990}
Chen C, Yin Y~Y and Elliott D~S 1990 {\em Phys. Rev. Lett.\/} {\bf 64} 507--510

\bibitem{ChenElliott1990}
Chen C and Elliott D~S 1990 {\em Phys. Rev. Lett.\/} {\bf 65} 1737--1740

\bibitem{ParkLuGordon1991}
Park S~M, Lu S~P and Gordon R~J 1991 {\em J. Chem. Phys.\/} {\bf 94} 8622--8624

\bibitem{LuParkXieGordon1992}
Lu S~P, Park S~M, Xie Y and Gordon R~J 1992 {\em J. Chem. Phys.\/} {\bf 96}
  6613--6620

\bibitem{XingBersohn1996}
Xing G, Wang X, Huang X, Bersohn R and Katz B 1996 {\em J. Chem. Phys.\/} {\bf
  104} 826--831

\bibitem{WangBersohn1996}
Wang X, Bersohn R, Takahashi K, Kawasaki M and Kim H~L 1996 {\em J. Chem.
  Phys.\/} {\bf 105} 2992--2997

\bibitem{MullerBucksbaum1990}
Muller H~G, Bucksbaum P~H, Schumacher D~W and Zavriyev A 1990 {\em J. Phys. B:
  Atom. Mol. Opt. Phys.\/} {\bf 23} 2761--2769

\bibitem{SchumacherBucksbaum1994}
Schumacher D~W, Weihe F, Muller H~G and Bucksbaum P~H 1994 {\em Phys. Rev.
  Lett.\/} {\bf 73} 1344--1347

\bibitem{YinChenElliottSmith1992}
Yin Y~Y, Chen C, Elliott D~S and Smith A~V 1992 {\em Phys. Rev. Lett.\/} {\bf
  69} 2353--2356

\bibitem{YinElliottShehadehGrant1995}
Yin Y~Y, Elliott D~S, Shehadeh R and Grant E~R 1995 {\em Chem. Phys. Lett.\/}
  {\bf 241} 591--596

\bibitem{Sheehy1995}
Sheehy B, Walker B and DiMauro L~F 1995 {\em Phys. Rev. Lett.\/} {\bf 74}
  4799--4802

\bibitem{KleimanZhuLiGordon1995}
Kleiman V~D, Zhu L, Li X and Gordon R~J 1995 {\em J. Chem. Phys.\/} {\bf 102}
  5863--5866

\bibitem{KleimanZhuAllenGordon1995}
Kleiman V~D, Zhu L, Allen J and Gordon R~J 1995 {\em J. Chem. Phys.\/} {\bf
  103} 10800--10803

\bibitem{ZhuKleimanLiLuTrentelmanGordon1995}
Zhu L, Kleiman V, Li X, Lu S~P, Trentelman K and Gordon R~J 1995 {\em
  Science\/} {\bf 270} 77--80

\bibitem{DupontCorkum1995}
Dupont E, Corkum P~B, Liu H~C, Buchanan M and Wasilewski Z~R 1995 {\em Phys.
  Rev. Lett.\/} {\bf 74} 3596--3599

\bibitem{Hache1997}
Hach\'e A, Kostoulas Y, Atanasov R, Hughes J~L~P, Sipe J~E and van Driel H~M
  1997 {\em Phys. Rev. Lett.\/} {\bf 78} 306--309

\bibitem{ChenElliott1996}
Chen C and Elliott D~S 1996 {\em Phys. Rev.\/} A {\bf 53} 272--279

\bibitem{TannorRice1985}
Tannor D~J and Rice S~A 1985 {\em J. Chem. Phys.\/} {\bf 83} 5013--5018

\bibitem{TannorKosloffRice1986}
Tannor D~J, Kosloff R and Rice S~A 1986 {\em J. Chem. Phys.\/} {\bf 85}
  5805--5820

\bibitem{Baumert1991a}
Baumert T, Grosser M, Thalweiser R and Gerber G 1991 {\em Phys. Rev. Lett.\/}
  {\bf 67} 3753--3756

\bibitem{Baumert1991b}
Baumert T, B\"{u}hler B, Grosser M, Thalweiser R, Weiss V, Wiedenmann E and
  Gerber G 1991 {\em J. Phys. Chem.\/} {\bf 95} 8103--8110

\bibitem{BaumertGerber1994}
Baumert T and Gerber G 1994 {\em Isr. J. Chem.\/} {\bf 34} 103--114

\bibitem{Potter1992}
Potter E~D, Herek J~L, Pedersen S, Liu Q and Zewail A~H 1992 {\em Nature\/}
  {\bf 355} 66--68

\bibitem{Herek1994}
Herek J~L, Materny A and Zewail A~H 1994 {\em Chem. Phys. Lett.\/} {\bf 228}
  15--25

\bibitem{GaiMcDonaldAnfinrud1997}
Gai F, McDonald J~C and Anfinrud P~A 1997 {\em J. Am. Chem. Soc.\/} {\bf 119}
  6201--6202

\bibitem{Logunov2001}
Logunov S~L, Volkov V~V, Braun M and El-Sayed M~A 2001 {\em Proc. Natl. Acad.
  Sci.\/} {\bf 98} 8475--8479

\bibitem{Ruhman2002}
Ruhman S, Hou B, Friedman N, Ottolenghi M and Sheves M 2002 {\em J. Am. Chem.
  Soc.\/} {\bf 124} 8854--8858

\bibitem{Larsen2004a}
Larsen D~S, Vengris M, van Stokkum I~H, van~der Horst M~A, de~Weerd F~L,
  Hellingwerf K~J and van Grondelle R 2004 {\em Biophys. J.\/} {\bf 86}
  2538--2550

\bibitem{Larsen2004b}
Larsen D~S, van Stokkum I~H, Vengris M, van~der Horst M~A, de~Weerd F~L,
  Hellingwerf K~J and van Grondelle R 2004 {\em Biophys. J.\/} {\bf 87}
  1858--1872

\bibitem{Larsen2005}
Larsen D~S and van Grondelle R 2005 {\em ChemPhysChem\/} {\bf 6} 828--837

\bibitem{Vengris2005}
Vengris M, Larsen D~S, van~der Horst M~A, Larsen O~F~A, Hellingwerf K~J and van
  Grondelle R 2005 {\em J. Phys. Chem.\/} B {\bf 109} 4197--4208

\bibitem{Vengris2004}
Vengris M, van Stokkum I~H~M, He X, Bell A~F, Tonge P~J, van Grondelle R and
  Larsen D~S 2004 {\em J. Phys. Chem. A,\/} {\bf 108} 4587--4598

\bibitem{Gaubatz1988}
Gaubatz U, Rudecki P, Becker M, Schiemann S, K\"ulz M and Bergmann K 1988 {\em
  Chem. Phys. Lett.\/} {\bf 149} 463--468

\bibitem{Kuklinski1989}
Kuklinski J~R, Gaubatz U, Hioe F~T and Bergmann K 1989 {\em Phys. Rev.\/} A
  {\bf 40} 6741--6744

\bibitem{Gaubatz1990}
Gaubatz U, Rudecki P, Schiemann S and Bergmann K 1990 {\em J. Chem. Phys.\/}
  {\bf 92} 5363--5376

\bibitem{ShoreBergmann1991}
Shore B~W, Bergmann K, Oreg J and Rosenwaks S 1991 {\em Phys. Rev.\/} A {\bf
  44} 7442--7447

\bibitem{Salour1977}
Salour M~M and Cohen-Tannoudji C 1977 {\em Phys. Rev. Lett.\/} {\bf 38}
  757--760

\bibitem{Teets1977}
Teets R, Eckstein J and H\"{a}nsch T~W 1977 {\em Phys. Rev. Lett.\/} {\bf 38}
  760--764

\bibitem{Noordam1992}
Noordam L~D, Duncan D~I and Gallagher T~F 1992 {\em Phys. Rev.\/} A {\bf 45}
  4734--4737

\bibitem{Jones1993}
Jones R~R, Raman C~S, Schumacher D~W and Bucksbaum P~H 1993 {\em Phys. Rev.
  Lett.\/} {\bf 71} 2575--2578

\bibitem{Jones1995a}
Jones R~R, Schumacher D~W, Gallagher T~F and Bucksbaum P~H 1995 {\em J. Phys.
  B: At. Mol. Opt. Phys.\/} {\bf 28} L405--L411

\bibitem{Blanchet1997}
Blanchet V, Nicole C, Bouchene M~A and Girard B 1997 {\em Phys. Rev. Lett.\/}
  {\bf 78} 2716--2719

\bibitem{Bouchene1998}
Bouchene M~A, Blanchet V, Nicole C, Melikechi N, Girard B, Ruppe H, Rutz S,
  Schreiber E and W\"{o}ste L 1998 {\em Eur. Phys. J. D\/} {\bf 2} 131--141

\bibitem{Scherer1991}
Scherer N~F, Carlson R~J, Matro A, Du M, Ruggiero A~J, Romero-Rochin V, Cina
  J~A, Fleming G~R and Rice S~A 1991 {\em J. Chem. Phys.\/} {\bf 95} 1487--1511

\bibitem{Scherer1992}
Scherer N~F, Matro A, Ziegler L~D, Du M, Carlson R~J, Cina J~A and Fleming G~R
  1992 {\em J. Chem. Phys.\/} {\bf 96} 4180--4194

\bibitem{Blanchet1998}
Blanchet V, Bouch\`{e}ne M~A and Girard B 1998 {\em J. Chem. Phys.\/} {\bf 108}
  4862--4876

\bibitem{Doule2001}
Doul\'{e} C, Hertz E, Berguiga L, Chaux R, Lavorel B and Faucher O 2001 {\em J.
  Phys. B: At. Mol. Opt. Phys.\/} {\bf 34} 1133--142

\bibitem{Ohmori2003}
Ohmori K, Sato Y, Nikitin E~E and Rice S~A 2003 {\em Phys. Rev. Lett.\/} {\bf
  91} 243003

\bibitem{Hertz2000}
Hertz E, Faucher O, Lavorel B and Chaux R 2000 {\em J. Chem. Phys.\/} {\bf 113}
  6132--6138

\bibitem{Bonadeo1998}
Bonadeo N~H, Erland J, Gammon D, Park D, Katzer D~S and Steel D~G 1998 {\em
  Science\/} {\bf 282} 1473--1476

\bibitem{Flissikowski2004}
Flissikowski T, Betke A, Akimov I~A and Henneberger F 2004 {\em Phys. Rev.
  Lett.\/} {\bf 92} 227401

\bibitem{Cina2008}
Cina J~A 2008 {\em Ann. Rev. Phys. Chem.\/} {\bf 59} 319--342

\bibitem{Shi1988}
Shi S, Woody A and Rabitz H 1988 {\em J. Chem. Phys.\/} {\bf 88} 6870--6883

\bibitem{Peirce1988}
Peirce A~P, Dahleh M~A and Rabitz H 1988 {\em Phys. Rev.\/} A {\bf 37}
  4950--4964

\bibitem{ShiRabitz1989}
Shi S and Rabitz H 1989 {\em Chem. Phys.\/} {\bf 139} 185--199

\bibitem{Kosloff1989}
Kosloff R, Rice S~A, Gaspard P, Tersigni S and Tannor D~J 1989 {\em Chem.
  Phys.\/} {\bf 139} 201--220

\bibitem{Jakubetz1990}
Jakubetz W, Manz J and Schreier H~J 1990 {\em Chem. Phys. Lett.\/} {\bf 165}
  100--106

\bibitem{ShiRabitz1990a}
Shi S and Rabitz H 1990 {\em J. Chem. Phys.\/} {\bf 92} 364--376

\bibitem{ShiRabitz1990b}
Shi S and Rabitz H 1990 {\em J. Chem. Phys.\/} {\bf 92} 2927--2937

\bibitem{Dahleh1990}
Dahleh M, Peirce A~P and Rabitz H 1990 {\em Phys. Rev.\/} A {\bf 42} 1065--1079

\bibitem{ShiRabitz1991}
Shi S and Rabitz H 1991 {\em Comp. Phys. Commun.\/} {\bf 63} 71--83

\bibitem{Gross1991}
Gross P, Neuhauser D and Rabitz H 1991 {\em J. Chem. Phys.\/} {\bf 94}
  1158--1166

\bibitem{Kaluza1994}
Kalu\v{z}a M, Muckerman J~T, Gross P and Rabitz H 1994 {\em J. Chem. Phys.\/}
  {\bf 100} 4211--4228

\bibitem{Sugawara1994}
Sugawara M and Fujimura Y 1994 {\em J. Chem. Phys.\/} {\bf 100} 5646--5655

\bibitem{BardeenYakovlevWilson1997}
Bardeen C~J, Yakovlev V~V, Wilson K~R, Carpenter S~D, Weber P~M and Warren W~S
  1997 {\em Chem. Phys. Lett.\/} {\bf 280} 151--158

\bibitem{Assion1998}
Assion A, Baumert T, Bergt M, Brixner T, Kiefer B, Seyfried V, Strehle M and
  Gerber G 1998 {\em Science\/} {\bf 282} 919--922

\bibitem{BardeenWangShank1995}
Bardeen C~J, Wang Q and Shank C~V 1995 {\em Phys. Rev. Lett.\/} {\bf 75}
  3410--3413

\bibitem{Kohler1995}
Kohler B, Yakovlev V~V, Che J, Krause J~L, Messina M, Wilson K~R, Schwentner N,
  Whitnell R~M and Yan Y 1995 {\em Phys. Rev. Lett.\/} {\bf 74} 3360--3363

\bibitem{BardeenCheWilson1997a}
Bardeen C~J, Che J, Wilson K~R, Yakovlev V~V, Apkarian V~A, Martens C~C,
  Zadoyan R, Kohler B and Messina M 1997 {\em J. Chem. Phys.\/} {\bf 106}
  8486--8503

\bibitem{BardeenCheWilson1997b}
Bardeen C~J, Che J, Wilson K~R, Yakovlev V~V, Cong P, Kohler B, Krause J~L and
  Messina M 1997 {\em J. Phys. Chem.\/} A {\bf 101} 3815--3822

\bibitem{BardeenWangShank1998}
Bardeen C~J, Wang Q and Shank C~V 1998 {\em J. Phys. Chem.\/} A {\bf 102}
  2759--2766

\bibitem{Misawa2000}
Misawa K and Kobayashi T 2000 {\em J. Chem. Phys.\/} {\bf 113} 7546--7553

\bibitem{Malkmus2005}
Malkmus S, D\"urr R, Sobotta C, Pulvermacher H, Zinth W and Braun M 2005 {\em
  J. Phys. Chem.\/} A {\bf 109} 10488--10492

\bibitem{Melinger1992}
Melinger J~S, Gandhi S~R, Hariharan A, Tull J~X and Warren W~S 1992 {\em Phys.
  Rev. Lett.\/} {\bf 68} 2000--2003

\bibitem{Broers1992}
Broers B, van Linden van~den Heuvell H~B and Noordam L~D 1992 {\em Phys. Rev.
  Lett.\/} {\bf 69} 2062--2065

\bibitem{Balling1994}
Balling P, Maas D~J and Noordam L~D 1994 {\em Phys. Rev.\/} A {\bf 50}
  4276--4285

\bibitem{KleimanArrivo1998}
Kleiman V~D, Arrivo S~M, Melinger J~S and Heilweil E~J 1998 {\em Chem. Phys.\/}
  {\bf 233} 207--216

\bibitem{Witte2003}
Witte T, Hornung T, Windhorn L, Proch D, de~Vivie-Riedle R, Motzkus M and Kompa
  K~L 2003 {\em J. Chem. Phys.\/} {\bf 118} 2021--2024

\bibitem{Witte2004}
Witte T, Yeston J~S, Motzkus M, Heilweil E~J and Kompa K~L 2004 {\em Chem.
  Phys. Lett.\/} {\bf 392} 156--161

\bibitem{AssionBaumert1996}
Assion A, Baumert T, Helbing J, Seyfried V and Gerber G 1996 {\em Chem. Phys.
  Lett.\/} {\bf 259} 488--494

\bibitem{CerulloBardeen1996}
Cerullo G, Bardeen C~J, Wang Q and Shank C~V 1996 {\em Chem. Phys. Lett.\/}
  {\bf 262} 362--368

\bibitem{BardeenYakovlev1998}
Bardeen C~J, Yakovlev V~V, Squier J~A and Wilson K~R 1998 {\em J. Am. Chem.
  Soc.\/} {\bf 120} 13023--13027

\bibitem{Brakenhoff1999}
Brakenhoff G~J, Buist A~H, M\"uller M, Squier J~A, Bardeen C~J, Yakovlev V~V
  and Wilson K~R 1999 {\em Proc. SPIE\/} {\bf 3605} 40--47

\bibitem{VogtNuernbergerSelle2006}
Vogt G, Nuernberger P, Selle R, Dimler F, Brixner T and Gerber G 2006 {\em
  Phys. Rev.\/} A {\bf 74} 033413

\bibitem{ChenMaterny2000}
Chen T, Vierheilig A, Waltner P, Heid M, Kiefer W and Materny A 2000 {\em Chem.
  Phys. Lett.\/} {\bf 326} 375--382

\bibitem{Hellerer2004}
Hellerer T, Enejder A~M and Zumbusch A 2004 {\em Appl. Phys. Lett.\/} {\bf 85}
  25--27

\bibitem{Knutsen2004}
Knutsen K~P, Johnson J~C, Miller A~E, Petersen P~B and Saykally R~J 2004 {\em
  Chem. Phys. Lett.\/} {\bf 387} 436--441

\bibitem{WollenhauptPrakelt2006}
Wollenhaupt M, Pr\"{a}kelt A, Sarpe-Tudoran C, Liese D and Baumert T 2006 {\em
  Appl. Phys.\/} B {\bf 82} 183--188

\bibitem{KrugBayerWollenhaupt2009}
Krug M, Bayer T, Wollenhaupt M, Sarpe-Tudoran C, Baumert T, Ivanov S~S and
  Vitanov N~V 2009 {\em New J. Phys.\/} {\bf 11} 105051

\bibitem{Branderhorst2008}
Branderhorst M~P~A, Londero P, Wasylczyk P, Brif C, Kosut R~L, Rabitz H and
  Walmsley I~A 2008 {\em Science\/} {\bf 320} 638--643

\bibitem{UnderwoodSpannerIvanov2003PRL}
Underwood J~G, Spanner M, Ivanov M~Y, Mottershead J, Sussman B~J and Stolow A
  2003 {\em Phys. Rev. Lett.\/} {\bf 90} 223001

\bibitem{SussmanUnderwood2006PRA}
Sussman B~J, Underwood J~G, Lausten R, Ivanov M~Y and Stolow A 2006 {\em Phys.
  Rev. A\/} {\bf 73} 053403

\bibitem{SussmanIvanovStolow2005PRA}
Sussman B~J, Ivanov M~Y and Stolow A 2005 {\em Phys. Rev. A\/} {\bf 71} 051401

\bibitem{SussmanTownsendIvanov2006S}
Sussman B~J, Townsend D, Ivanov M~Y and Stolow A 2006 {\em Science\/} {\bf 314}
  278--281

\bibitem{Abragam1983}
Abragam A 1983 {\em Principles of Nuclear Magnetism\/} (Oxford, UK: Oxford
  University Press)

\bibitem{Ernst1990}
Ernst R~R, Bodenhausen G and Wokaun A 1990 {\em Principles of Nuclear Magnetic
  Resonance in One and Two Dimensions\/} (Oxford, UK: Oxford University Press)

\bibitem{Freeman1998}
Freeman R 1998 {\em Spin Choreography: Basic Steps in High Resolution {NMR}\/}
  (Oxford, UK: Oxford University Press)

\bibitem{Levitt2008}
Levitt M~H 2008 {\em Spin Dynamics: Basics of Nuclear Magnetic Resonance\/} 2nd
  ed (Chichester, UK: Wiley)

\bibitem{Slichter2010}
Slichter C~P 2010 {\em Principles of Magnetic Resonance\/} (Berlin Heidelberg:
  Springer)

\bibitem{Freeman1998PNMRS}
Freeman R 1998 {\em Prog. Nucl. Magn. Reson. Spectrosc.\/} {\bf 32} 59--106

\bibitem{Cory2000FP}
Cory D~G, Laflamme R, Knill E, Viola L, Havel T~F, Boulant N, Boutis G,
  Fortunato E, Lloyd S, Martinez R, Negrevergne C, Pravia M, Sharf Y,
  Teklemariam G, Weinstein Y~S and Zurek W~H 2000 {\em Fortschr. Phys.\/} {\bf
  48} 875--907

\bibitem{Jones2000FP}
Jones J~A 2000 {\em Fortschr. Phys.\/} {\bf 48} 909--924

\bibitem{Jones2001PNMRS}
Jones J~A 2001 {\em Prog. Nucl. Magn. Reson. Spectrosc.\/} {\bf 38} 325--360

\bibitem{VandersypenChuang2005}
Vandersypen L~M~K and Chuang I~L 2005 {\em Rev. Mod. Phys.\/} {\bf 76}
  1037--1069

\bibitem{RyanNegrevergne2008}
Ryan C~A, Negrevergne C, Laforest M, Knill E and Laflamme R 2008 {\em Phys.
  Rev.\/} A {\bf 78} 012328

\bibitem{KhanejaReiss2005}
Khaneja N, Reiss T, Kehlet C, Schulte-Herbr\"{u}ggen T and Glaser S~J 2005 {\em
  J. Magn. Reson.\/} {\bf 172} 296--305

\bibitem{Bryson1975}
Bryson A~E and Ho Y~C 1975 {\em Applied Optimal Control: Optimization,
  Estimation and Control\/} (Boca Raton, FL: Taylor and Francis)

\bibitem{Stengel1994}
Stengel R~F 1994 {\em Optimal Control and Estimation\/} (Mineola, NY: Dover)

\bibitem{Kraus1983}
Kraus K 1983 {\em States, Effects and Operations: Fundamental Notions of
  Quantum Theory\/} ({\em Lecture Notes in Physics\/} vol 190) (Berlin:
  Springer)

\bibitem{BreuerPetruccione2002}
Breuer H~P and Petruccione F 2002 {\em The Theory of Open Quantum Systems\/}
  (New York: Oxford University Press)

\bibitem{Gardiner2004}
Gardiner C~W and Zoller P 2004 {\em Quantum Noise: A Handbook of Markovian and
  Non-Markovian Quantum Stochastic Methods with Applications to Quantum
  Optics\/} (Berlin: Springer)

\bibitem{Lindblad1976}
Lindblad G 1976 {\em Commun. Math. Phys.\/} {\bf 48} 119--130

\bibitem{YanGillilan1993}
Yan Y~J, Gillilan R~E, Whitnell R~M, Wilson K~R and Mukamel S 1993 {\em J.
  Phys. Chem.\/} {\bf 97} 2320--2333

\bibitem{BartanaKosloffTannor1993}
Bartana A, Kosloff R and Tannor D~J 1993 {\em J. Chem. Phys.\/} {\bf 99}
  196--210

\bibitem{BartanaKosloffTannor1997}
Bartana A, Kosloff R and Tannor D~J 1997 {\em J. Chem. Phys.\/} {\bf 106}
  1435--1448

\bibitem{OhtsukiZhuRabitz1999}
Ohtsuki Y, Zhu W~S and Rabitz H 1999 {\em J. Chem. Phys.\/} {\bf 110}
  9825--9832

\bibitem{NielsenChuang2000}
Nielsen M~A and Chuang I~L 2000 {\em Quantum Computation and Quantum
  Information\/} (Cambridge, UK: Cambridge University Press)

\bibitem{PalaoKosloff2002}
Palao J~P and Kosloff R 2002 {\em Phys. Rev. Lett.\/} {\bf 89} 188301

\bibitem{PalaoKosloff2003}
Palao J~P and Kosloff R 2003 {\em Phys. Rev.\/} A {\bf 68} 062308

\bibitem{HornJohnson1990}
Horn R~A and Johnson C~R 1990 {\em Matrix Analysis\/} (Cambridge, UK: Cambridge
  University Press)

\bibitem{RabitzHsiehRosenthal2005PRA}
Rabitz H, Hsieh M and Rosenthal C 2005 {\em Phys. Rev.\/} A {\bf 72} 052337

\bibitem{HsiehRabitz2008PRA}
Hsieh M and Rabitz H 2008 {\em Phys. Rev.\/} A {\bf 77} 042306

\bibitem{HoDominyRabitz2009PRA}
Ho T~S, Dominy J and Rabitz H 2009 {\em Phys. Rev.\/} A {\bf 79} 013422

\bibitem{Gilchrist2005}
Gilchrist A, Langford N~K and Nielsen M~A 2005 {\em Phys. Rev.\/} A {\bf 71}
  062310

\bibitem{KosutGrace2006}
Kosut R~L, Grace M, Brif C and Rabitz H 2006 On the distance between unitary
  propagators of quantum systems of differing dimensions arXiv:quant-ph/0606064

\bibitem{GraceDominy2010NJP}
Grace M~D, Dominy J, Kosut R~L, Brif C and Rabitz H 2010 {\em New J. Phys.\/}
  {\bf 12} 015001

\bibitem{DemiralpRabitz1993}
Demiralp M and Rabitz H 1993 {\em Phys. Rev.\/} A {\bf 47} 809--816

\bibitem{Jozsa1994}
Jozsa R 1994 {\em J. Mod. Opt.\/} {\bf 41} 2315--2323

\bibitem{FuchsGraaf1999}
Fuchs C~A and van~de Graaf J 1999 {\em IEEE Trans. Inf. Theory\/} {\bf 45}
  1216--1227

\bibitem{JirariPotz2005}
Jirari H and P\"otz W 2005 {\em Phys. Rev.\/} A {\bf 72} 013409

\bibitem{RabitzHsiehRosenthal2006JCP}
Rabitz H, Hsieh M and Rosenthal C 2006 {\em J. Chem. Phys.\/} {\bf 124} 204107

\bibitem{ShenHsiehRabitz2006JCP}
Shen Z, Hsieh M and Rabitz H 2006 {\em J. Chem. Phys.\/} {\bf 124} 204106

\bibitem{WuRabitzHsieh2008JPA}
Wu R~B, Rabitz H and Hsieh M 2008 {\em J. Phys. A: Math. Theor.\/} {\bf 41}
  015006

\bibitem{HsiehWuRabitz2009JCP}
Hsieh M, Wu R~B and Rabitz H 2009 {\em J. Chem. Phys.\/} {\bf 130} 104109

\bibitem{RabitzHsiehRosenthal2004}
Rabitz H, Hsieh M and Rosenthal C 2004 {\em Science\/} {\bf 303} 1998--2001

\bibitem{RabitzHoHsieh2006PRA}
Rabitz H, Ho T~S, Hsieh M, Kosut R and Demiralp M 2006 {\em Phys. Rev.\/} A
  {\bf 74} 012721

\bibitem{BertrandBertrand1987}
Bertrand J and Bertrand P 1987 {\em Found. Phys.\/} {\bf 17} 397--405

\bibitem{VogelRisken1989}
Vogel K and Risken H 1989 {\em Phys. Rev.\/} A {\bf 40} 2847--2849

\bibitem{Leonhardt1997}
Leonhardt U 1997 {\em Measuring the Quantum State of Light\/} (Cambridge, UK:
  Cambridge University Press)

\bibitem{Buzek1998}
Bu\v{z}ek V, Derka R, Adam G and Knight P~L 1998 {\em Ann. Phys. (N.Y.)\/} {\bf
  266} 454--496

\bibitem{BrifMann1999PRA}
Brif C and Mann A 1999 {\em Phys. Rev.\/} A {\bf 59} 971--987

\bibitem{BrifMann2000JOB245}
Brif C and Mann A 2000 {\em J. Opt. B: Quantum Semiclass. Opt.\/} {\bf 2}
  245--251

\bibitem{Rehacek2008}
\v{R}eh\'{a}\v{c}ek J, Mogilevtsev D and Hradil Z 2008 {\em New J. Phys.\/}
  {\bf 10} 043022

\bibitem{DarianoPresti2001}
D'Ariano G~M and Lo~Presti P 2001 {\em Phys. Rev. Lett.\/} {\bf 86} 4195--4198

\bibitem{KosutWalmsleyRabitz2004}
Kosut R, Walmsley I~A and Rabitz H 2004 Optimal experiment design for quantum
  state and process tomography and {Hamiltonian} parameter estimation
  arXiv:quant-ph/0411093

\bibitem{Branderhorst2008JPB}
Branderhorst M~P~A, Walmsley I~A, Kosut R~L and Rabitz H 2008 {\em J. Phys. B:
  At. Mol. Opt. Phys.\/} {\bf 41} 074004

\bibitem{Mohseni2008PRA}
Mohseni M, Rezakhani A~T and Lidar D~A 2008 {\em Phys. Rev.\/} A {\bf 77}
  032322

\bibitem{YoungWhaley2009}
Young K~C, Sarovar M, Kosut R and Whaley K~B 2009 {\em Phys. Rev.\/} A {\bf 79}
  062301

\bibitem{EmersonSilva2007}
Emerson J, Silva M, Moussa O, Ryan C, Laforest M, Baugh J, Cory D~G and
  Laflamme R 2007 {\em Science\/} {\bf 317} 1893--1896

\bibitem{Kosut2008L1norm}
Kosut R~L 2009 Quantum process tomography via {$\ell_1$-norm} minimization
  arXiv:0812.4323

\bibitem{BranderhorstNunn2009NJP}
Branderhorst M~P~A, Nunn J, Walmsley I~A and Kosut R~L 2009 {\em New J.
  Phys.\/} {\bf 11} 115010

\bibitem{ShabaniKosutRabitz2009}
Shabani A, Kosut R~L and Rabitz H 2009 Compressed quantum process tomography
  arXiv:0910.5498

\bibitem{BenderskyPastawskiPaz2008}
Bendersky A, Pastawski F and Paz J~P 2008 {\em Phys. Rev. Lett.\/} {\bf 100}
  190403

\bibitem{SchmiegelowLarotondaPaz2010}
Schmiegelow C~T, Larotonda M~A and Paz J~P 2010 Selective and efficient quantum
  process tomography with single photons arXiv:1002.4436

\bibitem{CramerPlenio2010}
Cramer M and Plenio M~B 2010 Reconstructing quantum states efficiently
  arXiv:1002.3780

\bibitem{FlammiaGross2010}
Flammia S~T, Gross D, Bartlett S~D and Somma R 2010 Heralded polynomial-time
  quantum state tomography arXiv:1002.3839

\bibitem{OhtsukiNakagami2001}
Ohtsuki Y, Nakagami K, Fujimura Y, Zhu W~S and Rabitz H 2001 {\em J. Chem.
  Phys.\/} {\bf 114} 8867--8876

\bibitem{Shir2007}
Shir O~M, Emmerich M, B\"{a}ck T and Vrakking M~J~J 2007 The application of
  evolutionary multi-criteria optimization to dynamic molecular alignment {\em
  Proceedings of IEEE Congress on Evolutionary Computation (CEC 2007)\/} pp
  4108--4115

\bibitem{RajWu2008a}
Chakrabarti R, Wu R~B and Rabitz H 2008 {\em Phys. Rev.\/} A {\bf 77} 063425

\bibitem{RajWu2008b}
Chakrabarti R, Wu R~B and Rabitz H 2008 {\em Phys. Rev.\/} A {\bf 78} 033414

\bibitem{BeltraniGhosh2009}
Beltrani V, Ghosh P and Rabitz H 2009 {\em J. Chem. Phys.\/} {\bf 130} 164112

\bibitem{ChankongHaimes1983}
Chankong V and Haimes Y~Y 1983 {\em Multiobjective Decision Making Theory and
  Methodology\/} (New York: North-Holland)

\bibitem{Steuer1986}
Steuer R~E 1986 {\em Multiple Criteria Optimization: Theory, Computation and
  Application\/} (New York: Wiley)

\bibitem{Miettinen1998}
Miettinen K~M 1998 {\em Nonlinear Multiobjective Optimization\/} (Norwell, MA:
  Kluwer)

\bibitem{Jurdjevic1997}
Jurdjevic V 1997 {\em Geometric Control Theory\/} (Cambridge, UK: Cambridge
  University Press)

\bibitem{DAlessandro2001a}
D'Alessandro D and Dahleh M 2001 {\em IEEE Trans. Autom. Control\/} {\bf 46}
  866--876

\bibitem{HuangTarnClark1983}
Huang G~M, Tarn T~J and Clark J~W 1983 {\em J. Math. Phys.\/} {\bf 24}
  2608--2618

\bibitem{Ramakrishna1995}
Ramakrishna V, Salapaka M~V, Dahleh M, Rabitz H and Peirce A 1995 {\em Phys.
  Rev.\/} A {\bf 51} 960--966

\bibitem{TuriniciRabitz2001}
Turinici G and Rabitz H 2001 {\em Chem. Phys.\/} {\bf 267} 1--9

\bibitem{TuriniciRabitz2003}
Turinici G and Rabitz H 2003 {\em J. Phys. A: Math. Gen.\/} {\bf 36} 2565--2576

\bibitem{Albertini2001}
Albertini F and D'Alessandro D 2001 Notions of controllability for quantum
  mechanical systems {\em Proceedings of the 40th IEEE Conference on Decision
  and Control\/} vol~2 pp 1589--1594

\bibitem{FuSchirmerSolomon2001}
Fu H, Schirmer S~G and Solomon A~I 2001 {\em J. Phys. A: Math. Gen.\/} {\bf 34}
  1679--1690

\bibitem{SchirmerFuSolomon2001}
Schirmer S~G, Fu H and Solomon A~I 2001 {\em Phys. Rev.\/} A {\bf 63} 063410

\bibitem{Altafini2002}
Altafini C 2002 {\em J. Math. Phys.\/} {\bf 43} 2051--2062

\bibitem{GirardeauKoch1998}
Girardeau M~D, Schirmer S~G, Leahy J~V and Koch R~M 1998 {\em Phys. Rev.\/} A
  {\bf 58} 2684--2689

\bibitem{SchirmerLeahy2001}
Schirmer S~G and Leahy J~V 2001 {\em Phys. Rev.\/} A {\bf 63} 025403

\bibitem{SchirmerSolomon2002a}
Schirmer S~G, Solomon A~I and Leahy J~V 2002 {\em J. Phys. A: Math. Gen.\/}
  {\bf 35} 4125--4141

\bibitem{SchirmerSolomon2002b}
Schirmer S~G, Solomon A~I and Leahy J~V 2002 {\em J. Phys. A: Math. Gen.\/}
  {\bf 35} 8551--8562

\bibitem{Albertini2003}
Albertini A and D'Alessandro D 2003 {\em IEEE Trans. Autom. Control\/} {\bf 48}
  1399--1403

\bibitem{ShahTannorRice2002PRA}
Shah S~P, Tannor D~J and Rice S~A 2002 {\em Phys. Rev.\/} A {\bf 66} 033405

\bibitem{GongRice2004PRA}
Gong J and Rice S~A 2004 {\em Phys. Rev.\/} A {\bf 69} 063410

\bibitem{SchirmerPullen2005}
Schirmer S~G, Pullen I~C~H and Solomon A~I 2005 {\em J. Opt. B: Quantum
  Semiclass. Opt.\/} {\bf 7} S293--S299

\bibitem{TuriniciRabitz2010JPA}
Turinici G and Rabitz H 2010 {\em J. Phys. A: Math. Theor.\/} {\bf 43} 105303

\bibitem{ClarkLucarelliTarn2003}
Clark J~W, Lucarelli D~G and Tarn T~J 2003 {\em Int. J. Mod. Phys.\/} B {\bf
  17} 5397--5411

\bibitem{WuTarnLi2006}
Wu R~B, Tarn T~J and Li C~W 2006 {\em Phys. Rev.\/} A {\bf 73} 012719

\bibitem{VilelaMendesManko2010}
Vilela~Mendes R and Man'ko V~I 2010 On the problem of quantum control in
  infinite dimensions arXiv:1004.3447

\bibitem{LloydViola2001}
Lloyd S and Viola L 2001 {\em Phys. Rev.\/} A {\bf 65} 010101

\bibitem{SolomonSchirmer2004eprint}
Solomon A~I and Schirmer S~G 2004 Dissipative quantum control
  arXiv:quant-ph/0401094

\bibitem{Altafini2003JMP}
Altafini C 2003 {\em J. Math. Phys.\/} {\bf 44} 2357--2372

\bibitem{Altafini2004PRA}
Altafini C 2004 {\em Phys. Rev.\/} A {\bf 70} 062321

\bibitem{Romano2005}
Romano R 2005 {\em J. Phys. A: Math. Gen.\/} {\bf 38} 9105--9114

\bibitem{WuPechenBrif2007}
Wu R, Pechen A, Brif C and Rabitz H 2007 {\em J. Phys. A: Math. Theor.\/} {\bf
  40} 5681--5693

\bibitem{VilelaMendes2009}
{Vilela Mendes} R 2009 {\em Phys. Lett.\/} A {\bf 373} 2529--2532

\bibitem{DirrHelmke2009}
Dirr G, Helmke U, Kurniawan I and Schulte-Herbr\"{u}ggen T 2009 {\em Rep. Math.
  Phys.\/} {\bf 64} 93--121

\bibitem{WuRaj2008}
Wu R~B, Chakrabarti R and Rabitz H 2008 {\em Phys. Rev.\/} A {\bf 77} 052303

\bibitem{OhtsukiNakagami2003}
Ohtsuki Y, Nakagami K, Zhu W~S and Rabitz H 2003 {\em Chem. Phys.\/} {\bf 287}
  197--216

\bibitem{XuYanOhtsuki2004}
Xu R, Yan Y~J, Ohtsuki Y, Fujimura Y and Rabitz H 2004 {\em J. Chem. Phys.\/}
  {\bf 120} 6600--6608

\bibitem{BeyversSaalfrank2008}
Beyvers S and Saalfrank P 2008 {\em J. Chem. Phys.\/} {\bf 128} 074104

\bibitem{CuiXiPan2008PRA}
Cui W, Xi Z~R and Pan Y 2008 {\em Phys. Rev.\/} A {\bf 77} 032117

\bibitem{MohseniRezakhani2009PRA}
Mohseni M and Rezakhani A~T 2009 {\em Phys. Rev.\/} A {\bf 80} 010101

\bibitem{WerschnikGross2005}
Werschnik J and Gross E~K~U 2005 {\em J. Opt. B: Quantum Semiclass. Opt.\/}
  {\bf 7} S300--S312

\bibitem{LapertTehini2009}
Lapert M, Tehini R, Turinici G and Sugny D 2009 {\em Phys. Rev.\/} A {\bf 79}
  063411

\bibitem{LapertTehini2008}
Lapert M, Tehini R, Turinici G and Sugny D 2008 {\em Phys. Rev.\/} A {\bf 78}
  023408

\bibitem{OhtsukiNakagami2008}
Ohtsuki Y and Nakagami K 2008 {\em Phys. Rev.\/} A {\bf 77} 033414

\bibitem{SerbanWerschnikGross2005}
Serban I, Werschnik J and Gross E~K~U 2005 {\em Phys. Rev.\/} A {\bf 71} 053810

\bibitem{KaiserMay2004}
Kaiser A and May V 2004 {\em J. Chem. Phys.\/} {\bf 121} 2528--2535

\bibitem{GrigorenkoGarcia2002}
Grigorenko I, Garcia M~E and Bennemann K~H 2002 {\em Phys. Rev. Lett.\/} {\bf
  89} 233003

\bibitem{MishimaYamashita2009a}
Mishima K and Yamashita K 2009 {\em J. Chem. Phys.\/} {\bf 130} 034108

\bibitem{MishimaYamashita2009b}
Mishima K and Yamashita K 2009 {\em J. Chem. Phys.\/} {\bf 131} 014109

\bibitem{Khaneja2001}
Khaneja N, Brockett R and Glaser S~J 2001 {\em Phys. Rev.\/} A {\bf 63} 032308

\bibitem{Khaneja2002}
Khaneja N, Glaser S~J and Brockett R 2002 {\em Phys. Rev.\/} A {\bf 65} 032301

\bibitem{ReissKhaneja2002}
Reiss T~O, Khaneja N and Glaser S~J 2002 {\em J. Magn. Reson.\/} {\bf 154}
  192--195

\bibitem{YuanKhaneja2005}
Yuan H and Khaneja N 2005 {\em Phys. Rev.\/} A {\bf 72} 040301

\bibitem{ZhaoRice1991}
Zhao M and Rice S~A 1991 {\em J. Chem. Phys.\/} {\bf 95} 2465--2472

\bibitem{WuPechenRabitz2008JMP}
Wu R~B, Pechen A, Rabitz H, Hsieh M and Tsou B 2008 {\em J. Math. Phys.\/} {\bf
  49} 022108

\bibitem{TannorKazakovOrlov1992}
Tannor D~J, Kazakov V and Orlov V 1992 Control of photochemical branching:
  Novel procedures for finding optimal pulses and global upper bounds {\em Time
  Dependent Quantum Molecular Dynamics\/} ed Broeckhove J and Lathouwers L (New
  York: Plenum) pp 347--360

\bibitem{SomloiKazakovTannor1993}
Soml\'{o}i J, Kazakov V~A and Tannor D~J 1993 {\em Chem. Phys.\/} {\bf 172}
  85--98

\bibitem{ZhuBotinaRabitz1998}
Zhu W~S, Botina J and Rabitz H 1998 {\em J. Chem. Phys.\/} {\bf 108} 1953--1963

\bibitem{ZhuRabitz1998}
Zhu W~S and Rabitz H 1998 {\em J. Chem. Phys.\/} {\bf 109} 385--391

\bibitem{Maday2003}
Maday G and Turinici G 2003 {\em J. Chem. Phys.\/} {\bf 118} 8191--8196

\bibitem{Ohtsuki2004}
Ohtsuki Y, Turinici G and Rabitz H 2004 {\em J. Chem. Phys.\/} {\bf 120}
  5509--5517

\bibitem{Ohtsuki2007}
Ohtsuki Y, Teranishi Y, Saalfrank P, Turinici G and Rabitz H 2007 {\em Phys.
  Rev.\/} A {\bf 75} 033407

\bibitem{BorziSalomon2008}
Borz\`{i} A, Salomon J and Volkwein S 2008 {\em J. Comput. Appl. Math.\/} {\bf
  216} 170--197

\bibitem{DitzBorzi2008}
Ditz P and Borz\`{i} A 2008 {\em Comp. Phys. Commun.\/} {\bf 178} 393--399

\bibitem{ZhuRabitz1999}
Zhu W~S and Rabitz H 1999 {\em J. Chem. Phys.\/} {\bf 110} 7142--7152

\bibitem{Hillermeier2001}
Hillermeier C 2001 {\em Nonlinear Multiobjective Optimization: A Generalized
  Homotopy Approach\/} (Basel: Birkh\"{a}user)

\bibitem{RothmanHoRabitz2005JCP}
Rothman A, Ho T~S and Rabitz H 2005 {\em J. Chem. Phys.\/} {\bf 123} 134104

\bibitem{RothmanHoRabitz2006PRA}
Rothman A, Ho T~S and Rabitz H 2006 {\em Phys. Rev.\/} A {\bf 73} 053401

\bibitem{CastroGross2009PRE}
Castro A and Gross E~K~U 2009 {\em Phys. Rev.\/} E {\bf 79} 056704

\bibitem{Yip2003}
Yip F, Mazziotti D and Rabitz H 2003 {\em J. Chem. Phys.\/} {\bf 118}
  8168--8172

\bibitem{BalintKurtiManby2005}
Balint-Kurti G~G, Manby F~R, Ren Q, Artamonov M, Ho T~S and Rabitz H 2005 {\em
  J. Chem. Phys.\/} {\bf 122} 084110

\bibitem{HsiehRabitz2008PRE}
Hsieh M and Rabitz H 2008 {\em Phys. Rev.\/} E {\bf 77} 037701

\bibitem{StroheckerRabitz2010JCC}
Strohecker T and Rabitz H 2010 {\em J. Comp. Chem.\/} {\bf 31} 151--153

\bibitem{ArtamonovHo2004CP}
Artamonov M, Ho T~S and Rabitz H 2004 {\em Chem. Phys.\/} {\bf 305} 213--222

\bibitem{ArtamonovHo2006CP}
Artamonov M, Ho T~S and Rabitz H 2006 {\em Chem. Phys.\/} {\bf 328} 147--155

\bibitem{ArtamonovHo2006JCP}
Artamonov M, Ho T~S and Rabitz H 2006 {\em J. Chem. Phys.\/} {\bf 124} 064306

\bibitem{KurosakiArtamonov2009}
Kurosaki Y, Artamonov M, Ho T~S and Rabitz H 2009 {\em J. Chem. Phys.\/} {\bf
  131} 044306

\bibitem{KannoHoki2007}
Kanno M, Hoki K, Kono H and Fujimura Y 2007 {\em J. Chem. Phys.\/} {\bf 127}
  204314

\bibitem{LiWelack2008}
Li G~Q, Welack S, Schreiber M and Kleinekath\"{o}fer U 2008 {\em Phys. Rev.\/}
  B {\bf 77} 075321

\bibitem{WangMay2009}
Wang L and May V 2009 {\em Chem. Phys.\/} {\bf 361} 1--8

\bibitem{Kosionis2007}
Kosionis S~G, Terzis A~F and Paspalakis E 2007 {\em Phys. Rev.\/} B {\bf 75}
  193305

\bibitem{RasanenCastro2007}
R\"{a}s\"{a}nen E, Castro A, Werschnik J, Rubio A and Gross E~K~U 2007 {\em
  Phys. Rev. Lett.\/} {\bf 98} 157404

\bibitem{RasanenCastro2008}
R\"{a}s\"{a}nen E, Castro A, Werschnik J, Rubio A and Gross E~K~U 2008 {\em
  Phys. Rev.\/} B {\bf 77} 085324

\bibitem{ChiaraCalarco2008}
Chiara G~D, Calarco T, Anderlini M, Montangero S, Lee P~J, Brown B~L, Phillips
  W~D and Porto J~V 2008 {\em Phys. Rev.\/} A {\bf 77} 052333

\bibitem{HohenesterRekdal2007}
Hohenester U, Rekdal P~K, Borz\`{i} A and Schmiedmayer J 2007 {\em Phys.
  Rev.\/} A {\bf 75} 023602

\bibitem{DoriaCalarcoMontangero2010}
Doria P, Calarco T and Montangero S 2010 Antiadiabatic control of many body
  quantum systems arXiv:1003.3750

\bibitem{JirariHekking2009}
Jirari H, Hekking F~W~J and Buisson O 2009 {\em Europhys. Lett.\/} {\bf 87}
  28004

\bibitem{GrigorenkoRabitz2009}
Grigorenko I and Rabitz H 2009 {\em Appl. Phys. Lett.\/} {\bf 94} 253107

\bibitem{HohenesterStadler2004}
Hohenester U and Stadler G 2004 {\em Phys. Rev. Lett.\/} {\bf 92} 196801

\bibitem{SklarzTannorKhaneja2004}
Sklarz S~E, Tannor D~J and Khaneja N 2004 {\em Phys. Rev.\/} A {\bf 69} 053408

\bibitem{Grigorenko2005a}
Grigorenko I~A and Khveshchenko D~V 2005 {\em Phys. Rev. Lett.\/} {\bf 94}
  040506

\bibitem{JirariPotz2006}
Jirari H and P\"{o}tz W 2006 {\em Phys. Rev.\/} A {\bf 74} 022306

\bibitem{WeninPotz2006}
Wenin M and P\"{o}tz W 2006 {\em Phys. Rev.\/} A {\bf 74} 022319

\bibitem{Potz2007JCE}
P\"{o}tz W 2007 {\em J. Comp. Electron.\/} {\bf 6} 171--174

\bibitem{Pelzer2007}
Pelzer A, Ramakrishna S and Seideman T 2007 {\em J. Chem. Phys.\/} {\bf 126}
  034503

\bibitem{Pelzer2008}
Pelzer A, Ramakrishna S and Seideman T 2008 {\em J. Chem. Phys.\/} {\bf 129}
  134301

\bibitem{PalaoKosloffKoch2008}
Palao J~P, Kosloff R and Koch C~P 2008 {\em Phys. Rev.\/} A {\bf 77} 063412

\bibitem{Jirari2009}
Jirari H 2009 {\em Europhys. Lett.\/} {\bf 87} 40003

\bibitem{Gordon2009}
Gordon G 2009 {\em J. Phys. B: At. Mol. Opt. Phys.\/} {\bf 42} 223001

\bibitem{SandersKimHolton1999}
Sanders G~D, Kim K~W and Holton W~C 1999 {\em Phys. Rev.\/} A {\bf 59}
  1098--1101

\bibitem{SklarzTannor2004}
Sklarz S~E and Tannor D~J 2004 Local control theory for unitary
  transformations: Application to quantum computing without leakage
  arXiv:quant-ph/0404081

\bibitem{SklarzTannor2006}
Sklarz S~E and Tannor D~J 2006 {\em Chem. Phys.\/} {\bf 322} 87--97

\bibitem{SchulteSporl2005}
Schulte-Herbr\"uggen T, Sp\"orl A, Khaneja N and Glaser S~J 2005 {\em Phys.
  Rev.\/} A {\bf 72} 042331

\bibitem{SporlSchulte2007}
Sp\"{o}rl A, Schulte-Herbr\"{u}ggen T, Glaser S~J, Bergholm V, Storcz M~J,
  Ferber J and Wilhelm F~K 2007 {\em Phys. Rev.\/} A {\bf 75} 012302

\bibitem{VivieRiedleTroppmann2007}
de~Vivie-Riedle R and Troppmann U 2007 {\em Chem. Rev.\/} {\bf 107} 5082--5100

\bibitem{DominyRabitz2008JPA}
Dominy J and Rabitz H 2008 {\em J. Phys. A: Math. Theor.\/} {\bf 41} 205305

\bibitem{SchroderBrown2009JCP}
Schr\"{o}der M and Brown A 2009 {\em J. Chem. Phys.\/} {\bf 131} 034101

\bibitem{Nebendahl2009}
Nebendahl V, H\"{a}ffner H and Roos C~F 2009 {\em Phys. Rev.\/} A {\bf 79}
  012312

\bibitem{Schirmer2009JMO}
Schirmer S 2009 {\em J. Mod. Opt.\/} {\bf 56} 831--839

\bibitem{NigmatullinSchirmer2009}
Nigmatullin R and Schirmer S~G 2009 {\em New J. Phys.\/} {\bf 11} 105032

\bibitem{FisherHelmerGlaser2010}
Fisher R, Helmer F, Glaser S~J, Marquardt F and Schulte-Herbr\"{u}ggen T 2010
  {\em Phys. Rev.\/} B {\bf 81} 085328

\bibitem{GollubKowalewski2008}
Gollub C, Kowalewski M and de~Vivie-Riedle R 2008 {\em Phys. Rev. Lett.\/} {\bf
  101} 073002

\bibitem{SchroderBrown2009NJP}
Schr\"{o}der M and Brown A 2009 {\em New J. Phys.\/} {\bf 11} 105031

\bibitem{LiGaitan2010}
Li R and Gaitan F 2010 High-fidelity universal quantum gates through
  group-symmetrized rapid passage arXiv:1004.0710

\bibitem{Grigorenko2005b}
Grigorenko I~A and Khveshchenko D~V 2005 {\em Phys. Rev. Lett.\/} {\bf 95}
  110501

\bibitem{SchulteSporlKhaneja2006}
Schulte-Herbr\"{u}ggen T, Sp\"{o}rl A, Khaneja N and Glaser S~J 2009 Optimal
  control for generating quantum gates in open dissipative systems
  arXiv:quant-ph/0609037

\bibitem{Hohenester2006}
Hohenester U 2006 {\em Phys. Rev.\/} B {\bf 74} 161307

\bibitem{MontangeroCalarcoFazio2007PRL}
Montangero S, Calarco T and Fazio R 2007 {\em Phys. Rev. Lett.\/} {\bf 99}
  170501

\bibitem{GraceBrif2007JPB}
Grace M, Brif C, Rabitz H, Walmsley I~A, Kosut R~L and Lidar D~A 2007 {\em J.
  Phys. B: At. Mol. Opt. Phys.\/} {\bf 40} S103--S125

\bibitem{GraceBrif2007JMO}
Grace M~D, Brif C, Rabitz H, Lidar D~A, Walmsley I~A and Kosut R~L 2007 {\em J.
  Mod. Opt.\/} {\bf 54} 2339--2349

\bibitem{WeninPotz2008PRA}
Wenin M and P\"{o}tz W 2008 {\em Phys. Rev.\/} A {\bf 78} 012358

\bibitem{WeninPotz2008PRB}
Wenin M and P\"{o}tz W 2008 {\em Phys. Rev.\/} B {\bf 78} 165118

\bibitem{Rebentrost2009PRL}
Rebentrost P, Serban I, Schulte-Herbr\"{u}ggen T and Wilhelm F~K 2009 {\em
  Phys. Rev. Lett.\/} {\bf 102} 090401

\bibitem{RebentrostWilhelm2009}
Rebentrost P and Wilhelm F~K 2009 {\em Phys. Rev.\/} B {\bf 79} 060507

\bibitem{MotzoiGambetta2009PRL}
Motzoi F, Gambetta J~M, Rebentrost P and Wilhelm F~K 2009 {\em Phys. Rev.
  Lett.\/} {\bf 103} 110501

\bibitem{SafaeiMontangero2009}
Safaei S, Montangero S, Taddei F and Fazio R 2009 {\em Phys. Rev.\/} B {\bf 79}
  064524

\bibitem{RoloffPotz2009PRB}
Roloff R and P\"otz W 2009 {\em Phys. Rev.\/} B {\bf 79} 224516

\bibitem{WeninRoloffPotz2009}
Wenin M, Roloff R and P\"{o}tz W 2009 {\em J. Appl. Phys.\/} {\bf 105} 084504

\bibitem{RoloffWeninPotz2009JCE}
Roloff R, Wenin M and P\"{o}tz W 2009 {\em J. Comp. Electron.\/} {\bf 8} 29--34

\bibitem{RoloffWeninPotz2009JCTN}
Roloff R, Wenin M and P\"{o}tz W 2009 {\em J. Comput. Theor. Nanosci.\/} {\bf
  6} 1837--1863

\bibitem{GalveLutz2009}
Galve F and Lutz E 2009 {\em Phys. Rev.\/} A {\bf 79} 032327

\bibitem{FisherYuan2009}
Fisher R, Yuan H, Sp\"{o}rl A and Glaser S 2009 {\em Phys. Rev.\/} A {\bf 79}
  042304

\bibitem{WangSchirmer2009PRA}
Wang X and Schirmer S~G 2009 {\em Phys. Rev.\/} A {\bf 80} 042305

\bibitem{MurphyMontangero2010}
Murphy M, Montangero S, Giovannetti V and Calarco T 2010 Communication at the
  quantum speed limit along a spin chain arXiv:1004.3445

\bibitem{TimoneyElman2008}
Timoney N, Elman V, Glaser S, Weiss C, Johanning M, Neuhauser W and Wunderlich
  C 2008 {\em Phys. Rev.\/} A {\bf 77} 052334

\bibitem{NunnWalmsley2007}
Nunn J, Walmsley I~A, Raymer M~G, Surmacz K, Waldermann F~C, Wang Z and Jaksch
  D 2007 {\em Phys. Rev.\/} A {\bf 75} 011401

\bibitem{GorshkovAndreLukin2007}
Gorshkov A~V, Andr\'{e} A, Fleischhauer M, S{\o}rensen A~S and Lukin M~D 2007
  {\em Phys. Rev. Lett.\/} {\bf 98} 123601

\bibitem{GorshkovCalarcoLukin2008}
Gorshkov A~V, Calarco T, Lukin M~D and S{\o}rensen A~S 2008 {\em Phys. Rev.\/}
  A {\bf 77} 043806

\bibitem{NovikovaGorshkovPhillips2007}
Novikova I, Gorshkov A~V, Phillips D~F, S{\o}rensen A~S, Lukin M~D and
  Walsworth R~L 2007 {\em Phys. Rev. Lett.\/} {\bf 98} 243602

\bibitem{NovikovaPhillipsGorshkov2008}
Novikova I, Phillips N~B and Gorshkov A~V 2008 {\em Phys. Rev.\/} A {\bf 78}
  021802

\bibitem{PhillipsGorshkovNovikova2008}
Phillips N~B, Gorshkov A~V and Novikova I 2008 {\em Phys. Rev.\/} A {\bf 78}
  023801

\bibitem{Chakrabarti2009}
Chakrabarti R and Ghosh A 2009 Asymptotic efficiency and finite sample
  performance of frequentist quantum state estimation arXiv:0904.1628

\bibitem{HoRab2006a}
Ho T~S and Rabitz H 2006 {\em J. Photochem. Photobiol.\/} A {\bf 180} 226--240

\bibitem{Glaser1998}
Glaser S~J, Schulte-Herbr\"{u}ggen T, Sieveking M, Schedletzky O, Nielsen N~C,
  S{\o}rensen O~W and Griesinger C 1998 {\em Science\/} {\bf 280} 421--424

\bibitem{BonnardChyba2003}
Bonnard B and Chyba M 2003 {\em Singular Trajectories and Their Role in Control
  Theory\/} (Berlin: Springer)

\bibitem{WuDominyHo2009}
Wu R~B, Dominy J, Ho T~S and Rabitz H 2009 Singularities of quantum control
  landscapes arXiv:0907.2354

\bibitem{LapertZhangBraun2010}
Lapert M, Zhang Y, Braun M, Glaser S~J and Sugny D 2010 {\em Phys. Rev.
  Lett.\/} {\bf 104} 083001

\bibitem{Brockett1991}
Brockett R~W 1991 {\em Linear Alg. Appl.\/} {\bf 146} 79--91

\bibitem{VonNeumann1937a}
von Neumann J 1937 {\em Tomsk Univ. Rev.\/} {\bf 1} 286--300

\bibitem{RajWu2007b}
Chakrabarti R, Wu R~B and Rabitz H 2008 Computational complexity of quantum
  optimal control landscapes arXiv:0708.3513

\bibitem{MooreHsiehRabitz2008JCP}
Moore K, Hsieh M and Rabitz H 2008 {\em J. Chem. Phys.\/} {\bf 128} 154117

\bibitem{OzaPechen2009JPA}
Oza A, Pechen A, Dominy J, Beltrani V, Moore K and Rabitz H 2009 {\em J. Phys.
  A: Math. Theor.\/} {\bf 42} 205305

\bibitem{RoslundRabitz2009}
Roslund J and Rabitz H 2009 {\em Phys. Rev.\/} A {\bf 79} 053417

\bibitem{RoslundRoth2006}
Roslund J, Roth M and Rabitz H 2006 {\em Phys. Rev.\/} A {\bf 74} 043414

\bibitem{WollenhauptPrakelt2005JMO}
Wollenhaupt M, Pr\"{a}kelt A, Sarpe-Tudoran C, Liese D and Baumert T 2005 {\em
  J. Mod. Opt.\/} {\bf 52} 2187--2195

\bibitem{Bayer2008}
Bayer T, Wollenhaupt M and Baumert T 2008 {\em J. Phys. B: At. Mol. Opt.
  Phys.\/} {\bf 41} 074007

\bibitem{MarquetandNuernberger2007}
Marquetand P, Nuernberger P, Vogt G, Brixner T and Engel V 2007 {\em Europhys.
  Lett.\/} {\bf 80} 53001

\bibitem{RoslundRabitz2009b}
Roslund J and Rabitz H 2009 {\em Phys. Rev.\/} A {\bf 80} 013408

\bibitem{JudsonRabitz1992}
Judson R~S and Rabitz H 1992 {\em Phys. Rev. Lett.\/} {\bf 68} 1500--1503

\bibitem{GeremiaZhu2000}
Geremia J~M, Zhu W~S and Rabitz H 2000 {\em J. Chem. Phys.\/} {\bf 113}
  10841--10848

\bibitem{OmenettoLuceTaylor1999}
Omenetto F~G, Luce B~P and Taylor A~J 1999 {\em J. Opt. Soc. Am.\/} B {\bf 16}
  2005--2009

\bibitem{BrixnerAbajo2005PRL}
Brixner T, Garc\'{i}a~de Abajo F~J, Schneider J and Pfeiffer W 2005 {\em Phys.
  Rev. Lett.\/} {\bf 95} 093901

\bibitem{BrixnerAbajo2006PRB}
Brixner T, de~Abajo F~J~G, Schneider J, Spindler C and Pfeiffer W 2006 {\em
  Phys. Rev.\/} B {\bf 73} 125437

\bibitem{BrixnerAbajo2006APB}
Brixner T, Garc\'{\i}a~de Abajo F, Spindler C and Pfeiffer W 2006 {\em Appl.
  Phys.\/} B {\bf 84} 89--95

\bibitem{HertzRouzee2007}
Hertz E, Rouz\'{e}e A, Gu\'{e}rin S, Lavorel B and Faucher O 2007 {\em Phys.
  Rev.\/} A {\bf 75} 031403

\bibitem{Voronine2006JCP}
Voronine D, Abramavicius D and Mukamel S 2006 {\em J. Chem. Phys.\/} {\bf 124}
  034104

\bibitem{Voronine2007JCP}
Voronine D~V, Abramavicius D and Mukamel S 2007 {\em J. Chem. Phys.\/} {\bf
  126} 044508

\bibitem{Tuchscherer2009}
Tuchscherer P, Rewitz C, Voronine D~V, Garc\'{i}a~de Abajo F~J, Pfeiffer W and
  Brixner T 2009 {\em Opt. Expr.\/} {\bf 17} 14235--14259

\bibitem{ZhuRabitz2003}
Zhu W~S and Rabitz H 2003 {\em J. Chem. Phys.\/} {\bf 118} 6751--6757

\bibitem{GraceBrif2006}
Grace M, Brif C, Rabitz H, Walmsley I, Kosut R and Lidar D 2006 {\em New J.
  Phys.\/} {\bf 8} 35

\bibitem{GollubVivieRiedle2008PRA}
Gollub C and de~Vivie-Riedle R 2008 {\em Phys. Rev.\/} A {\bf 78} 033424

\bibitem{GollubVivieRiedle2009}
Gollub C and de~Vivie-Riedle R 2009 {\em Phys. Rev.\/} A {\bf 79} 021401

\bibitem{Belavkin1983}
Belavkin V~P 1983 {\em Autom. Remote Control\/} {\bf 44} 178--188
  arXiv:quant-ph/0408003

\bibitem{WisemanMilburn1993PRL}
Wiseman H~M and Milburn G~J 1993 {\em Phys. Rev. Lett.\/} {\bf 70} 548--551

\bibitem{Wiseman1994PRA}
Wiseman H~M 1994 {\em Phys. Rev.\/} A {\bf 49} 2133--2150

\bibitem{DohertyHabib2000}
Doherty A~C, Habib S, Jacobs K, Mabuchi H and Tan S~M 2000 {\em Phys. Rev.\/} A
  {\bf 62} 012105

\bibitem{DohertyDoyle2000}
Doherty A, Doyle J, Mabuchi H, Jacobs K and Habib S 2000 Robust control in the
  quantum domain {\em Proceedings of the 39th IEEE Conference on Decision and
  Control\/} vol~1 pp 949--954 arXiv:quant-ph/0105018

\bibitem{WisemanMilburn2010}
Wiseman H~M and Milburn G~J 2010 {\em Quantum Measurement and Control\/}
  (Cambridge, UK: Cambridge University Press)

\bibitem{WeinerLeaird1990}
Weiner A~M, Leaird D~E, Patel J~S and Wullert J~R 1990 {\em Opt. Lett.\/} {\bf
  15} 326

\bibitem{WefersNelson1993}
Wefers M~M and Nelson K~A 1993 {\em Opt. Lett.\/} {\bf 18} 2032--2034

\bibitem{WefersNelson1995}
Wefers M~M and Nelson K~A 1995 {\em Opt. Lett.\/} {\bf 20} 1047

\bibitem{HillegasWarren1994}
Hillegas C~W, Tull J~X, Goswami D, Strickland D and Warren W~S 1994 {\em Opt.
  Lett.\/} {\bf 19} 737--739

\bibitem{VerluiseLaude2000}
Verluise F, Laude V, Cheng Z, Spielmann C and Tournois P 2000 {\em Opt.
  Lett.\/} {\bf 25} 575--577

\bibitem{Stobrawa2001}
Stobrawa G, Hacker M, Feurer T, Zeidler D, Motzkus M and Reichel F 2001 {\em
  Appl. Phys.\/} B {\bf 72} 627--630

\bibitem{Monmayrant2004}
Monmayrant A and Chatel B 2004 {\em Review of Scientific Instruments\/} {\bf
  75} 2668--2671

\bibitem{PrakeltWollenhaupt2003}
Pr\"{a}kelt A, Wollenhaupt M, Assion A, Horn C, Sarpe-Tudoran C, Winter M and
  Baumert T 2003 {\em Rev. Sci. Instrum.\/} {\bf 74} 4950--4953

\bibitem{FrumkerTal2005}
Frumker E, Tal E, Silberberg Y and Majer D 2005 {\em Opt. Lett.\/} {\bf 30}
  2796--2798

\bibitem{JiangWeiner2007a}
Jiang Z, Huang C~B, Leaird D~E and Weiner A~M 2007 {\em Nature Photon.\/} {\bf
  1} 463--467

\bibitem{JiangWeiner2007b}
Jiang Z, Huang C~B, Leaird D~E and Weiner A~M 2007 {\em J. Opt. Soc. Am.\/} B
  {\bf 24} 2124--2128

\bibitem{BrixnerGerber2001}
Brixner T and Gerber G 2001 {\em Opt. Lett.\/} {\bf 26} 557--559

\bibitem{BrixnerKrampert2002APB}
Brixner T, Krampert G, Niklaus P and Gerber G 2002 {\em Appl. Phys.\/} B {\bf
  74} S133--S144

\bibitem{PolachekOron2006}
Polachek L, Oron D and Silberberg Y 2006 {\em Opt. Lett.\/} {\bf 31} 631--633

\bibitem{PlewickiWeise2006}
Plewicki M, Weise F, Weber S~M and Lindinger A 2006 {\em Appl. Opt.\/} {\bf 45}
  8354--8359

\bibitem{NinckGaller2007}
Ninck M, Galler A, Feurer T and Brixner T 2007 {\em Opt. Lett.\/} {\bf 32}
  3379--3381

\bibitem{MasihzadehSchlup2007}
Masihzadeh O, Schlup P and Bartels R~A 2007 {\em Opt. Expr.\/} {\bf 15}
  18025--18032

\bibitem{PlewickiWeber2007}
Plewicki M, Weber S~M, Weise F and Lindinger A 2007 {\em Appl. Phys.\/} B {\bf
  86} 259--263

\bibitem{WeiseLindinger2009}
Weise F and Lindinger A 2009 {\em Opt. Lett.\/} {\bf 34} 1258--1260

\bibitem{KupkaSchlup2009}
Kupka D, Schlup P and Bartels R~A 2009 {\em Review of Scientific Instruments\/}
  {\bf 80} 053110

\bibitem{NuernbergerVogtSelle2007}
Nuernberger P, Vogt G, Selle R, Fechner S, Brixner T and Gerber G 2007 {\em
  Appl. Phys.\/} B {\bf 88} 519--526

\bibitem{ParkerNunn2009}
Parker D~S~N, Nunn A~D~G, Minns R~S and Fielding H~H 2009 {\em Appl. Phys.\/} B
  {\bf 94} 181--186

\bibitem{SelleNuernberger2008}
Selle R, Nuernberger P, Langhojer F, Dimler F, Fechner S, Gerber G and Brixner
  T 2008 {\em Opt. Lett.\/} {\bf 33} 803--805

\bibitem{NuernbergerSelle2009}
Nuernberger P, Selle R, Langhojer F, Dimler F, Fechner S, Gerber G and Brixner
  T 2009 {\em J. Opt. A: Pure Appl. Opt.\/} {\bf 11} 085202

\bibitem{BaumertBrixner1997}
Baumert T, Brixner T, Seyfried V, Strehle M and Gerber G 1997 {\em Appl.
  Phys.\/} B {\bf 65} 779--782

\bibitem{Yelin1997}
Yelin D, Meshulach D and Silberberg Y 1997 {\em Opt. Lett.\/} {\bf 22}
  1793--1795

\bibitem{BrixnerStrehleGerber1999}
Brixner T, Strehle M and Gerber G 1999 {\em Appl. Phys.\/} B {\bf 68} 281--284

\bibitem{Zeek1999}
Zeek E, Maginnis K, Backus S, Russek U, Murnane M, Mourou G, Kapteyn H and
  Vdovin G 1999 {\em Opt. Lett.\/} {\bf 24} 493--495

\bibitem{Zeek2000}
Zeek E, Bartels R, Murnane M~M, Kapteyn H~C, Backus S and Vdovin G 2000 {\em
  Opt. Lett.\/} {\bf 25} 587--589

\bibitem{ZeidlerHornung2000}
Zeidler D, Hornung T, Proch D and Motzkus M 2000 {\em Appl. Phys.\/} B {\bf 70}
  S125--S131

\bibitem{SiegnerHaimlKunde2002}
Siegner U, Haiml M, Kunde J and Keller U 2002 {\em Opt. Lett.\/} {\bf 27}
  315--317

\bibitem{Efimov1998}
Efimov A, Moores M~D, Beach N~M, Krause J~L and Reitze D~H 1998 {\em Opt.
  Lett.\/} {\bf 23} 1915--1917

\bibitem{Efimov2000}
Efimov A, Moores M~D, Mei B, Krause J~L, Siders C~W and Reitze D~H 2000 {\em
  Appl. Phys.\/} B {\bf 70} S133--S141

\bibitem{MeshulachYelinSilberberg1998}
Meshulach D, Yelin D and Silberberg Y 1998 {\em J. Opt. Soc. Am.\/} B {\bf 15}
  1615--1619

\bibitem{BrixnerOehrlein2000}
Brixner T, Oehrlein A, Strehle M and Gerber G 2000 {\em Appl. Phys.\/} B {\bf
  70} S119--S124

\bibitem{BrixnerDamrauerKrampert2003}
Brixner T, Damrauer N~H, Krampert G, Niklaus P and Gerber G 2003 {\em J. Opt.
  Soc. Am.\/} B {\bf 20} 878--881

\bibitem{Suzuki2004ApplOpt}
Suzuki T, Minemoto S and Sakai H 2004 {\em Appl. Opt.\/} {\bf 43} 6047--6050

\bibitem{Aeschlimann2007}
Aeschlimann M, Bauer M, Bayer D, Brixner T, Garc\'{i}a~de Abajo F~J, Pfeiffer
  W, Rohmer M, Spindler C and Steeb F 2007 {\em Nature\/} {\bf 446} 301--304

\bibitem{Bartels2000}
Bartels R, Backus S, Zeek E, Misoguti L, Vdovin G, Christov I~P, Murnane M~M
  and Kapteyn H~C 2000 {\em Nature\/} {\bf 406} 164--166

\bibitem{Bartels2001}
Bartels R, Backus S, Christov I, Kapteyn H and Murnane M 2001 {\em Chem.
  Phys.\/} {\bf 267} 277--289

\bibitem{Bartels2004}
Bartels R~A, Murnane M~M, Kapteyn H~C, Christov I and Rabitz H 2004 {\em Phys.
  Rev.\/} A {\bf 70} 043404

\bibitem{Reitze2004}
Reitze D~H, Kazamias S, Weihe F, Mullot G, Douillet D, Aug\'{e} F, Albert O,
  Ramanathan V, Chambaret J~P, Hulin D and Balcou P 2004 {\em Opt. Lett.\/}
  {\bf 29} 86--88

\bibitem{PfeiferKemmer2005}
Pfeifer T, Kemmer R, Spitzenpfeil R, Walter D, Winterfeldt C, Gerber G and
  Spielmann C 2005 {\em Opt. Lett.\/} {\bf 30} 1497--1499

\bibitem{WalterPfeifer2006}
Walter D, Pfeifer T, Winterfeldt C, Kemmer R, Spitzenpfeil R, Gerber G and
  Spielmann C 2006 {\em Opt. Expr.\/} {\bf 14} 3433--3442

\bibitem{SpitzenpfeilEyring2009}
Spitzenpfeil R, Eyring S, Kern C, Ott C, Lohbreier J, Henneberger J, Franke N,
  Jung S, Walter D, Weger M, Winterfeldt C, Pfeifer T and Spielmann C 2009 {\em
  Appl. Phys.\/} A {\bf 96} 69--81

\bibitem{PfeiferWalter2005}
Pfeifer T, Walter D, Winterfeldt C, Spielmann C and Gerber G 2005 {\em Appl.
  Phys.\/} B {\bf 80} 277--280

\bibitem{PfeiferSpitzenpfeil2007}
Pfeifer T, Spitzenpfeil R, Walter D, Winterfeldt C, Dimler F, Gerber G and
  Spielmann C 2007 {\em Opt. Expr.\/} {\bf 15} 3409--3416

\bibitem{MeshulachSilberberg1998}
Meshulach D and Silberberg Y 1998 {\em Nature\/} {\bf 396} 239--242

\bibitem{MeshulachSilberberg1999}
Meshulach D and Silberberg Y 1999 {\em Phys. Rev.\/} A {\bf 60} 1287--1292

\bibitem{HornungMeierZeidler2000}
Hornung T, Meier R, Zeidler D, Kompa K~L, Proch D and Motzkus M 2000 {\em Appl.
  Phys.\/} B {\bf 71} 277--284

\bibitem{Dudovich2001}
Dudovich N, Dayan B, Gallagher~Faeder S~M and Silberberg Y 2001 {\em Phys. Rev.
  Lett.\/} {\bf 86} 47--50

\bibitem{TralleroHerreroCohen2006}
Trallero-Herrero C, Cohen J~L and Weinacht T 2006 {\em Phys. Rev. Lett.\/} {\bf
  96} 063603

\bibitem{Papastathopoulos2005}
Papastathopoulos E, Strehle M and Gerber G 2005 {\em Chem. Phys. Lett.\/} {\bf
  408} 65--70

\bibitem{WollenhauptPrakelt2005JOptB}
Wollenhaupt M, Pr\"{a}kelt A, Sarpe-Tudoran C, Liese D and Baumert T 2005 {\em
  J. Opt. B: Quantum Semiclass. Opt.\/} {\bf 7} S270--S276

\bibitem{Weinacht1999a}
Weinacht T~C, Ahn J and Bucksbaum P~H 1999 {\em Nature\/} {\bf 397} 233--235

\bibitem{Leichtle1998}
Leichtle C, Schleich W~P, Averbukh I~S and Shapiro M 1998 {\em Phys. Rev.
  Lett.\/} {\bf 80} 1418--1421

\bibitem{Nahmias2005}
Nahmias O, Bismuth O, Shoshana O and Ruhman S 2005 {\em J. Phys. Chem.\/} A
  {\bf 109} 8246--8253

\bibitem{LeeJungSungHongNam2002}
Lee S~H, Jung K~H, Sung J~H, Hong K~H and Nam C~H 2002 {\em J. Chem. Phys.\/}
  {\bf 117} 9858--9861

\bibitem{Prokhorenko2005}
Prokhorenko V~I, Nagy A~M and Miller R~J~D 2005 {\em J. Chem. Phys.\/} {\bf
  122} 184502

\bibitem{ZhangSun2005}
Zhang S, Sun Z, Zhang X, Xu Y, Wang Z, Xu Z and Li R 2005 {\em Chem. Phys.
  Lett.\/} {\bf 415} 346--350

\bibitem{vanderWalleHerek2009}
van~der Walle P, Milder M~T~W, Kuipers L and Herek J~L 2009 {\em Proc. Natl.
  Acad. Sci.\/} {\bf 106} 7714--7717

\bibitem{BrixnerDamrauer2003}
Brixner T, Damrauer N~H, Kiefer B and Gerber G 2003 {\em J. Chem. Phys.\/} {\bf
  118} 3692--3701

\bibitem{MontgomeryMeglen2006}
Montgomery M~A, Meglen R~R and Damrauer N~H 2006 {\em J. Phys. Chem.\/} A {\bf
  110} 6391--6394

\bibitem{MontgomeryMeglen2007}
Montgomery M~A, Meglen R~R and Damrauer N~H 2007 {\em J. Phys. Chem.\/} A {\bf
  111} 5126--5129

\bibitem{MontgomeryDamrauer2007}
Montgomery M~A and Damrauer N~H 2007 {\em J. Phys. Chem.\/} A {\bf 111}
  1426--1433

\bibitem{KurodaKleiman2009}
Kuroda D~G, Singh C~P, Peng Z and Kleiman V~D 2009 {\em Science\/} {\bf 326}
  263--267

\bibitem{OtakeKanoWada2006}
Otake I, Kano S~S and Wada A 2006 {\em J. Chem. Phys.\/} {\bf 124} 014501

\bibitem{BonacinaWolf2007}
Bonacina L, Extermann J, Rondi A, Boutou V and Wolf J~P 2007 {\em Phys. Rev.\/}
  A {\bf 76} 023408

\bibitem{Okada2004}
Okada T, Otake I, Mizoguchi R, Onda K, Kano S~S and Wada A 2004 {\em J. Chem.
  Phys.\/} {\bf 121} 6386--6391

\bibitem{RothGuyonRoslund2009}
Roth M, Guyon L, Roslund J, Boutou V, Courvoisier F, Wolf J~P and Rabitz H 2009
  {\em Phys. Rev. Lett.\/} {\bf 102} 253001

\bibitem{BergtBrixner1999}
Bergt M, Brixner T, Kiefer B, Strehle M and Gerber G 1999 {\em J. Phys.
  Chem.\/} A {\bf 103} 10381--10387

\bibitem{BrixnerKiefer2001}
Brixner T, Kiefer B and Gerber G 2001 {\em Chem. Phys.\/} {\bf 267} 241--246

\bibitem{Damrauer2002}
Damrauer N~H, Dietl C, Krampert G, Lee S~H, Jung K~H and Gerber G 2002 {\em
  Europ. Phys. J.\/} D {\bf 20} 71--76

\bibitem{BergtBrixnerDietl2002}
Bergt M, Brixner T, Dietl C, Kiefer B and Gerber G 2002 {\em J. Organomet.
  Chem.\/} {\bf 661} 199--209

\bibitem{Levis2001}
Levis R~J, Menkir G~M and Rabitz H 2001 {\em Science\/} {\bf 292} 709--713

\bibitem{Daniel2001}
Daniel C, Full J, Gonz\'{a}lez L, Kaposta C, Krenz M, Lupulescu C, Manz J,
  Minemoto S, Oppel M, Rosendo-Francisco P, Vajda {\v{S}} and W\"{o}ste L 2001
  {\em Chem. Phys.\/} {\bf 267} 247--260

\bibitem{VajdaRosendo2001}
Vajda {\v{S}}, Rosendo-Francisco P, Kaposta C, Krenz M, Lupulescu C and
  W\"{o}ste L 2001 {\em Eur. Phys. J.\/} D {\bf 16} 161--164

\bibitem{Daniel2003}
Daniel C, Full J, Gonz\'{a}lez L, Lupulescu C, Manz J, Merli A, Vajda {\v{S}}
  and W\"{o}ste L 2003 {\em Science\/} {\bf 299} 536--539

\bibitem{Cardoza2005JCP}
Cardoza D, Baertschy M and Weinacht T 2005 {\em J. Chem. Phys.\/} {\bf 123}
  074315

\bibitem{Cardoza2005CPL}
Cardoza D, Baertschy M and Weinacht T 2005 {\em Chem. Phys. Lett.\/} {\bf 411}
  311--315

\bibitem{Cardoza2004}
Cardoza D, Langhojer F, Trallero-Herrero C, Monti O~L~A and Weinacht T 2004
  {\em Phys. Rev.\/} A {\bf 70} 053406

\bibitem{LanghojerCardoza2005}
Langhojer F, Cardoza D, Baertschy M and Weinacht T 2005 {\em J. Chem. Phys.\/}
  {\bf 122} 014102

\bibitem{Cardoza2005a}
Cardoza D, Trallero-Herrero C, Langhojer F, Rabitz H and Weinacht T 2005 {\em
  J. Chem. Phys.\/} {\bf 122} 124306

\bibitem{CardozaPearson2006}
Cardoza D, Pearson B~J, Baertschy M and Weinacht T 2006 {\em J. Photochem.
  Photobiol.\/} A {\bf 180} 277--281

\bibitem{VajdaBartelt2001}
Vajda {\v{S}}, Bartelt A, Kaposta E~C, Leisner T, Lupulescu C, Minemoto S,
  Rosendo-Francisco P and W\"{o}ste L 2001 {\em Chem. Phys.\/} {\bf 267}
  231--239

\bibitem{BarteltMinemoto2001EPJD}
Bartelt A, Minemoto S, Lupulescu C, Vajda {\v{S}} and W\"{o}ste L 2001 {\em
  Eur. Phys. J.\/} D {\bf 16} 127--131

\bibitem{LindingerLupulescu2003SAB}
Lindinger A, Lupulescu C, Bartelt A, Vajda {\v{S}} and W\"{o}ste L 2003 {\em
  Spectrochim. Acta B: At. Spectrosc.\/} {\bf 58} 1109--1124

\bibitem{BarteltLindinger2004PCCP}
Bartelt A, Lindinger A, Lupulescu C, Vajda {\v{S}} and W\"{o}ste L 2004 {\em
  Phys. Chem. Chem. Phys.\/} {\bf 6} 1679--1686

\bibitem{WellsBetsch2005}
Wells E, Betsch K~J, Conover C~W~S, DeWitt M~J, Pinkham D and Jones R~R 2005
  {\em Phys. Rev.\/} A {\bf 72} 063406

\bibitem{WellsMcKenna2010JPB}
Wells E, McKenna J, Sayler A~M, Jochim B, Gregerson N, Averin R, Zohrabi M,
  Carnes K~D and Ben-Itzhak I 2010 {\em J. Phys. B: At. Mol. Opt. Phys.\/} {\bf
  43} 015101

\bibitem{ChenWangHill2009}
Chen G~Y, Wang Z~W and Hill W~T 2009 {\em Phys. Rev.\/} A {\bf 79} 011401

\bibitem{LaarmannShchatsinin2007JCP}
Laarmann T, Shchatsinin I, Singh P, Zhavoronkov N, Gerhards M, Schulz C~P and
  Hertel I~V 2007 {\em J. Chem. Phys.\/} {\bf 127} 201101

\bibitem{LaarmannShchatsinin2008JPB}
Laarmann T, Shchatsinin I, Singh P, Zhavoronkov N, Schulz C~P and Hertel I~V
  2008 {\em J. Phys. B: At. Mol. Opt. Phys.\/} {\bf 41} 074005

\bibitem{Palliyaguru2008}
Palliyaguru L, Sloss J, Rabitz H and Levis R~J 2008 {\em J. Mod. Opt.\/} {\bf
  55} 177--185

\bibitem{LozovoyZhuDantus2008}
Lozovoy V~V, Zhu X, Gunaratne T~C, Harris D~A, Shane J~C and Dantus M 2008 {\em
  J. Phys. Chem.\/} A {\bf 112} 3789--3812

\bibitem{ZhuGunaratneLozovoy2009}
Zhu X, Gunaratne T~C, Lozovoy V~V and Dantus M 2009 {\em J. Phys. Chem.\/} A
  {\bf 113} 5264--5266

\bibitem{Levis2009}
Levis R~J 2009 {\em J. Phys. Chem.\/} A {\bf 113} 5267--5268

\bibitem{BrixnerKrampert2004}
Brixner T, Krampert G, Pfeifer T, Selle R, Gerber G, Wollenhaupt M, Graefe O,
  Horn C, Liese D and Baumert T 2004 {\em Phys. Rev. Lett.\/} {\bf 92} 208301

\bibitem{Suzuki2004PRL}
Suzuki T, Minemoto S, Kanai T and Sakai H 2004 {\em Phys. Rev. Lett.\/} {\bf
  92} 133005

\bibitem{Weber2008}
Weber S~M, Plewicki M, Weise F and Lindinger A 2008 {\em J. Chem. Phys.\/} {\bf
  128} 174306

\bibitem{LupulescuLindingerPlewicki2004CP}
Lupulescu C, Lindinger A, Plewicki M, Merli A, Weber S~M and W\"{o}ste L 2004
  {\em Chem. Phys.\/} {\bf 296} 63--69

\bibitem{WeberLindinger2004CP}
Weber S~M, Lindinger A, Plewicki M, Lupulescu C, Vetter F and W\"{o}ste L 2004
  {\em Chem. Phys.\/} {\bf 306} 287--293

\bibitem{SchaferBungMitric2004}
Sch\'{a}fer-Bung B, Mitri\'{c} R, Bona\v{c}i\'{c}-Kouteck\'{y} V, Bartelt A,
  Lupulescu C, Lindinger A, Vajda {\v{S}}, Weber S~M and W\"{o}ste L 2004 {\em
  J. Phys. Chem.\/} A {\bf 108} 4175--4179

\bibitem{LindingerWeber2005}
Lindinger A, Weber S~M, Lupulescu C, Vetter F, Plewicki M, Merli A, W\"{o}ste
  L, Bartelt A~F and Rabitz H 2005 {\em Phys. Rev.\/} A {\bf 71} 013419

\bibitem{BarteltFeurer2005}
Bartelt A~F, Feurer T and W\"{o}ste L 2005 {\em Chem. Phys.\/} {\bf 318}
  207--216

\bibitem{LindingerWeberMerli2006}
Lindinger A, Weber S~M, Merli A, Sauer F, Plewicki M and W\"{o}ste L 2006 {\em
  J. Photochem. Photobiol.\/} A {\bf 180} 256--261

\bibitem{BallardStauffer2002}
Ballard J~B, Stauffer H~U, Amitay Z and Leone S~R 2002 {\em J. Chem. Phys.\/}
  {\bf 116} 1350--1360

\bibitem{LindingerVetter2004CPL}
Lindinger A, Vetter F, Lupulescu C, Plewicki M, Weber S~M, Merli A and
  W\"{o}ste L 2004 {\em Chem. Phys. Lett.\/} {\bf 397} 123--127

\bibitem{Lindinger2004}
Lindinger A, Lupulescu C, Plewicki M, Vetter F, Merli A, Weber S~M and
  W\"{o}ste L 2004 {\em Phys. Rev. Lett.\/} {\bf 93} 033001

\bibitem{LindingerLupulescu2005}
Lindinger A, Lupulescu C, Vetter F, Plewicki M, Weber S~M, Merli A and
  W\"{o}ste L 2005 {\em J. Chem. Phys.\/} {\bf 122} 024312

\bibitem{SiedschlagShir2006OC}
Siedschlag C, Shir O~M, B\"{a}ck T and Vrakking M~J~J 2006 {\em Opt. Commun.\/}
  {\bf 264} 511--518

\bibitem{ShirBeltrani2008JPB}
Shir O~M, Beltrani V, B\"{a}ck T, Rabitz H and Vrakking M~J~J 2008 {\em J.
  Phys. B: At. Mol. Opt. Phys.\/} {\bf 41} 074021

\bibitem{RouzeeGijsbertsen2009NJP}
Rouz\'{e}e A, Gijsbertsen A, Ghafur O, Shir O~M, B\"{a}ck T, Stolte S and
  Vrakking M~J~J 2009 {\em New J. Phys.\/} {\bf 11} 105040

\bibitem{LeibscherAverbukhRabitz2003}
Leibscher M, Averbukh I~S and Rabitz H 2003 {\em Phys. Rev. Lett.\/} {\bf 90}
  213001

\bibitem{LeibscherAverbukhRabitz2004}
Leibscher M, Averbukh I~S and Rabitz H 2004 {\em Phys. Rev.\/} A {\bf 69}
  013402

\bibitem{StapelfeldtSeideman2003}
Stapelfeldt H and Seideman T 2003 {\em Rev. Mod. Phys.\/} {\bf 75} 543--557

\bibitem{BisgaardPoulsen2004}
Bisgaard C~Z, Poulsen M~D, P\'{e}ronne E, Viftrup S~S and Stapelfeldt H 2004
  {\em Phys. Rev. Lett.\/} {\bf 92} 173004

\bibitem{RenardHertz2004}
Renard M, Hertz E, Lavorel B and Faucher O 2004 {\em Phys. Rev.\/} A {\bf 69}
  043401

\bibitem{RenardHertz2005}
Renard M, Hertz E, Gu\'{e}rin S, Jauslin H~R, Lavorel B and Faucher O 2005 {\em
  Phys. Rev.\/} A {\bf 72} 025401

\bibitem{LeeVilleneuve2006}
Lee K~F, Villeneuve D~M, Corkum P~B, Stolow A and Underwood J~G 2006 {\em Phys.
  Rev. Lett.\/} {\bf 97} 173001

\bibitem{HornWollenhaupt2006}
Horn C, Wollenhaupt M, Krug M, Baumert T, de~Nalda R and {Ba\~{n}ares} L 2006
  {\em Phys. Rev.\/} A {\bf 73} 031401

\bibitem{deNaldaHorn2007}
de~Nalda R, Horn C, Wollenhaupt M, Krug M, {Ba\~{n}ares} L and Baumert T 2007
  {\em J. Raman Spectrosc.\/} {\bf 38} 543--550

\bibitem{PinkhamMooneyJones2007}
Pinkham D, Mooney K~E and Jones R~R 2007 {\em Phys. Rev.\/} A {\bf 75} 013422

\bibitem{HornungMeierMotzkus2000}
Hornung T, Meier R and Motzkus M 2000 {\em Chem. Phys. Lett.\/} {\bf 326}
  445--453

\bibitem{WeinachtBartels2001CPL}
Weinacht T~C, Bartels R, Backus S, Bucksbaum P~H, Pearson B, Geremia J~M,
  Rabitz H, Kapteyn H~C and Murnane M~M 2001 {\em Chem. Phys. Lett.\/} {\bf
  344} 333--338

\bibitem{BartelsWeinacht2002PRL}
Bartels R~A, Weinacht T~C, Leone S~R, Kapteyn H~C and Murnane M~M 2002 {\em
  Phys. Rev. Lett.\/} {\bf 88} 033001

\bibitem{Weinacht1999b}
Weinacht T~C, White J~L and Bucksbaum P~H 1999 {\em J. Phys. Chem.\/} A {\bf
  103} 10166--10168

\bibitem{Pearson2001}
Pearson B~J, White J~L, Weinacht T~C and Bucksbaum P~H 2001 {\em Phys. Rev.\/}
  A {\bf 63} 063412

\bibitem{WhitePearson2004}
White J~L, Pearson B~J and Bucksbaum P~H 2004 {\em J. Phys. B: At. Mol. Opt.
  Phys.\/} {\bf 37} L399--L405

\bibitem{PearsonBucksbaum2004PRL}
Pearson B~J and Bucksbaum P~H 2004 {\em Phys. Rev. Lett.\/} {\bf 92} 243003

\bibitem{SpannerBrumer2006a}
Spanner M and Brumer P 2006 {\em Phys. Rev.\/} A {\bf 73} 023809

\bibitem{SpannerBrumer2006b}
Spanner M and Brumer P 2006 {\em Phys. Rev.\/} A {\bf 73} 023810

\bibitem{ZeidlerFrey2002}
Zeidler D, Frey S, Wohlleben W, Motzkus M, Busch F, Chen T, Kiefer W and
  Materny A 2002 {\em J. Chem. Phys.\/} {\bf 116} 5231--5235

\bibitem{KonradiSingh2005}
Konradi J, Singh A~K and Materny A 2005 {\em Phys. Chem. Chem. Phys.\/} {\bf 7}
  3574--3579

\bibitem{KonradiScaria2007}
Konradi J, Scaria A, Namboodiri V and Materny A 2007 {\em J. Raman
  Spectrosc.\/} {\bf 38} 1006--1021

\bibitem{KonradiSingh2006JRS}
Konradi J, Singh A~K, Scaria A~V and Materny A 2006 {\em J. Raman Spectrosc.\/}
  {\bf 37} 697--704

\bibitem{KonradiSingh2006JPPA}
Konradi J, Singh A~K and Materny A 2006 {\em J. Photochem. Photobiol.\/} A {\bf
  180} 289--299

\bibitem{ScariaKonradi2008}
Scaria A, Konradi J, Namboodiri V and Materny A 2008 {\em J. Raman
  Spectrosc.\/} {\bf 39} 739--749

\bibitem{ZhangZhang2007}
Zhang S, Zhang L, Zhang X, Ding L, Chen G, Sun Z and Wang Z 2007 {\em Chem.
  Phys. Lett.\/} {\bf 433} 416--421

\bibitem{VacanoWohllebenMotzkus2006}
von Vacano B, Wohlleben W and Motzkus M 2006 {\em Opt. Lett.\/} {\bf 31}
  413--415

\bibitem{StrasfeldShim2007}
Strasfeld D~B, Shim S~H and Zanni M~T 2007 {\em Phys. Rev. Lett.\/} {\bf 99}
  038102

\bibitem{StrasfeldMiddleton2009}
Strasfeld D~B, Middleton C~T and Zanni M~T 2009 {\em New J. Phys.\/} {\bf 11}
  105046

\bibitem{Kawano2003}
Kawano H, Nabekawa Y, Suda A, Oishi Y, Mizuno H, Miyawaki A and Midorikawa K
  2003 {\em Biochem. Biophys. Res. Commun.\/} {\bf 311} 592--596

\bibitem{ChenKawano2004}
Chen J, Kawano H, Nabekawa Y, Mizuno H, Miyawaki A, Tanabe T, Kannari F and
  Midorikawa K 2004 {\em Opt. Expr.\/} {\bf 12} 3408--3414

\bibitem{TadaKono2007}
Tada J, Kono T, Suda A, Mizuno H, Miyawaki A, Midorikawa K and Kannari F 2007
  {\em Appl. Opt.\/} {\bf 46} 3023--3030

\bibitem{IsobeSudaTanaka2009a}
Isobe K, Suda A, Tanaka M, Kannari F, Kawano H, Mizuno H, Miyawaki A and
  Midorikawa K 2009 {\em Opt. Expr.\/} {\bf 17} 13737--13746

\bibitem{LiTuriniciRamakhrishna2002}
Li B~Q, Turinici G, Ramakrishna V and Rabitz H 2002 {\em J. Phys. Chem.\/} B
  {\bf 106} 8125--8131

\bibitem{TuriniciRamakhrishnaLi2004}
Turinici G, Ramakhrishna V, Li B~Q and Rabitz H 2004 {\em J. Phys. A: Math.
  Gen.\/} {\bf 37} 273--282

\bibitem{LiRabitzWolf2005}
Li B, Rabitz H and Wolf J~P 2005 {\em J. Chem. Phys.\/} {\bf 122} 154103

\bibitem{LiZhuRabitz2006}
Li B~Q, Zhu W~S and Rabitz H 2006 {\em J. Chem. Phys.\/} {\bf 124} 024101

\bibitem{BrixnerDamrauer2001}
Brixner T, Damrauer N~H, Niklaus P and Gerber G 2001 {\em Nature\/} {\bf 414}
  57--60

\bibitem{Tkaczyk2010JLumin}
Tkaczyk E~R, Tkaczyk A~H, Mauring K, Ye J~Y, Baker J~R and Norris T~B 2010 {\em
  J. Lumin.\/} {\bf 130} 29--34

\bibitem{Herek2002}
Herek J~L, Wohlleben W, Cogdell R~J, Zeidler D and Motzkus M 2002 {\em
  Nature\/} {\bf 417} 533--535

\bibitem{WohllebenBuckup2003}
Wohlleben W, Buckup T, Herek J~L, Cogdell R~J and Motzkus M 2003 {\em Biophys.
  J.\/} {\bf 85} 442--450

\bibitem{BuckupLebold2006}
Buckup T, Lebold T, Weigel A, Wohlleben W and Motzkus M 2006 {\em J. Photochem.
  Photobiol.\/} A {\bf 180} 314--321

\bibitem{BruggemannYartsev2006}
Br\"{u}ggemann B, Organero J~A, Pascher T, Pullerits T and Yartsev A 2006 {\em
  Phys. Rev. Lett.\/} {\bf 97} 208301

\bibitem{VogtKrampert2005}
Vogt G, Krampert G, Niklaus P, Nuernberger P and Gerber G 2005 {\em Phys. Rev.
  Lett.\/} {\bf 94} 068305

\bibitem{HokiBrumer2005}
Hoki K and Brumer P 2005 {\em Phys. Rev. Lett.\/} {\bf 95} 168305

\bibitem{HuntRobb2005}
Hunt P~A and Robb M~A 2005 {\em J. Am. Chem. Soc.\/} {\bf 127} 5720--5726

\bibitem{ImprotaSantoro2005}
Improta R and Santoro F 2005 {\em J. Chem. Theory Comput.\/} {\bf 1} 215--229

\bibitem{DietzekYartsev2006}
Dietzek B, Br\"{u}ggemann B, Pascher T and Yartsev A 2006 {\em Phys. Rev.
  Lett.\/} {\bf 97} 258301

\bibitem{DietzekYartsev2007}
Dietzek B, Br\"{u}ggemann B, Pascher T and Yartsev A 2007 {\em J. Am. Chem.
  Soc.\/} {\bf 129} 13014--13021

\bibitem{ProkhorenkoNagy2006}
Prokhorenko V~I, Nagy A~M, Waschuk S~A, Brown L~S, Birge R~R and Miller R~J~D
  2006 {\em Science\/} {\bf 313} 1257--1261

\bibitem{Prokhorenko2007}
Prokhorenko V~I, Nagy A~M, Brown L~S and Miller R~J~D 2007 {\em Chem. Phys.\/}
  {\bf 341} 296--309

\bibitem{VogtNuernberger2006}
Vogt G, Nuernberger P, Brixner T and Gerber G 2006 {\em Chem. Phys. Lett.\/}
  {\bf 433} 211--215

\bibitem{FloreanCardoza2009}
Florean A~C, Cardoza D, White J~L, Lanyi J~K, Sension R~J and Bucksbaum P~H
  2009 {\em Proc. Natl. Acad. Sci.\/} {\bf 106} 10896--10900

\bibitem{HokiBrumer2009CPL}
Hoki K and Brumer P 2009 {\em Chem. Phys. Lett.\/} {\bf 468} 23--27

\bibitem{KatzRatnerKosloff2010NJP}
Katz G, Ratner M~A and Kosloff R 2010 {\em New J. Phys.\/} {\bf 12} 015003

\bibitem{ShuangRabitz2006JCP}
Shuang F and Rabitz H 2006 {\em J. Chem. Phys.\/} {\bf 124} 154105

\bibitem{CarrollPearson2006}
Carroll E~C, Pearson B~J, Florean A~C, Bucksbaum P~H and Sension R~J 2006 {\em
  J. Chem. Phys.\/} {\bf 124} 114506

\bibitem{CarrollWhite2008}
Carroll E~C, White J~L, Florean A~C, Bucksbaum P~H and Sension R~J 2008 {\em J.
  Phys. Chem.\/} A {\bf 112} 6811--6822

\bibitem{KoturWeinacht2009}
Kotur M, Weinacht T, Pearson B~J and Matsika S 2009 {\em J. Chem. Phys.\/} {\bf
  130} 134311

\bibitem{GreenfieldMcGrane2009}
Greenfield M, McGrane S~D and Moore D~S 2009 {\em J. Phys. Chem.\/} A {\bf 113}
  2333--2339

\bibitem{Laarmann2007}
Laarmann T, Shchatsinin I, Stalmashonak A, Boyle M, Zhavoronkov N, Handt J,
  Schmidt R, Schulz C~P and Hertel I~V 2007 {\em Phys. Rev. Lett.\/} {\bf 98}
  058302

\bibitem{Kunde2000}
Kunde J, Baumann B, Arlt S, Morier-Genoud F, Siegner U and Keller U 2000 {\em
  Appl. Phys. Lett.\/} {\bf 77} 924--926

\bibitem{Kunde2001}
Kunde J, Baumann B, Arlt S, Morier-Genoud F, Siegner U and Keller U 2001 {\em
  J. Opt. Soc. Am.\/} B {\bf 18} 872--881

\bibitem{ChungWeiner2006}
Chung J~H and Weiner A 2006 {\em IEEE J. Select. Top. Quantum Electron.\/} {\bf
  12} 297--306

\bibitem{Brif2001}
Brif C, Rabitz H, Wallentowitz S and Walmsley I~A 2001 {\em Phys. Rev.\/} A
  {\bf 63} 063404

\bibitem{Haeberlen1976}
Haeberlen U 1976 {\em High Resolution NMR in Solids\/} (New York: Academic
  Press)

\bibitem{ViolaLloyd1998}
Viola L and Lloyd S 1998 {\em Phys. Rev.\/} A {\bf 58} 2733--2744

\bibitem{ViolaKnillLloyd1999}
Viola L, Knill E and Lloyd S 1999 {\em Phys. Rev. Lett.\/} {\bf 82} 2417--2421

\bibitem{Zanardi1999}
Zanardi P 1999 {\em Phys. Lett.\/} A {\bf 258} 77--82

\bibitem{VitaliTombesi1999}
Vitali D and Tombesi P 1999 {\em Phys. Rev.\/} A {\bf 59} 4178--4186

\bibitem{VitaliTombesi2001}
Vitali D and Tombesi P 2001 {\em Phys. Rev.\/} A {\bf 65} 012305

\bibitem{ByrdLidar2003}
Byrd M~S and Lidar D~A 2003 {\em Phys. Rev.\/} A {\bf 67} 012324

\bibitem{KhodjastehLidar2005}
Khodjasteh K and Lidar D~A 2005 {\em Phys. Rev. Lett.\/} {\bf 95} 180501

\bibitem{FacchiTasaki2005}
Facchi P, Tasaki S, Pascazio S, Nakazato H, Tokuse A and Lidar D~A 2005 {\em
  Phys. Rev.\/} A {\bf 71} 022302

\bibitem{ViolaKnill2005}
Viola L and Knill E 2005 {\em Phys. Rev. Lett.\/} {\bf 94} 060502

\bibitem{Uhrig2007}
Uhrig G~S 2007 {\em Phys. Rev. Lett.\/} {\bf 98} 100504

\bibitem{PasiniUhrig2010}
Pasini S and Uhrig G~S 2010 {\em Phys. Rev.\/} A {\bf 81} 012309

\bibitem{UhrigPasini2010NJP}
Uhrig G~S and Pasini S 2010 {\em New J. Phys.\/} {\bf 12} 045001

\bibitem{FravalSellars2005}
Fraval E, Sellars M~J and Longdell J~J 2005 {\em Phys. Rev. Lett.\/} {\bf 95}
  030506

\bibitem{MortonTyryshkin2006}
Morton J~J~L, Tyryshkin A~M, Ardavan A, Benjamin S~C, Porfyrakis K, Lyon S~A
  and Briggs G~A~D 2006 {\em Nature Phys.\/} {\bf 2} 40--43

\bibitem{MortonTyryshkin2008}
Morton J~J~L, Tyryshkin A~M, Brown R~M, Shankar S, Lovett B~W, Ardavan A,
  Schenkel T, Haller E~E, Ager J~W and Lyon S~A 2008 {\em Nature\/} {\bf 455}
  1085--1088

\bibitem{DamodarakurupLucamarini2009}
Damodarakurup S, Lucamarini M, Di~Giuseppe G, Vitali D and Tombesi P 2009 {\em
  Phys. Rev. Lett.\/} {\bf 103} 040502

\bibitem{SagiAlmogDavidson2009}
Sagi Y, Almog I and Davidson N 2009 Suppression of collisional decoherence
  arXiv:0905.0286

\bibitem{SagiAlmogDavidson2010}
Sagi Y, Almog I and Davidson N 2010 Process tomography of dynamical decoupling
  in a dense optically trapped atomic ensemble arXiv:1004.1011

\bibitem{BiercukBollinger2009}
Biercuk M~J, Uys H, VanDevender A~P, Shiga N, Itano W~M and Bollinger J~J 2009
  {\em Nature\/} {\bf 458} 996--1000

\bibitem{Schwefel1995}
Schwefel H~P 1995 {\em Evolution and Optimum Seeking\/} (New York: Wiley)

\bibitem{Goldberg2007}
Goldberg D~E 2007 {\em Genetic Algorithms in Search, Optimization, and Machine
  Learning\/} (Reading, MA: Addison-Wesley)

\bibitem{BarteltRoth2005}
Bartelt A~F, Roth M, Mehendale M and Rabitz H 2005 {\em Phys. Rev.\/} A {\bf
  71} 063806

\bibitem{ZeidlerFreyKompa2001}
Zeidler D, Frey S, Kompa K~L and Motzkus M 2001 {\em Phys. Rev.\/} A {\bf 64}
  023420

\bibitem{ShirSiedschlag2006}
Shir O~M, Siedschlag C, B\"{a}ck T and Vrakking M~J~J 2006 Niching in evolution
  strategies and its application to laser pulse shaping {\em Artificial
  Evolution\/} ({\em Lecture Notes in Computer Science\/} vol 3871) (Berlin:
  Springer) pp 85--96

\bibitem{FonsecaFleming1995}
Fonseca C~M and Fleming P~J 1995 {\em Evol. Comput.\/} {\bf 3} 1--16

\bibitem{Deb1999}
Deb K 1999 {\em Evolutionary Computation\/} {\bf 7} 205--230

\bibitem{Gollub2009NJP}
Gollub C and de~Vivie-Riedle R 2009 {\em New J. Phys.\/} {\bf 11} 013019

\bibitem{Kirkpatrick1983}
Kirkpatrick S, Gelatt C~D and Vecchi M~P 1983 {\em Science\/} {\bf 220}
  671--680

\bibitem{Dorigo1996}
Dorigo M, Maniezzo V and Colorni A 1996 {\em IEEE Trans. Syst. Man Cybern.\/} B
  {\bf 26} 29--41

\bibitem{Bonabeau2000}
Bonabeau E, Dorigo M and Theraulaz G 2000 {\em Nature\/} {\bf 406} 39--42

\bibitem{Feurer1999}
Feurer T 1999 {\em Appl. Phys.\/} B {\bf 68} 55--60

\bibitem{GlassRozgonyi2000}
Gla{\ss} A, Rozgonyi T, Feurer T, Sauerbrey R and Szab\'{o} G 2000 {\em Appl.
  Phys.\/} B {\bf 71} 267--276

\bibitem{RoslundShir2009}
Roslund J, Shir O~M, B\"{a}ck T and Rabitz H 2009 {\em Phys. Rev.\/} A {\bf 80}
  043415

\bibitem{DudovichOron2002Nature}
Dudovich N, Oron D and Silberberg Y 2002 {\em Nature\/} {\bf 418} 512--514

\bibitem{OronDudovichYelin2002PRL}
Oron D, Dudovich N, Yelin D and Silberberg Y 2002 {\em Phys. Rev. Lett.\/} {\bf
  88} 063004

\bibitem{OronDudovich2002PRL}
Oron D, Dudovich N and Silberberg Y 2002 {\em Phys. Rev. Lett.\/} {\bf 89}
  273001

\bibitem{OronDudovichYelin2002PRA}
Oron D, Dudovich N, Yelin D and Silberberg Y 2002 {\em Phys. Rev.\/} A {\bf 65}
  043408

\bibitem{DudovichOron2003JCP}
Dudovich N, Oron D and Silberberg Y 2003 {\em J. Chem. Phys.\/} {\bf 118}
  9208--9215

\bibitem{OronDudovich2003PRL}
Oron D, Dudovich N and Silberberg Y 2003 {\em Phys. Rev. Lett.\/} {\bf 90}
  213902

\bibitem{GershgorenBartels2003}
Gershgoren E, Bartels R~A, Fourkas J~T, Tobey R, Murnane M~M and Kapteyn H~C
  2003 {\em Opt. Lett.\/} {\bf 28} 361--363

\bibitem{PastirkDantus2003}
Pastirk I, Cruz J~D, Walowicz K, Lozovoy V and Dantus M 2003 {\em Opt. Expr.\/}
  {\bf 11} 1695--1701

\bibitem{LimCasterLeone2005}
Lim S~H, Caster A~G and Leone S~R 2005 {\em Phys. Rev.\/} A {\bf 72} 041803

\bibitem{OgilvieDebarre2006}
Ogilvie J~P, D\'{e}barre D, Solinas X, Martin J~L, Beaurepaire E and Joffre M
  2006 {\em Opt. Expr.\/} {\bf 14} 759--766

\bibitem{VacanoMotzkus2007}
von Vacano B and Motzkus M 2007 {\em J. Chem. Phys.\/} {\bf 127} 144514

\bibitem{PestovWang2008}
Pestov D, Wang X, Murawski R~K, Ariunbold G~O, Sautenkov V~A and Sokolov A~V
  2008 {\em J. Opt. Soc. Am.\/} B {\bf 25} 768--772

\bibitem{Postma2008}
Postma S, van Rhijn A~C~W, Korterik J~P, Gross P, Herek J~L and Offerhaus H~L
  2008 {\em Opt. Expr.\/} {\bf 16} 7985--7996

\bibitem{IsobeSudaTanaka2009b}
Isobe K, Suda A, Tanaka M, Hashimoto H, Kannari F, Kawano H, Mizuno H, Miyawaki
  A and Midorikawa K 2009 {\em Opt. Expr.\/} {\bf 17} 11259--11266

\bibitem{DudovichOron2004}
Dudovich N, Oron D and Silberberg Y 2004 {\em Phys. Rev. Lett.\/} {\bf 92}
  103003

\bibitem{WollenhauptKrug2009}
Wollenhaupt M, Krug M, K\"{o}hler J, Bayer T, Sarpe-Tudoran C and Baumert T
  2009 {\em Appl. Phys.\/} B {\bf 95} 245--259

\bibitem{PrakeltWollenhaupt2004}
Pr\"akelt A, Wollenhaupt M, Sarpe-Tudoran C and Baumert T 2004 {\em Phys.
  Rev.\/} A {\bf 70} 063407

\bibitem{BarrosLozano2005OL}
Barros H~G, B W~L, Vianna S~S and Acioli L~H 2005 {\em Opt. Lett.\/} {\bf 30}
  3081--3083

\bibitem{BarrosFerraz2006PRA}
Barros H~G, Ferraz J, B W~L, Acioli L~H and Vianna S~S 2006 {\em Phys. Rev.\/}
  A {\bf 74} 055402

\bibitem{WollenhauptPrakeltBaumert2006PRA}
Wollenhaupt M, Pr\"{a}kelt A, Sarpe-Tudoran C, Liese D, Bayer T and Baumert T
  2006 {\em Phys. Rev.\/} A {\bf 73} 063409

\bibitem{DudovichPolack2005}
Dudovich N, Polack T, Pe'er A and Silberberg Y 2005 {\em Phys. Rev. Lett.\/}
  {\bf 94} 083002

\bibitem{Amitay2008}
Amitay Z, Gandman A, Chuntonov L and Rybak L 2008 {\em Phys. Rev. Lett.\/} {\bf
  100} 193002

\bibitem{ZhdanovichShapiroHepburn2009}
Zhdanovich S, Shapiro E~A, Hepburn J~W, Shapiro M and Milner V 2009 {\em Phys.
  Rev.\/} A {\bf 80} 063405

\bibitem{ViftrupKumarappan2009}
Viftrup S~S, Kumarappan V, Holmegaard L, Bisgaard C~Z, Stapelfeldt H, Artamonov
  M, Hamilton E and Seideman T 2009 {\em Phys. Rev.\/} A {\bf 79} 023404

\bibitem{IbrahimSchwentner2009}
Ibrahim H, H\'{e}jjas M, Fushitani M and Schwentner N 2009 {\em J. Phys.
  Chem.\/} A {\bf 113} 7439--7450

\bibitem{NakamuraPashkinTsai1999}
Nakamura Y, Pashkin Y~A and Tsai J~S 1999 {\em Nature\/} {\bf 398} 786--788

\bibitem{FeurerVaughanNelson2003}
Feurer T, Vaughan J~C and Nelson K~A 2003 {\em Science\/} {\bf 299} 374--377

\bibitem{FanciulliWeiner2005PRB}
Fanciulli R, Weiner A~M, Dignam M~M, Meinhold D and Leo K 2005 {\em Phys.
  Rev.\/} B {\bf 71} 153304

\bibitem{GolanFradkin2009}
Golan B, Fradkin Z, Kopnov G, Oron D and Naaman R 2009 {\em J. Chem. Phys.\/}
  {\bf 130} 064705

\bibitem{Lloyd2000}
Lloyd S 2000 {\em Phys. Rev.\/} A {\bf 62} 022108

\bibitem{Belavkin1999}
Belavkin V~P 1999 {\em Rep. Math. Phys.\/} {\bf 43} A405--A425

\bibitem{BoutenHandelJames2007}
Bouten L, van Handel R and James M~R 2007 {\em SIAM J. Control Optim.\/} {\bf
  46} 2199--2241

\bibitem{BoutenHandelJames2009}
Bouten L, van Handel R and James M~R 2009 {\em SIAM Rev.\/} {\bf 51} 239--316

\bibitem{NelsonWeinsteinCory2000}
Nelson R~J, Weinstein Y, Cory D and Lloyd S 2000 {\em Phys. Rev. Lett.\/} {\bf
  85} 3045--3048

\bibitem{YanagisawaKimura2003I}
Yanagisawa M and Kimura H 2003 {\em IEEE Trans. Autom. Control\/} {\bf 48}
  2107--2120

\bibitem{YanagisawaKimura2003II}
Yanagisawa M and Kimura H 2003 {\em IEEE Trans. Autom. Control\/} {\bf 48}
  2121--2132

\bibitem{WisemanMilburn1994}
Wiseman H~M and Milburn G~J 1994 {\em Phys. Rev.\/} A {\bf 49} 4110--4125

\bibitem{DHelonJames2006}
D'Helon C and James M~R 2006 {\em Phys. Rev.\/} A {\bf 73} 053803

\bibitem{JamesNurdinPetersen2008TAC}
James M~R, Nurdin H~I and Petersen I~R 2008 {\em IEEE Trans. Autom. Control\/}
  {\bf 53} 1787--1803

\bibitem{Mabuchi2008PRA}
Mabuchi H 2008 {\em Phys. Rev.\/} A {\bf 78} 032323

\bibitem{KallushKosloff2006}
Kallush S and Kosloff R 2006 {\em Phys. Rev.\/} A {\bf 73} 032324

\bibitem{PechenBrif2010}
Pechen A, Brif C, Wu R~B, Chakrabarti R and Rabitz H 2010 General unifying
  features of controlled quantum phenomena arXiv:1003.3506

\bibitem{BushevRotterWilson2006}
Bushev P, Rotter D, Wilson A, Dubin F, Becher C, Eschner J, Blatt R, Steixner
  V, Rabl P and Zoller P 2006 {\em Phys. Rev. Lett.\/} {\bf 96} 043003

\bibitem{BerglundMcHaleMabuchi2007}
Berglund A~J, McHale K and Mabuchi H 2007 {\em Opt. Lett.\/} {\bf 32} 145--147

\bibitem{GillettDaltonLanyon2010}
Gillett G~G, Dalton R~B, Lanyon B~P, Almeida M~P, Barbieri M, Pryde G~J,
  O'Brien J~L, Resch K~J, Bartlett S~D and White A~G 2010 {\em Phys. Rev.
  Lett.\/} {\bf 104} 080503

\bibitem{DohertyJacobs2001}
Doherty A~C, Jacobs K and Jungman G 2001 {\em Phys. Rev.\/} A {\bf 63} 062306

\bibitem{ThomsenManciniWiseman2002PRA}
Thomsen L~K, Mancini S and Wiseman H~M 2002 {\em Phys. Rev.\/} A {\bf 65}
  061801

\bibitem{ThomsenManciniWiseman2002JPB}
Thomsen L~K, Mancini S and Wiseman H~M 2002 {\em J. Phys. B: At. Mol. Opt.
  Phys.\/} {\bf 35} 4937--4952

\bibitem{StocktonGeremiaDoherty2004PRA}
Stockton J~K, Geremia J~M, Doherty A~C and Mabuchi H 2004 {\em Phys. Rev.\/} A
  {\bf 69} 032109

\bibitem{BerglundMabuchi2004APB}
Berglund A~J and Mabuchi H 2004 {\em Appl. Phys.\/} B {\bf 78} 653--659

\bibitem{MabuchiKhaneja2005}
Mabuchi H and Khaneja N 2005 {\em Int. J. Robust Nonlinear Control\/} {\bf 15}
  647--667

\bibitem{WisemanDoherty2005}
Wiseman H~M and Doherty A~C 2005 {\em Phys. Rev. Lett.\/} {\bf 94} 070405

\bibitem{GoughBelavkinSmolyanov2005}
Gough J, Belavkin V~P and Smolyanov O~G 2005 {\em J. Opt. B: Quantum Semiclass.
  Opt.\/} {\bf 7} S237--S244

\bibitem{BelavkinNegrettiMolmer2009}
Belavkin V~P, Negretti A and M{\o}lmer K 2009 {\em Phys. Rev.\/} A {\bf 79}
  022123

\bibitem{JacobsShabani2008}
Jacobs K and Shabani A 2008 {\em Contemp. Phys.\/} {\bf 49} 435--448

\bibitem{NielsenHopkinsMabuchi2009}
Nielsen A~E~B, Hopkins A~S and Mabuchi H 2009 {\em New J. Phys.\/} {\bf 11}
  105043

\bibitem{Jacobs2009}
Jacobs K 2009 Feedback control using only quantum back-action arXiv:0904.3745

\bibitem{MirrahimiHandel2007}
Mirrahimi M and van Handel R 2007 {\em SIAM J. Control Optim.\/} {\bf 46}
  445--467

\bibitem{JacobsLund2007}
Jacobs K and Lund A~P 2007 {\em Phys. Rev. Lett.\/} {\bf 99} 020501

\bibitem{YamamotoTsumuraHara2007}
Yamamoto N, Tsumura K and Hara S 2007 {\em Automatica\/} {\bf 43} 981--992

\bibitem{DotsenkoMirrahimi2009}
Dotsenko I, Mirrahimi M, Brune M, Haroche S, Raimond J~M and Rouchon P 2009
  {\em Phys. Rev.\/} A {\bf 80} 013805

\bibitem{DongPetersen2009NJP}
Dong D and Petersen I~R 2009 {\em New J. Phys.\/} {\bf 11} 105033

\bibitem{AhnDohertyLandahl2002}
Ahn C, Doherty A~C and Landahl A~J 2002 {\em Phys. Rev.\/} A {\bf 65} 042301

\bibitem{AhnWisemanMilburn2003}
Ahn C, Wiseman H~M and Milburn G~J 2003 {\em Phys. Rev.\/} A {\bf 67} 052310

\bibitem{AhnWisemanJacobs2004}
Ahn C, Wiseman H and Jacobs K 2004 {\em Phys. Rev.\/} A {\bf 70} 024302

\bibitem{SarovarAhnJacobs2004}
Sarovar M, Ahn C, Jacobs K and Milburn G~J 2004 {\em Phys. Rev.\/} A {\bf 69}
  052324

\bibitem{ChaseLandahlGeremia2008}
Chase B~A, Landahl A~J and Geremia J~M 2008 {\em Phys. Rev.\/} A {\bf 77}
  032304

\bibitem{KerckhoffNurdin2009}
Kerckhoff J, Nurdin H~I, Pavlichin D~S and Mabuchi H 2010 Designing quantum
  memories with embedded control: photonic circuits for autonomous quantum
  error correction arXiv:0907.0236

\bibitem{StocktonHandelMabuchi2004}
Stockton J~K, van Handel R and Mabuchi H 2004 {\em Phys. Rev.\/} A {\bf 70}
  022106

\bibitem{HandelStocktonMabuchi2005JOB}
van Handel R, Stockton J~K and Mabuchi H 2005 {\em J. Opt. B: Quantum
  Semiclass. Opt.\/} {\bf 7} S179--S197

\bibitem{Mancini2006}
Mancini S 2006 {\em Phys. Rev.\/} A {\bf 73} 010304

\bibitem{ManciniWiseman2007}
Mancini S and Wiseman H~M 2007 {\em Phys. Rev.\/} A {\bf 75} 012330

\bibitem{YamamotoNurdinJames2008}
Yamamoto N, Nurdin H~I, James M~R and Petersen I~R 2008 {\em Phys. Rev.\/} A
  {\bf 78} 042339

\bibitem{Nielsen2010PRA}
Nielsen A~E~B 2010 {\em Phys. Rev.\/} A {\bf 81} 012307

\bibitem{HopkinsJacobs2003PRB}
Hopkins A, Jacobs K, Habib S and Schwab K 2003 {\em Phys. Rev.\/} B {\bf 68}
  235328

\bibitem{SteckJacobsMabuchi2004}
Steck D~A, Jacobs K, Mabuchi H, Bhattacharya T and Habib S 2004 {\em Phys. Rev.
  Lett.\/} {\bf 92} 223004

\bibitem{SteckJacobsMabuchi2006}
Steck D~A, Jacobs K, Mabuchi H, Habib S and Bhattacharya T 2006 {\em Phys.
  Rev.\/} A {\bf 74} 012322

\bibitem{WilsonCarvalhoHope2007}
Wilson S~D, Carvalho A~R~R, Hope J~J and James M~R 2007 {\em Phys. Rev.\/} A
  {\bf 76} 013610

\bibitem{Jacobs2003PRA}
Jacobs K 2003 {\em Phys. Rev.\/} A {\bf 67} 030301

\bibitem{CombesJacobs2006}
Combes J and Jacobs K 2006 {\em Phys. Rev. Lett.\/} {\bf 96} 010504

\bibitem{HandelStocktonMabuchi2005}
van Handel R, Stockton J~K and Mabuchi H 2005 {\em IEEE Trans. Autom.
  Control\/} {\bf 50} 768--780

\bibitem{WisemanRalph2006NJP}
Wiseman H~M and Ralph J~F 2006 {\em New J. Phys.\/} {\bf 8} 90

\bibitem{WisemanBouten2008}
Wiseman H~M and Bouten L 2008 {\em Quant. Inf. Process.\/} {\bf 7} 71--83

\bibitem{ChiruvelliJacobs2008}
Chiruvelli A and Jacobs K 2008 {\em Phys. Rev.\/} A {\bf 77} 012102

\bibitem{ShabaniJacobs2008PRL}
Shabani A and Jacobs K 2008 {\em Phys. Rev. Lett.\/} {\bf 101} 230403

\bibitem{ArmenAuStockton2002PRL}
Armen M~A, Au J~K, Stockton J~K, Doherty A~C and Mabuchi H 2002 {\em Phys. Rev.
  Lett.\/} {\bf 89} 133602

\bibitem{CookMartinGeremia2007N}
Cook R~L, Martin P~J and Geremia J~M 2007 {\em Nature\/} {\bf 446} 774--777

\bibitem{ShorterIpRabitz1999}
Shorter J~A, Ip P~C and Rabitz H~A 1999 {\em J. Phys. Chem.\/} A {\bf 103}
  7192--7198

\bibitem{RabitzAlis1999}
Rabitz H and Al\i\c{s} {\"{O}}~F 1999 {\em J. Math. Chem.\/} {\bf 25} 197--233

\bibitem{RabitzAlisShorter1999}
Rabitz H, Al\i\c{s} {\"{O}}~F, Shorter J and Shim K 1999 {\em Comp. Phys.
  Commun.\/} {\bf 117} 11--20

\bibitem{AlisRabitz2001}
Al\i\c{s} {\"{O}}~F and Rabitz H 2001 {\em J. Math. Chem.\/} {\bf 29} 127--142

\bibitem{LiRosenthalRabitz2001}
Li G, Rosenthal C and Rabitz H 2001 {\em J. Phys. Chem.\/} A {\bf 105}
  7765--7777

\bibitem{GeremiaWeissRabitz2001}
Geremia J~M, Weiss E and Rabitz H 2001 {\em Chem. Phys.\/} {\bf 267} 209--222

\bibitem{HertzKroghPalmer1991}
Hertz J, Krogh A and Palmer R~G 1991 {\em Introduction to the Theory of Neural
  Computation\/} (New York: Westview Press)

\bibitem{FreemanSkapura1991}
Freeman J~A and Skapura D~M 1991 {\em Neural Networks: Algorithms,
  Applications, and Programming Techniques\/} (Redwood City: Addison-Wesley)

\bibitem{SelleVogt2007}
Selle R, Vogt G, Brixner T, Gerber G, Metzler R and Kinzel W 2007 {\em Phys.
  Rev.\/} A {\bf 76} 023810

\bibitem{SelleBrixner2008}
Selle R, Brixner T, Bayer T, Wollenhaupt M and Baumert T 2008 {\em J. Phys. B:
  At. Mol. Opt. Phys.\/} {\bf 41} 074019

\bibitem{McGraneScharff2009}
McGrane S~D, Scharff R~J, Greenfield M and Moore D~S 2009 {\em New J. Phys.\/}
  {\bf 11} 105047

\bibitem{RossiKuhn2002RMP}
Rossi F and Kuhn T 2002 {\em Rev. Mod. Phys.\/} {\bf 74} 895--950

\bibitem{Hanson2007RMP}
Hanson R, Kouwenhoven L~P, Petta J~R, Tarucha S and Vandersypen L~M~K 2007 {\em
  Rev. Mod. Phys.\/} {\bf 79} 1217

\bibitem{AppelFigueroaLvovsky2008}
Appel J, Figueroa E, Korystov D, Lobino M and Lvovsky A~I 2008 {\em Phys. Rev.
  Lett.\/} {\bf 100} 093602

\bibitem{ChoiDengLauratKimble2008}
Choi K~S, Deng H, Laurat J and Kimble H~J 2008 {\em Nature\/} {\bf 452} 67--71

\bibitem{StaudtGisin2007}
Staudt M~U, Hastings-Simon S~R, Nilsson M, Afzelius M, Scarani V, Ricken R,
  Suche H, Sohler W, Tittel W and Gisin N 2007 {\em Phys. Rev. Lett.\/} {\bf
  98} 113601

\bibitem{ReimNunnLorenz2009}
Reim K~F, Nunn J, Lorenz V~O, Sussman B~J, Lee K, Langford N~K, Jaksch D and
  Walmsley I~A 2009 Towards high-speed optical quantum memories arXiv:0912.2970

\bibitem{LvovskySandersTittel2009NP}
Lvovsky A~I, Sanders B~C and Tittel W 2009 {\em Nature Photon.\/} {\bf 3}
  706--714

\bibitem{SimonAfzeliusAppel2010}
Simon C, Afzelius M, Appel J, de~la Giroday A~B, Dewhurst S~J, Gisin N, Hu C~Y,
  Jelezko F, Kroll S, Muller J~H, Nunn J, Polzik E, Rarity J, de~Riedmatten H,
  Rosenfeld W, Shields A~J, Skold N, Stevenson R~M, Thew R, Walmsley I, Weber
  M, Weinfurter H, Wrachtrup J and Young R~J 2010 {Quantum memories. A review
  based on the European integrated project ``Qubit applications (QAP)''}
  arXiv:1003.1107

\bibitem{SolasAshton2009}
Solas F, Ashton J~M, Markmann A and Rabitz H~A 2009 {\em J. Chem. Phys.\/} {\bf
  130} 214702

\bibitem{MakhlinSchonShnirman2001RMP}
Makhlin Y, Sch\"{o}n G and Shnirman A 2001 {\em Rev. Mod. Phys.\/} {\bf 73}
  357--400

\bibitem{ClarkeWilhelm2008N}
Clarke J and Wilhelm F~K 2008 {\em Nature\/} {\bf 453} 1031--1042

\bibitem{Walmsley2009privatecommun}
Walmsley I~A 2009 Private communication

\bibitem{LiuKohlerKeitel2009}
Liu C, Kohler M~C, Hatsagortsyan K~Z, Muller C and Keitel C~H 2009 {\em New J.
  Phys.\/} {\bf 11} 105045

\bibitem{BeltraniDominyHoRabitz2007}
Beltrani V, Dominy J, Ho T~S and Rabitz H 2007 {\em J. Chem. Phys.\/} {\bf 126}
  094105

\bibitem{MitraRabitz2003}
Mitra A and Rabitz H 2003 {\em Phys. Rev.\/} A {\bf 67} 033407

\bibitem{MitraSolaRabitz2003}
Mitra A, Sol\'{a} I~R and Rabitz H 2003 {\em Phys. Rev.\/} A {\bf 67} 043409

\bibitem{MitraRabitz2004}
Mitra A and Rabitz H 2004 {\em J. Phys. Chem.\/} A {\bf 108} 4778--4785

\bibitem{SharpRabitz2004}
Sharp R~W and Rabitz H 2004 {\em J. Chem. Phys.\/} {\bf 121} 4516--4527

\bibitem{MitraRabitz2006}
Mitra A and Rabitz H 2006 {\em J. Chem. Phys.\/} {\bf 125} 194107

\bibitem{MitraRabitz2008}
Mitra A and Rabitz H 2008 {\em J. Chem. Phys.\/} {\bf 128} 044112

\bibitem{MitraSolaRabitz2008}
Mitra A, Sola I~R and Rabitz H 2008 {\em Phys. Rev.\/} A {\bf 77} 043415

\bibitem{SharpMitraRabitz2008}
Sharp R, Mitra A and Rabitz H 2008 {\em J. Math. Chem.\/} {\bf 44} 142--171

\bibitem{ReydeCastroRabitz2010}
{Rey-de-Castro} R and Rabitz H 2010 Laboratory implementation of quantum
  control mechanism identification through {Hamiltonian}-encoding and
  observable-decoding (to be published in Phys. Rev. A)

\bibitem{EngelFleming2007Nat}
Engel G~S, Calhoun T~R, Read E~L, Ahn T~K, Man\v{c}al T, Cheng Y~C, Blankenship
  R~E and Fleming G~R 2007 {\em Nature\/} {\bf 446} 782--786

\bibitem{LeeChengFleming2007Sci}
Lee H, Cheng Y~C and Fleming G~R 2007 {\em Science\/} {\bf 316} 1462--1465

\bibitem{MercerElTahaKajumba2009PRL}
Mercer I~P, El-Taha Y~C, Kajumba N, Marangos J~P, Tisch J~W~G, Gabrielsen M,
  Cogdell R~J, Springate E and Turcu E 2009 {\em Phys. Rev. Lett.\/} {\bf 102}
  057402

\bibitem{ColliniWongWilk2010}
Collini E, Wong C~Y, Wilk K~E, Curmi P~M~G, Brumer P and Scholes G~D 2010 {\em
  Nature\/} {\bf 463} 644--647

\bibitem{Engel2010}
Panitchayangkoon G, Hayes D, Fransted K~A, Caram J~R, Harel E, Wen J,
  Blankenship R~E and Engel G~S 2010 Long-lived quantum coherence in
  photosynthetic complexes at physiological temperature arXiv:1001.5108

\bibitem{ColliniScholes2009Sci}
Collini E and Scholes G~D 2009 {\em Science\/} {\bf 323} 369--373

\bibitem{ColliniScholes2009JPCA}
Collini E and Scholes G~D 2009 {\em J. Phys. Chem.\/} A {\bf 113} 4223--4241

\bibitem{ChengFleming2009ARPC}
Cheng Y~C and Fleming G~R 2009 {\em Ann. Rev. Phys. Chem.\/} {\bf 60} 241--262

\bibitem{BeljonneCurutchetScholes2009}
Beljonne D, Curutchet C, Scholes G~D and Silbey R~J 2009 {\em J. Phys. Chem.\/}
  B {\bf 113} 6583--6599

\bibitem{AbramaviciusPalmieriMukamel2009}
Abramavicius D, Palmieri B, Voronine D~V, \v{S}anda F and Mukamel S 2009 {\em
  Chem. Rev.\/} {\bf 109} 2350--2408

\bibitem{ArndtJuffmannVedral2009}
Arndt M, Juffmann T and Vedral V 2009 {\em HFSP J.\/} {\bf 3} 386--400

\bibitem{ChengFleming2008JPCA}
Cheng Y~C and Fleming G~R 2008 {\em J. Phys. Chem.\/} A {\bf 112} 4254--4260

\bibitem{IshizakiFleming2009PNAS}
Ishizaki A and Fleming G~R 2009 {\em Proc. Natl. Acad. Sci.\/} {\bf 106}
  17255--17260

\bibitem{IshizakiFleming2009a}
Ishizaki A and Fleming G~R 2009 {\em J. Chem. Phys.\/} {\bf 130} 234110

\bibitem{IshizakiFleming2009b}
Ishizaki A and Fleming G~R 2009 {\em J. Chem. Phys.\/} {\bf 130} 234111

\bibitem{OlayaCastro2008PRB}
Olaya-Castro A, Lee C~F, Olsen F~F and Johnson N~F 2008 {\em Phys. Rev.\/} B
  {\bf 78} 085115

\bibitem{YuBerdingWang2008}
Yu Z~G, Berding M~A and Wang H 2008 {\em Phys. Rev.\/} E {\bf 78} 050902

\bibitem{JangCheng2008JCP}
Jang S, Cheng Y~C, Reichman D~R and Eaves J~D 2008 {\em J. Chem. Phys.\/} {\bf
  129} 101104

\bibitem{Jang2009JCP}
Jang S 2009 {\em J. Chem. Phys.\/} {\bf 131} 164101

\bibitem{MohseniRebentrost2008JCP}
Mohseni M, Rebentrost P, Lloyd S and Aspuru-Guzik A 2008 {\em J. Chem. Phys.\/}
  {\bf 129} 174106

\bibitem{RebentrostMohseni2009NJP}
Rebentrost P, Mohseni M, Kassal I, Lloyd S and Aspuru-Guzik A 2009 {\em New J.
  Phys.\/} {\bf 11} 033003

\bibitem{RebentrostMohseni2009JPCB}
Rebentrost P, Mohseni M and Aspuru-Guzik A 2009 {\em J. Phys. Chem.\/} B {\bf
  113} 9942--9947

\bibitem{RebentrostChakraborty2009JCP}
Rebentrost P, Chakraborty R and Aspuru-Guzik A 2009 {\em J. Chem. Phys.\/} {\bf
  131} 184102

\bibitem{PlenioHuelga2008}
Plenio M~B and Huelga S~F 2008 {\em New J. Phys.\/} {\bf 10} 113019

\bibitem{CarusoChinPlenio2009}
Caruso F, Chin A~W, Datta A, Huelga S~F and Plenio M~B 2009 {\em J. Chem.
  Phys.\/} {\bf 131} 105106

\bibitem{PalmieriAbramavicius2009}
Palmieri B, Abramavicius D and Mukamel S 2009 {\em J. Chem. Phys.\/} {\bf 130}
  204512

\bibitem{ThorwartEckel2009}
Thorwart M, Eckel J, Reina J, Nalbach P and Weiss S 2009 {\em Chem. Phys.
  Lett.\/} {\bf 478} 234--237

\bibitem{Nazir2009PRL}
Nazir A 2009 {\em Phys. Rev. Lett.\/} {\bf 103} 146404

\bibitem{CarusoChinDatta2009}
Caruso F, Chin A~W, Datta A, Huelga S~F and Plenio M~B 2009 Entanglement and
  entangling power of the dynamics in light-harvesting complexes
  arXiv:0912.0122

\bibitem{BradlerWilde2009}
Br\'{a}dler K, Wilde M~M, Vinjanampathy S and Uskov D~B 2009 Identifying the
  quantum correlations in light-harvesting complexes arXiv:0912.5112

\bibitem{PerdomoVogt2010}
Perdomo A, Vogt L, Najmaie A and Aspuru-Guzik A 2010 Engineering directed
  excitonic energy transfer arXiv:1001.2602

\bibitem{FassioliOlayaCastro2010}
Fassioli F and Olaya-Castro A 2010 Distribution of entanglement in
  light-harvesting complexes and their quantum efficiency arXiv:1003.3610

\bibitem{LuRabitz1995PRA}
Lu Z~M and Rabitz H 1995 {\em Phys. Rev.\/} A {\bf 52} 1961--1967

\bibitem{LuRabitz1995JPC}
Lu Z~M and Rabitz H 1995 {\em J. Phys. Chem.\/} {\bf 99} 13731--13735

\bibitem{ZhuRabitz1999b}
Zhu W~S and Rabitz H 1999 {\em J. Chem. Phys.\/} {\bf 111} 472--480

\bibitem{ZhuRabitz1999c}
Zhu W~S and Rabitz H 1999 {\em J. Phys. Chem.\/} A {\bf 103} 10187--10193

\bibitem{BrifRabitz2000}
Brif C and Rabitz H 2000 {\em J. Phys. B: At. Mol. Opt. Phys.\/} {\bf 33}
  L519--L525

\bibitem{KurtzRabitz2002}
Kurtz L, Rabitz H and de~Vivie-Riedle R 2002 {\em Phys. Rev.\/} A {\bf 65}
  032514

\bibitem{GeremiaRabitz2001JCP}
Geremia J~M and Rabitz H 2001 {\em J. Chem. Phys.\/} {\bf 115} 8899--8912

\bibitem{GeremiaRabitz2004PRA}
Geremia J~M and Rabitz H~A 2004 {\em Phys. Rev.\/} A {\bf 70} 023804

\bibitem{GeremiaRabitz2002PRL}
Geremia J~M and Rabitz H 2002 {\em Phys. Rev. Lett.\/} {\bf 89} 263902

\bibitem{GeremiaRabitz2003JCP}
Geremia J~M and Rabitz H 2003 {\em J. Chem. Phys.\/} {\bf 118} 5369--5382

\end{thebibliography}
\bibliographystyle{iopart-num}

\end{document}